\documentclass[acmsmall]{acmart}

\usepackage{csquotes}
\usepackage[export]{adjustbox}
\usepackage{caption}
\usepackage{subcaption}
\usepackage{tabularx}
\usepackage{extarrows}

\newtheorem{Proposition}{Proposition}
\usepackage{url} % simple URL typesetting

\def\BibTeX{{\rm B\kern-.05em{\sc i\kern-.025em b}\kern-.08emT\kern-.1667em\lower.7ex\hbox{E}\kern-.125emX}}
    
\setcopyright{acmlicensed}
\acmJournal{TWEB}
\acmYear{2019}

\begin{document}

%
% The "title" command has an optional parameter, allowing the author to define a "short title" to be used in page headers.
\title{Decoupled Variational Embedding for Signed Directed Networks}

%
% The "author" command and its associated commands are used to define the authors and their affiliations.
% Of note is the shared affiliation of the first two authors, and the "authornote" and "authornotemark" commands
% used to denote shared contribution to the research.
\author{Xu Chen}
\affiliation{%
  \institution{Cooperative Medianet Innovation Center, Shanghai Jiao Tong University}
  \streetaddress{800 Dongchuan Rd}
  \city{Shanghai}
  \country{China}}
\email{xuchen2016@sjtu.edu.cn}

\author{Jiangchao Yao}
\affiliation{%
  \institution{Cooperative Medianet Innovation Center, Shanghai Jiao Tong University}
  \city{Shanghai}
  \country{China}}
\email{Sunarker@sjtu.edu.cn}

\author{Maosen Li}
\affiliation{%
  \institution{Cooperative Medianet Innovation Center, Shanghai Jiao Tong University}
  \city{Shanghai}
  \country{China}}
\email{maosen_li@sjtu.edu.cn}

\author{Ya Zhang}
\authornote{Prof. Ya Zhang is the corresponding author.}
\affiliation{%
  \institution{Cooperative Medianet Innovation Center, Shanghai Jiao Tong University}
  \city{Shanghai}
  \country{China}}
\email{ya_zhang@sjtu.edu.cn}

\author{Yanfeng Wang}
\affiliation{%
  \institution{Cooperative Medianet Innovation Center, Shanghai Jiao Tong University}
  \city{Shanghai}
  \country{China}}
\email{wangyanfeng@sjtu.edu.cn}

%
% By default, the full list of authors will be used in the page headers. Often, this list is too long, and will overlap
% other information printed in the page headers. This command allows the author to define a more concise list
% of authors' names for this purpose.
\renewcommand{\shortauthors}{Xu Chen, et al.}

%
% The abstract is a short summary of the work to be presented in the article.
\begin{abstract}
Node representation learning for signed directed networks has received considerable attention in many real-world applications such as link sign prediction, node classification and node recommendation.
The challenge lies in how to adequately encode the complex topological information of the networks.
Recent studies mainly focus on preserving the \emph{first-order} network topology which indicates the closeness relationships of nodes.
However, these methods generally fail to capture the \emph{high-order} topology which indicates the local structures of nodes and serves as an essential characteristic of the network topology.
In addition, for the \emph{first-order} topology, the additional value of non-existent links is largely ignored. 
In this paper, we propose to learn more representative node embeddings by simultaneously capturing the \emph{first-order} and \emph{high-order} topology in signed directed networks.
In particular, we reformulate the representation learning problem on signed directed networks from a variational auto-encoding perspective and further develop a decoupled variational embedding (DVE) method. 
DVE leverages a specially designed auto-encoder structure to capture both the \emph{first-order} and \emph{high-order} topology of signed directed networks, and thus learns more representative node embeddings.
Extensive experiments are conducted on three widely used real-world datasets. Comprehensive results on both link sign prediction and node recommendation task demonstrate the effectiveness of DVE. Qualitative results and analysis are also given to provide a better understanding of DVE. Codes are available online:~\url{https://github.com/xuChenSJTU/DVE-master}
\end{abstract}

%
% The code below is generated by the tool at http://dl.acm.org/ccs.cfm.
% Please copy and paste the code instead of the example below.
%


%
% End generated code
%


%
% A "teaser" image appears between the author and affiliation information and the body 
% of the document, and typically spans the page. 
%%\begin{teaserfigure}
%%  \includegraphics[width=\textwidth]{sampleteaser}
%%  \caption{Seattle Mariners at Spring Training, 2010.}
%%  \Description{Enjoying the baseball game from the third-base seats. Ichiro Suzuki preparing to bat.}
%%  \label{fig:teaser}
%%\end{teaserfigure}

%
% This command processes the author and affiliation and title information and builds
% the first part of the formatted document.
\maketitle

\section{Introduction}
In recent years, learning node representation on graphs, which is called network embedding or graph embedding, has drawn great interest among various academic topics. Study on this field benefits many learning paradigms, such as semi-supervised learning~\cite{kipf2016semi, hamilton2017representation} and relational inference~\cite{battaglia2018relational, chen2018variational, kipf2018neural} as well as some practical data mining tasks, such as link prediction~\cite{liben2007link,Gaeta:2018:MID:3176641.3160000}, community detection~\cite{dong2015coupledlp,papadopoulos2012community,Zhang:2016:DSP:2870642.2846102} and node classification~\cite{bhagat2011node,wang2016linked}.
% \footnote{Rewrite this paragraph. It seems that you merely mention the real-world scenarios but some professional tasks. It will be hard for readers to understand its practical value.}

\begin{figure*}[h!]
\centering
\begin{minipage}[t]{0.30\textwidth}
\centering
\includegraphics[width=\textwidth]{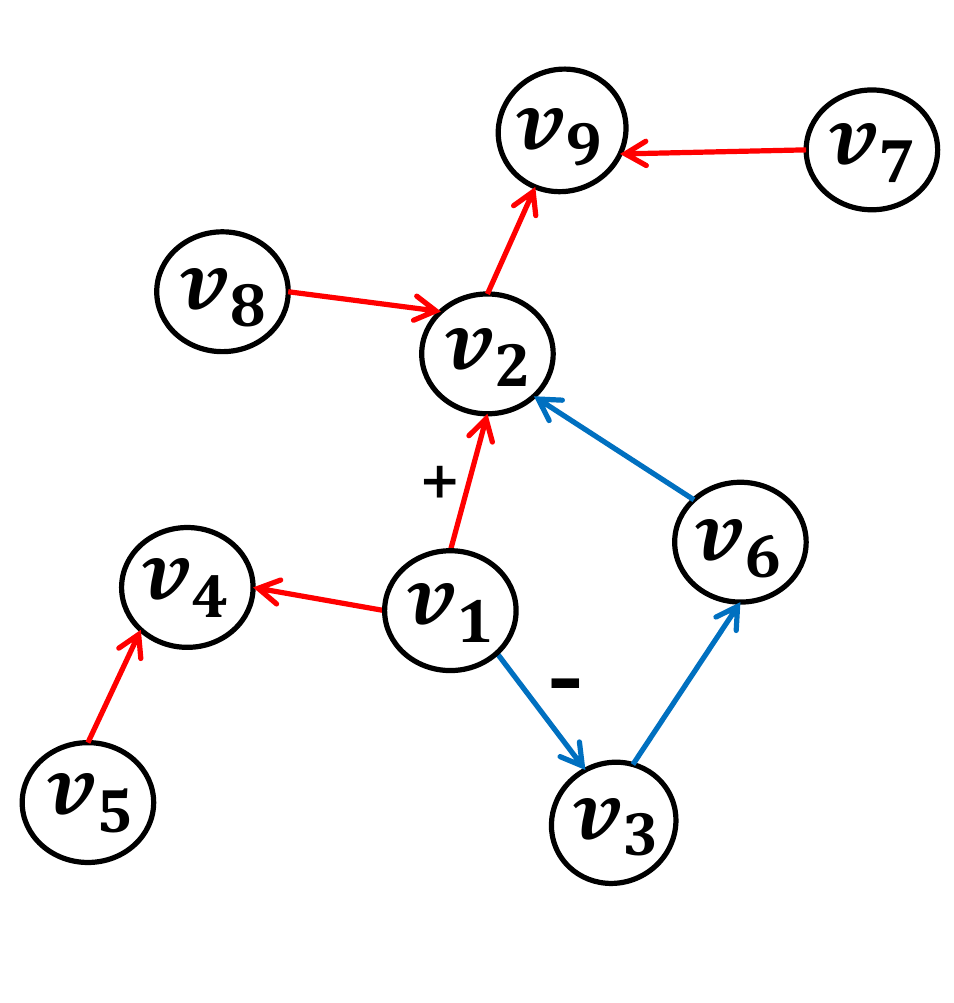}
\captionsetup{justification=centering}
\caption*{(a) One signed directed network example $\mathcal{G}_{e}$}
\end{minipage}
\begin{minipage}[t]{0.30\textwidth}
\centering
\includegraphics[width=\textwidth]{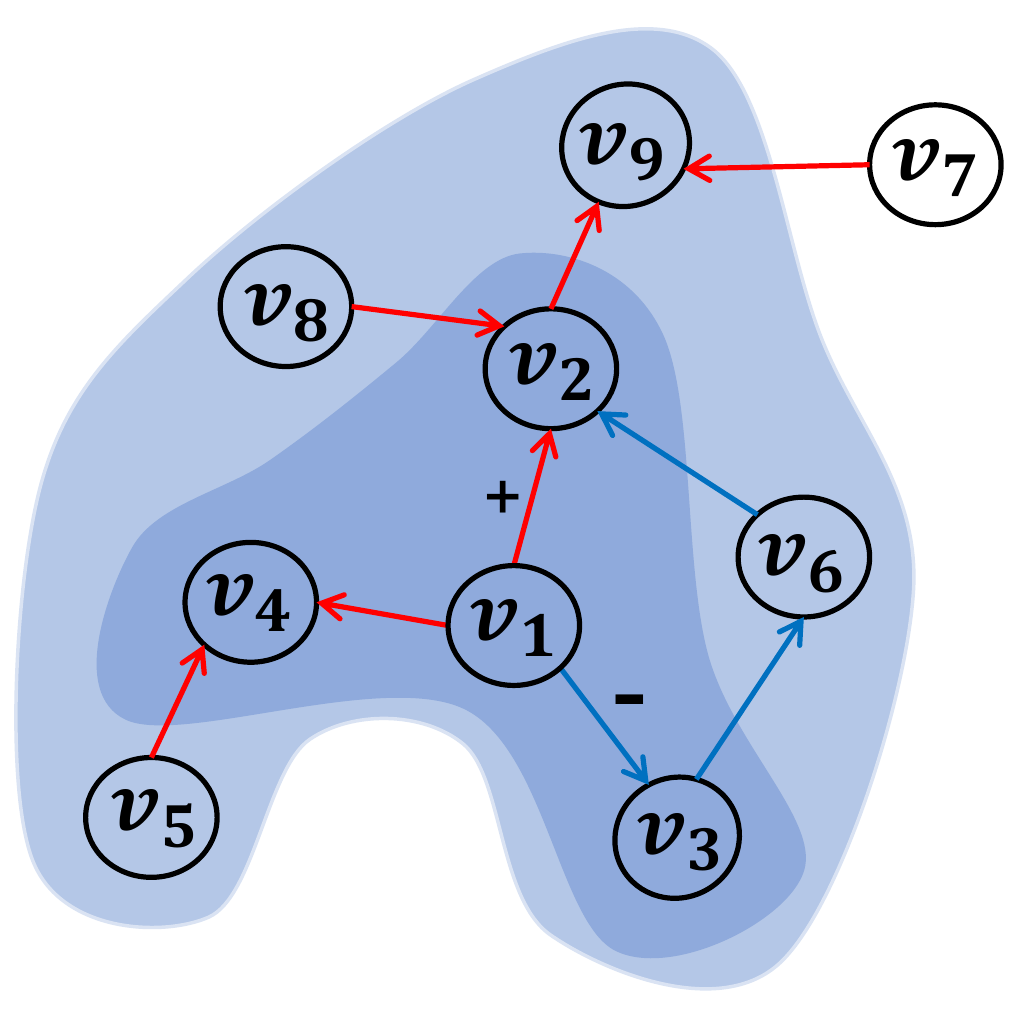}
\captionsetup{justification=centering}
\caption*{(b) The \emph{high-order} topology}
\end{minipage} 
\begin{minipage}[t]{0.30\textwidth}
\centering
\includegraphics[width=\textwidth]{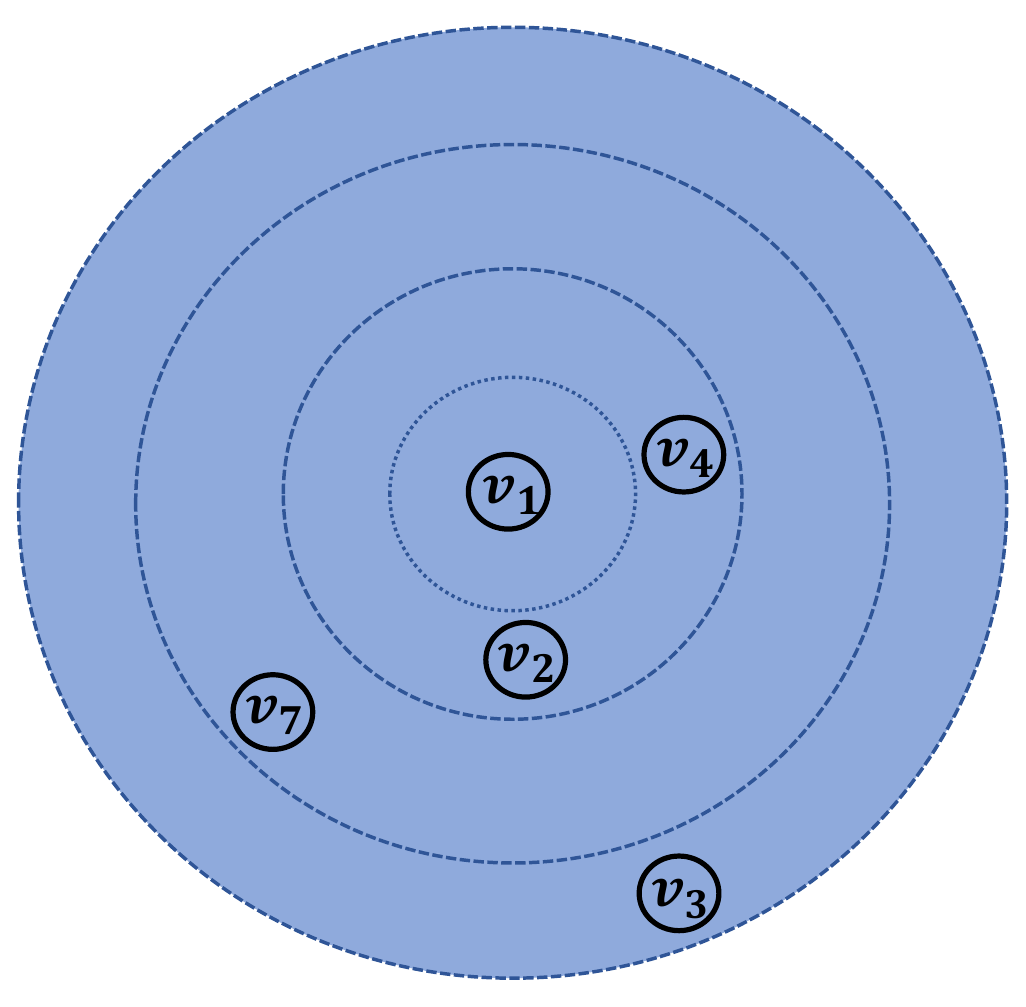}
\captionsetup{justification=centering}
\caption*{(c) The \emph{first-order} topology}
\end{minipage}
\caption{An example to illustrate the \emph{high-order} topology and \emph{first-order} topology in signed directed networks. Red arrows mean positive directed edges and blue arrows indicate negative directed edges. (a) is a signed directed network example $\mathcal{G}_{e}$. (b) indicates the \emph{high-order} topology of node $v_{1}$ in $\mathcal{G}_{e}$, namely the local structures of node $v_{1}$. Different depth of colored shades represent different local structure orders for node $v_{1}$. (c) shows the \emph{first-order} topology of node $v_{1}$ in $\mathcal{G}_{e}$, namely the closeness relationships between node $v_{1}$ and its directly linked neighbors. The concentric circles with node $v_{1}$ as the center indicate the closeness between the center node $v_{1}$ and positively linked nodes $v_{2},v_{4}$, non-linked node $v_{7}$, negatively linked node $v_{3}$.}
\label{figure:motivation}
\end{figure*}

Many social networks have both directed and signed (positive and negative) links, such as Epinions\footnote{http://www.epinions.com/?sb=1} and Slashdot\footnote{https://slashdot.org/}, which are called signed directed networks.
Negative links in social networks hold opposite semantic meaning and contain additional information~\cite{kunegis2013added,Cacheda:2018:CPU:3176641.3157059,Victor:2013:ETR:2460383.2460385} that helps many tasks, e.g. link sign prediction and node classification. In addition, link direction indicates the asymmetric relationship between two nodes, which is important for node (i.e. user) recommendation in social networks~\cite{ou2016asymmetric,Aiello:2012:FPH:2180861.2180866}. For example, stars may not follow common people while common people tend to follow stars.
% \footnote{You give two examples about the value of signed directed links. One is enough. Merge or drop.} Two examples are introduced to illustrate sign and direction respectively. Neither merge or drop is a good choice.
The key of representation learning on signed directed networks lies in how to encode the complex topological information into low-dimensional embeddings for nodes. In particular, the topological information is composed of both the \emph{high-order} and the \emph{first-order} topology.
The \emph{high-order} topology indicates the local structures since it is generated by information propagation of a node's neighbors and the \emph{first-order} topology indicates the closeness relationships between a node and its directly linked neighbors. Both the \emph{high-order} and \emph{first-order} topology are intrinsic characteristics of signed directed networks. To make it more clearly, we give an example in Figure~\ref{figure:motivation}.

However, existing embedding methods fail to capture both the \emph{first-order} and \emph{high-order} topology for signed directed networks. Firstly, the majority of them concentrate on how to mine the \emph{first-order} topology, namely preserving the closeness relationships of nodes. For example, MF~\cite{hsieh2012low} performs matrix factorization on the signed directed adjacent matrix to learn low-dimensional embeddings for nodes. SNE~\cite{yuan2017sne} exploits random walk and log-bilinear model to learn node embeddings with signed links. SiNE~\cite{wang2017signed} learns node embeddings through a deep neural network model based on social theory. These works model the closeness relationships in restrictive distance metrics or usually ignore the additional value of non-existent links. Secondly, the \emph{high-order} topology, indicating the local structures of nodes, is difficult to be extracted in signed directed networks, because it is coupled with signed directed links. Different signs and directions have distinctive information propagation influence. SNE~\cite{yuan2017sne} with random walk applies homophily effects on different signs and fail to capture the \emph{high-order} pattern in signed directed networks. How to encode both the intrinsic \emph{high-order} and \emph{first-order} topological information is an important problem for representation learning on signed directed networks.

% \begin{figure*}[h!]
% \centering
% \includegraphics[width=10cm]{images/motivation.png}
% \caption{Left: an example to demonstrate the local structure pattern with colored shades in signed directed network. Different depth of colored shades represent different neighborhood orders for node $v_{1}$. Right: an instance to illustrate the general closeness relationships among nodes in signed directed network. The concentric circles with node $v_{1}$ as the center indicate the closeness between the center node $v_{1}$ and positively linked nodes $v_{2},v_{4}$, non-connected node $v_{7}$, negatively linked node $v_{3}$.}
% \label{figure:motivation}
% \end{figure*}

\begin{figure*}[h!]
\centering
\includegraphics[width=10cm]{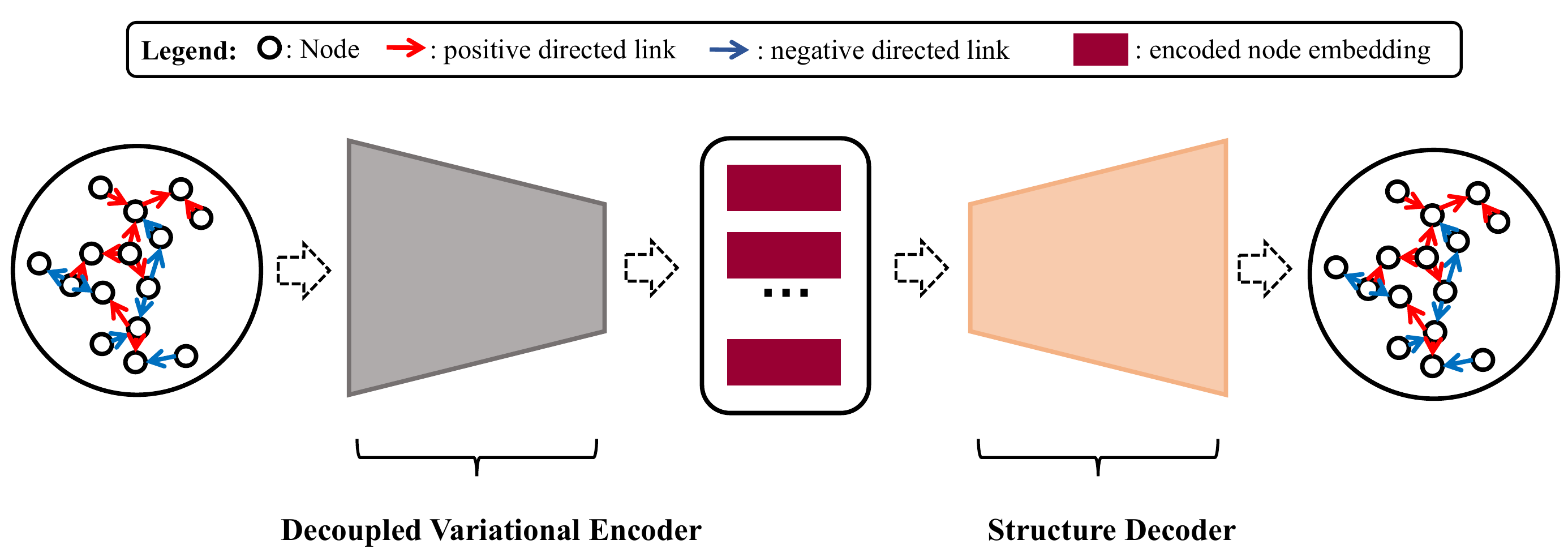}
\caption{General architecture of DVE.}
\label{figure:genral_architecture}
% \vspace{-10pt}
\end{figure*}
In this paper, we propose to learn more representative node embeddings by simultaneously capturing the \emph{first-order} and \emph{high-order} topology in signed directed networks.
In particular, we reformulate the representation learning on signed directed networks from a variational auto-encoding perspective and further propose a decoupled variational embedding (DVE).
DVE is a specially designed auto-encoder structure which contains a decoupled variational encoder and a structure decoder. The general architecture of DVE is shown in Figure~\ref{figure:genral_architecture}.
In the decoupled variational encoder, representation of a node is decoupled into source node embeddings and target node embeddings according to link direction. 
Both the source node embeddings and the target node embeddings contain the local structure pattern that is extracted by graph convolution on the decoupled positive and negative graph according to link sign.
The structured decoder is formulated as a novel Balance Pair-wise Ranking (BPWR) loss, which is developed from the Extended Structural Balance Theory~\cite{qian2013extended, Qian:2014:FTD:2639948.2628438}.
BPWR extracts the closeness relationships among positive links, negative links and non-existent links in a Bayesian personalized ranking manner, as well as refines embeddings learned from the former encoder. The auto-encoding formulation encourages DVE to preserve the network topology in an end-to-end manner. 
In brief, our contributions are summarized as follows:
\begin{itemize}
\item We propose a variational auto-encoding based method named DVE to learn more representative node embeddings for signed directed networks. To the best of our knowledge, DVE is the first model that simultaneously models both the \emph{first-order} and the \emph{high-order} topology in signed directed networks;
\item Developed from Extended Structural Balance Theory, we develop a novel Balance Pair-wise Ranking (BPWR) loss which also works as the decoder of DVE. BPWR adequately mines the closeness relationships of nodes indicated by positive links, negative links and non-existent links in a ranking form rather than existing point-wise or distance-level metrics; 
\item Extensive experiments are conducted on three widely used real-world datasets. The superior performance of DVE compared with recent competitive baselines illustrates the effectiveness of DVE both quantitatively and qualitatively.
\end{itemize}

The rest of this paper is organized as follows. Section 2 introduces the related works. Section 3 gives the problem definition and a concise introduction of graph convolutional networks. Section 4 demonstrates the details about the proposed DVE. Section 5 provides the experiments and analysis on both link sign prediction task and node recommendation task, as well as the qualitative results. Finally, conclusion and future work are given in Section 6.

\section{Related Work}
This section is the related work part where network embedding methods for both unsigned undirected networks and signed directed networks are introduced. Furthermore, some works of variational auto-encoding are introduced to better illustrate the proposed model in Section~\ref{sec:model}.
\subsection{Network Embedding}
Network embedding arises as one hot research topic to learn representative node embeddings for a given network.
It benefits many network analysis tasks such as link prediction~\cite{liben2007link,Gaeta:2018:MID:3176641.3160000}, node classification~\cite{bhagat2011node,wang2016linked,shen2018discrete}, online voting~\cite{wang2017joint}
% community detection~\cite{dong2015coupledlp,papadopoulos2012community} 
and sentiment analysis~\cite{wang2018shine}. Various methods have been proposed for network embedding. For example, spectral analysis is performed on Laplacian matrix decomposition~\cite{belkin2002laplacian}. Similarity based node embedding methods such as Adamic/Adar and Katz are utilized in~\citet{liben2007link}. Recently, inspired by the skip-gram model for word representation in Natural Language Processing (NLP)~\cite{mikolov2013distributed}, DeepWalk~\cite{perozzi2014deepwalk} learns node embeddings from random walk sequences in social networks. LINE~\cite{tang2015line} defines first-order and second-order proximity to describe the context of a node and trains node embeddings via negative sampling. Node2Vec~\cite{grover2016node2vec} extends DeepWalk by designing a biased random walk to control the Bread First Search (BFS) and Deep First Search (DFS). Embedding methods for directed networks are studied in HOPE~\cite{ou2016asymmetric} and APP~\cite{zhou2017scalable}. Qiu et.al~\cite{qiu2018network} unified DeepWalk, LINE and Node2Vec into one matrix factorization framework. SDNE in~\cite{wang2016structural} is a semi-supervised deep model that captures the highly non-linear graph structure. 

Recently, Graph Convolutional Networks (GCN)~\cite{defferrard2016convolutional} is proposed and it analyses graph signal processing in spectral domain. Kipf et.al~\cite{kipf2016semi} simplified GCNs in~\citet{defferrard2016convolutional} into a deep learning method by stacking multiple graph convolutional layers.
Some variants of GCN have been proposed such as GAT~\cite{velivckovic2017graph} and GraphSage~\cite{hamilton2017inductive}. GAT applies multi-head attention to GCN. GraphSAGE introduces neighborhood sampling and different aggregation manners to make inductive graph convolution on large graphs. Some works also study how to accelerate GCN via importance sampling~\cite{chen2018fastgcn} and variance reduction~\cite{chen2017stochastic}. In general, GCN based methods are superior over random walk based methods both on performance and end-to-end training, which has introduced a new perspective for graph representation learning. 
% We do not introduce more about GCN here since most GCN works are targeted for homogeneous or heterogeneous graphs rather than the signed directed graphs. 
For more details about GCN, we recommend~\cite{zhang2018deep}.

\subsection{Signed Directed Networks}
The network embedding methods discussed above are designed for unsigned or undirected networks. In reality, both the existence of directed and signed (positive and negative) links in social media are ubiquitous. The negative links have been proven to have distinct properties and added value over positive links~\cite{leskovec2010predicting,tang2015negative}.
Several works have studied how to distinctly model the positive and negative links in signed directed networks. Degree based features like the number of positive-incoming and negative-incoming links are explored in~\citet{leskovec2010predicting}. While these hand-crafted features are limited and not capable in many situations. Instead, in~\citet{kunegis2010spectral}, spectral analysis is extended for signed network. Matrix Factorization (MF)~\cite{hsieh2012low} is also adopted to learn low-dimensional embeddings for signed directed networks. To reduce the computation burden of matrix decomposition, a specific aggregation manner for learning node embeddings in signed networks is proposed in~\citet{derr2018signed}. It follows the principle that the enemy of a friend is an enemy and the enemy of an enemy is a friend, which extends the positive and negative neighbors for a specific node. 

Due to the superior representation learning ability of deep learning, researchers attempt to use deep learning techniques to learn more representative node embeddings on networks. A deep learning framework for signed network named SiNE is proposed in~\citet{wang2017signed}, where the objective function is guided by social theory. Although the framework leverages non-linearity to learn node representation, it does not model link direction which is an important factor for some asymmetric tasks. SNE is proposed in~\citet{yuan2017sne} and log-bilinear model is extended to support sign and direction modeling. SNE trains node embeddings based on a uniform random walk and node context rather than social theory. However, random walk in SNE applies homophily effects on different signed links and fail to capture the local structures in signed directed networks, as well as does not support end-to-end training. SIDE~\cite{kim2018side} is another random walk based method based on social balance theory~\cite{cartwright1956structural}. SNEA~\cite{wang2017attributed} exploits both the network structures and node attributes simultaneously for network embedding on attributed signed networks. Specifically, a margin ranking loss is proposed in SNEA. However, the margin ranking loss is non-smooth and difficult to be optimized by gradient based algorithms. Bayesian Personalized Ranking~\cite{rendle2009bpr} derived from maximizing the posterior of observations is also a ranking method, which has some advantages such as flexibility and easy optimization by gradient based algorithms. It has been successfully applied in many areas such as recommendation~\cite{rendle2010pairwise,liu2017deepstyle,he2016vbpr}. Despite the great potential of BPR to model the pairwise relationships, it has not been explored in signed directed networks. In this paper, based on the Extended Structural Balance Theory and BPR, we develop an objective function called Balance Pair-wise Ranking (BPWR) to mine the \emph{first-order} topology in signed directed networks.

From the above, we see that most existing works focus on capturing the \emph{first-order} topology, namely learning the closeness relationships of nodes. From this aspect, these methods~\cite{hsieh2012low,yuan2017sne,wang2017signed,derr2018signed} extract the \emph{first-order} topology in restrictive distance metrics and some of them ignore the additional value of non-existent links.
Besides, although some methods have introduced random walk~\cite{yuan2017sne}, they fail to capture the \emph{high-order} topology for signed directed networks since they apply homophily effects on different signs. The proposed DVE reformulates the representation learning problem on signed directed networks from a variational auto-encoding perspective and simultaneously models the \emph{first-order} and \emph{high-order} topology. 
% Furthermore, BPWR loss of DVE has better ability to mine the \emph{first-order} topology, because it works in a Bayesian Personalized Ranking scheme and can be easily 
% optimized by gradient based algorithms. 

\subsection{Variational Auto-encoding}
Variational auto-encoding (VAE) has attracted enormous attention in recent years and has become one of the most popular techniques in unsupervised representation learning~\cite{doersch2016tutorial}. VAE theory is appealing since it is built based on standard Bayes theory and meanwhile can be trained with stochastic gradient descent. VAE first emerged in~\citet{kingma2013auto} where the authors aim to perform efficient inference and learning in directed probabilistic graphic models even with the intractable posteriors or large datasets. In~\citet{kingma2013auto}, the authors first derive the variational \textit{evidence lower bound} (\textit{ELBO}) of the marginal log-likelihood of observed datapoints. Then a reparameterization trick is applied to approximate the intractable posteriors, which also enables VAE to be straightforwardly optimized using standard stochastic gradient based methods. 

After~\cite{kingma2013auto}, an enormous amounts of researchers have studied VAE from various perspectives, which advances the whole community of variational auto-encoding. Recent advances of VAE theory could be categorized into two aspects. First, from more expressive likelihood aspect, the standard VAE~\cite{kingma2013auto} makes an assumption that the likelihoods factorizes over dimensions, which may cause poor approximation for tasks involving images. Thereby, a sequential auto-encoding framework is proposed in DRAW~\cite{gregor2015draw} to perform image generation. Also, Gulrajani et.al proposed to model the dependencies within an image and further developed an auto-regressive decoder in VAE for fine-grained image generation. Moreover, there are also some works trying to deriving more expressive likelihoods from information theory such as~\cite{zhao2017infovae,zheng2018degeneration,zheng2019understanding}. Second, from more expressive posterior aspect, the main idea is that the standard VAE uses mean field approach, which lacks expressiveness for modeling complex posteriors. Thus, IWAE~\cite{burda2015importance} weights the samples in the posterior approximation process, which increases the model's flexibility to capture complex posteriors. Also, normalizing flows~\cite{jimenez2015variational} is introduced in VAE to transform a simple approximate posterior into a more expressive one through multiple successive invertible transformations. Apart from the advances in VAE theory, there are various applications involving VAE such as hand-written digits~\cite{salimans2015markov}, segmentation~\cite{sohn2015learning} and graph representation learning~\cite{kipf2016variational}.
Since variational auto-encoding is a huge topic and we mainly concentrate on signed directed networks, we cannot cover comprehensively here. For more details about variational auto-encoding, we recommend~\cite{doersch2016tutorial,zhao2017towards} 

\section{Preliminary}
In this section, we give the problem definition of node representation learning on signed directed networks, as well as an introduction of graph convolutional networks (GCNs). The introduction of GCNs illustrates how the signal on graphs are convolved and builds a foundation to demonstrate the propose model. 
\subsection{Problem Definition}
A signed directed network is defined as $\mathcal{G}=(\mathcal{V}, \mathcal{E}^{p}, \mathcal{E}^{n})$, where $\mathcal{V}$ is the set of all nodes and $\mathcal{E}^{p}$ ($\mathcal{E}^{n}$) represents positive (negative) links. Let $\mathcal{E}=\mathcal{E}^{p}\bigcup \mathcal{E}^{n}$ be the observed links in $\mathcal{G}$. For each link $e\in \mathcal{E}$, it is represented as $e_{u\rightarrow v}=(u,v,\epsilon_{u\rightarrow v})$, where $u\rightarrow v$ denotes the direction from source node $u$ to target node $v$. And $\epsilon_{u\rightarrow v}$ indicates the sign value of link $e_{u\rightarrow v}$, i.e. $\epsilon_{u\rightarrow v}=1$ if $e_{u\rightarrow v}\in \mathcal{E}^{p}$ or $\epsilon_{u\rightarrow v}=-1$ if $e_{u\rightarrow v}\in \mathcal{E}^{n}$. When the nodes $\mathcal{V}$ have raw features, the node feature matrix of $\mathcal{G}$ is represented as $X\in \mathbb{R}^{N\times F}$, where $F$ indicates the raw feature dimension. Given $\mathcal{G}$, the objective of node representation learning on signed directed networks is to embed nodes into low-dimensional embeddings $Z\in \mathbb{R}^{N\times d}$ that facilitate downstream tasks such as node recommendation, node classification and link prediction. 
% In the real scenario of signed directed networks, link direction $u\to v$ means the asymmetric relationship between two nodes, like follower and followee in social media. Link sign $\epsilon_{u\rightarrow v}$ indicates the opposite semantic meaning such as trust and distrust in social networks. Our goal is to learn representative embeddings for nodes in signed directed networks. 
The notations in this paper are summarized in Table~\ref{table:notations}.
\begin{table}[]
\caption{Notations in this paper.}
\label{table:notations}
\begin{tabular}{ll}
\hline
Notation & Description \\ \hline
$\mathcal{G}$        & signed directed graph            \\
$\mathcal{V}$        & node set of $\mathcal{G}$            \\
$\mathcal{E}^{p}$             & positive link set of $\mathcal{G}$   \\
$\mathcal{E}^{n}$            & negative link set of $\mathcal{G}$  \\
$\mathcal{E}$                  & observed links in $\mathcal{G}$           \\ 
$L$    & the symmetric normalized Laplacian matrix of unsigned undirected graph \\
$I_{N}$        & an identity matrix of size $N$ \\
$d$        & the latent embedding dimension \\ 
$\phi_{s}$          & parameters of the source node encoder               \\
$\phi_{t}$              & parameters of the target node encoder              \\
$Z_{s}$     & source node embeddings          \\
$Z_{t}$    & target node embeddings   \\
$A^{p}$    & adjacent matrix of the undirected positive graph   \\
$A^{n}$    & adjacent matrix of the undirected negative graph   \\
$X$    & the raw feature matrix of nodes   \\
$L_{KL}^{s}$    & KL divergence loss for the source node encoder \\
$L_{BPWR}$    & Balance Pair-wise Ranking (BPWR) loss as the structure decoder  \\
$L_{DVE}$    & loss of DVE method  \\
\hline
\end{tabular}
\end{table}

\subsection{Graph Convolutional Networks}
% \footnote{Can we remove this subsection? Is it tightly related to your description below or will it simplify your description below?} It should be preserved, or it would seem too hard to understand what the inference procedure in our method in case reviewers do not clearly know about this.
Graph Convolutional Networks (GCN) is one essential ingredient for DVE, thus we give a concise introduction about it. GCN is one type of neural network that learns superior node representations by capturing local structures of nodes. GCNs can be regarded as a feature extractor working on graphs. It can be equipped with a variety of models and applied in a variety of tasks~\cite{kipf2016semi,velickovic2017graph,hamilton2017inductive}.

GCN is firstly derived from the spectral convolution on graphs that is defined as the multiplication of a signal $x\in \mathbb{R}^{N}$ with a parameterized filter $g_{\theta}$ in the Fourier domain. Let $\star$ be the convolution operation, the convolution of GCN can be expressed as:
\begin{equation}
    g_{\theta}\star x=Ug_{\theta}(\Lambda)U^{T}x
\label{eq:decomposition}
\end{equation}
where $U$ is the eigenvector matrix and $\Lambda$ is the eigenvalue matrix of the graph Laplacian $L=I_{N}-D^{-\frac{1}{2}}AD^{-\frac{1}{2}}=U\Lambda U^{T}$. And $U^{T}x$ indicates the graph Fourier transform of $x$. According to~\citet{defferrard2016convolutional}, a polynomial filter is usually taken as $ g_{\theta}(\Lambda)=\sum_{k=0}^{K}\theta_{k}\Lambda^{k}$.
% \begin{equation}
%     g_{\theta}(\Lambda)=\sum_{k=0}^{K}\theta_{k}\Lambda^{k}
% \label{eq:multi_polynomial}
% \end{equation}

While the convolution filter defined in Eq.~\ref{eq:decomposition} involves the eigen-decomposition of $L$ and might be computationally expensive for large graphs.  To circumvent this problem, according to~\citet{defferrard2016convolutional}, $g_{\theta}(\Lambda)$ with the polynomial filter can be well-approximated by a truncated expansion in terms of Chebshev polynomials $T_{k}(x)$ up to $K^{th}$ order:
\begin{equation}
\label{eq:approx}
    g_{\theta}(\Lambda)=\sum_{k=0}^{K}\theta_{k}\Lambda^{k}\approx \sum_{k=0}^{K}\theta_{k}T_{k}(\widetilde{\Lambda})
\end{equation}
where $\widetilde{\Lambda}=\frac{2}{\lambda_{max}}\Lambda-I_{N}$ is a rescaled version of $\Lambda$ and $\lambda_{max}$ indicates the largest eigenvalue of $L$. The Chebshev polynomials are recursively defined as $T_{k}(x)=2xT_{k-1}(x)-T_{k-2}(x)$ with $T_{0}(x)=1$ and $T_{1}(x)=x$. Then taking Eq.~\ref{eq:approx} into consideration, Eq.~\ref{eq:decomposition} can be written as:
\begin{equation}
\label{eq:deffered}
    g_{\theta}\star x\approx \sum_{k=0}^{K}\theta_{k} T_{k}(\widetilde{L})x
\end{equation}
where $\widetilde{L}=\frac{2}{\lambda_{max}}L-I_{N}$. Eq.~\ref{eq:deffered} is also called as $K$-localized convolution on graphs since it is a $K$-th order polynomial in the Laplacian.

Inspired by the idea that high-order convolutions can be built by stacking multiple convolutional layers~\cite{karpathy2016cs231n}, Kipf et.al~\cite{kipf2016semi} achieves $K^{th}$ convolution by stacking multiple convolutional layers of Eq.~\ref{eq:deffered}, and each layer is followed by a point-wise non-linear function. In particular, the layer-wise convolution in Eq.~\ref{eq:deffered} is defined as $K=1$, which indicates a linear function on the graph Laplacian spectrum. Additionally, Kipf et.al~\citet{kipf2016semi} approximate $\lambda_{max}\approx 2$ by assuming the neural network can adapt to this change in scale during training, which simplifies Eq.~\ref{eq:deffered} as:
\begin{equation}
g_{\theta}\star x \approx \theta_{0}x+\theta_{1}(L-I_{N})x=\theta_{0}x-\theta_{1}D^{-\frac{1}{2}}AD^{-\frac{1}{2}}x
\end{equation}
where $\theta_{0}$ and $\theta_{1}$ are two free parameters. In practice, GCNs constrain $\theta=\theta_{0}=-\theta_{1}$ to avoid over-fitting and this leads to the following expression:
\begin{equation}
\label{eq:step1}
    g_{\theta}\star x \approx \theta(I_{N}+D^{-\frac{1}{2}}AD^{-\frac{1}{2}})x
\end{equation}
Note that $I_{N}+D^{-\frac{1}{2}}AD^{-\frac{1}{2}}$ now has eigenvalues that range in $[0,2]$. Repeating the calculation in Eq.~\ref{eq:step1} will lead to numerical instabilities and even exploding/vanishing gradients when stacking number of layers. To alleviate this problem, a renormalization trick is introduced which is: $I_{N}+D^{-\frac{1}{2}}AD^{-\frac{1}{2}}$ $\rightarrow$ $\widehat{D}^{-\frac{1}{2}} (A+I_{N}) \widehat{D}^{-\frac{1}{2}}$.

Then when giving a signal matrix $X\in \mathbb{R}^{N\times F}$ where $N$ denotes the number of samples and $F$ denotes the feature dimension, the layer-wise graph convolution in~\citet{kipf2016semi} is defined as follows:
\begin{equation}
Z^{l+1}=\widetilde{A} H^{l}\Theta^{l},~H^{l}=h(Z^{l}),~H^0=X
\label{eq:GCN}
\end{equation}
where the propagation matrix $\widetilde{A}=\widehat{D}^{-\frac{1}{2}} (A+I_{N}) \widehat{D}^{-\frac{1}{2}}$ and $\widehat{D}$ is the degree matrix of $A+I_{N}$. $H^{l}$ is the activation matrix in the $l$-th layer, whose each row is the vector representation of a node. $\Theta^{l}$ is now a matrix of filter parameters in the $l$-th layer and $h(\cdot)$ is the non-linear $ReLu$ function. $Z^{l+1}$ is the node representation of $(l+1)$-th layer. This layer-wise convolution as well connects the graph convolution operation in spectral domain to that in the vertex domain. For more details about GCN, we recommend~\cite{kipf2016semi}. 
% We do not introduce more detailed derivation here since the derivation is not our contribution.

From above we can see that GCN learns a node's representation by aggregating its neighbors which are also called the \textit{receptive field}. The \textit{receptive field} is enlarged through stacking layers like the $L$-hop in a graph. When $\widetilde{A}=I_{N}$, GCN degrades to a multi-layer perceptron (MLP) model, which does not consider the graph structures and the \textit{receptive field} of a node is just itself. In our model, we highlight that the intrinsic \emph{high-order} local structures in signed directed networks, we thus have $\widetilde{A}\neq I_{N}$, namely GCN will not degrade to a MLP here.

\section{DVE: Decoupled Variational Embedding}
\label{sec:model}
\begin{figure*}[h!]
\centering
\includegraphics[width=12.5cm]{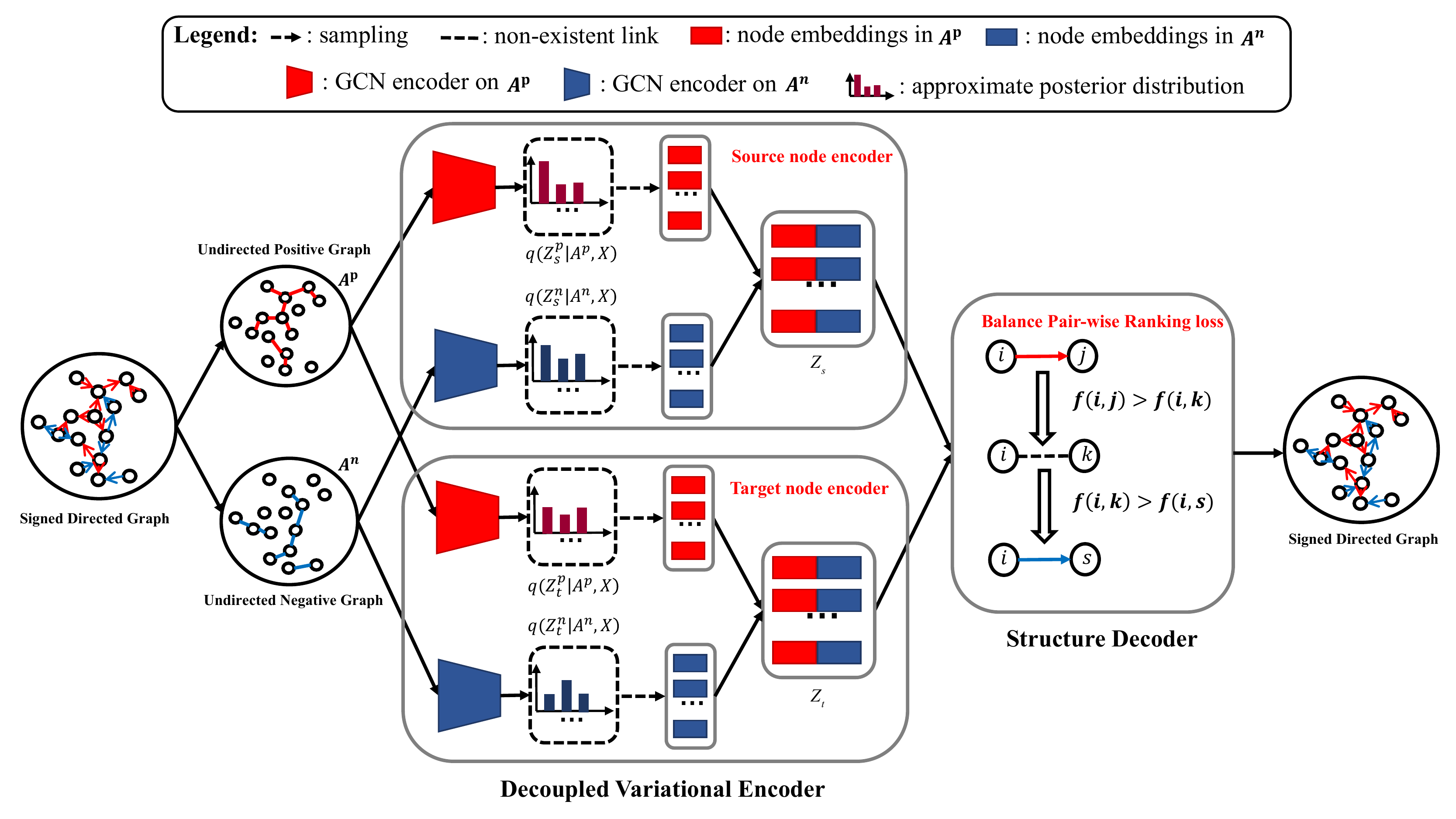}
\caption{Model architecture of DVE (best view in color). We first decouple the signed directed graph into an undirected positive graph which is indicated by $A^{p}$ and an undirected negative graph which is indicated by $A^{n}$. Then our decoupled variational encoder encodes $A^{p}$ and $A^{n}$ as the source node representation $Z_{s}$ and target node representation $Z_{t}$, respectively. Finally, $Z_{s}$ and $Z_{t}$ are used to perform the balance pair-wise ranking loss which is also the structure decoder of DVE. $i$ as source node is from $Z_{s}$ and $j,k,r$ as target nodes are from $Z_{t}$. $f(\cdot,\cdot)$ represents positive link existence score defined in our paper.}
\label{figure:model}
\end{figure*}

In this section, our decoupled variational embedding method for signed directed networks is introduced. Model architecture is shown in Figure~\ref{figure:model}. Details are illustrated as follows.

\subsection{Variational Auto-Encoding Formulation}
In this subsection, we formulate the node representation learning problem on signed directed networks from a variational auto-encoding perspective.
Link direction and sign are two key elements when describing signed directed networks. Link direction between two nodes indicates the asymmetric relationship that implies the different roles of two nodes in an interaction.
This asymmetric information is an essential factor that facilitates information propagation in signed directed networks. However, it is inappropriate to apply some GNNs methods such as~\cite{defferrard2016convolutional,kipf2016semi,xu2019graph} on directed graphs since they require a symmetric Laplacian matrix for graph convolutions. Since a node in a directed relationship may both present as the source node and the target node, we thus try to leverage the asymmetric information by decoupling node embeddings into source node embeddings $Z_{s}$ and target node embeddings $Z_{t}$. From the variational auto-encoding perspective, we assume the semantics of edges are drawn from some underlying distributions.
To clarify, we denote $\theta$ as the parameter symbol for all non-specified models and $p_{\theta}(\mathcal{E})$ as the probability density function of $\mathcal{E}$. The probability distribution function of observed signed directed links $\mathcal{E}$ is represented as $P(\mathcal{E})$ and can be written as Eq.~\ref{equation:data_joint_probability}:
\begin{equation}
P(\mathcal{E})=\int_{Z_{s},Z_{t}} p_{\psi}(\mathcal{E}|Z_{s},Z_{t})p_{\theta}(Z_{s},Z_{t})d{Z_{s}}d{Z_{t}}
\label{equation:data_joint_probability}
\end{equation}
where $Z_{s}$ and $Z_{t}$ also indicate the latent variables of source nodes and target nodes respectively. 
% The graphical model is shown in the upper right part of Figure~\ref{figure:model}. 
By modeling node embeddings through two different latent variables, the asymmetric relationship can be well captured. The true posterior distribution of $Z_{s},Z_{t}$ can be written as:
\begin{equation}
    p_{\theta}(Z_{s},Z_{t}|\mathcal{E})=\frac{p_{\psi}(\mathcal{E}|Z_{s},Z_{t})p_{\theta}(Z_{s},Z_{t})}{p_{\theta}(\mathcal{E})}
\label{eq:true_posters}
\end{equation}
% A typical scheme for learning $Z_{s}$ and $Z_{t}$ is to maximize the sum of the marginal log-likelihood of all individual datapoints in $\mathcal{E}$, namely maximizing $\log p_{\theta}(\mathcal{E})$. We thus have:
% \begin{equation}
% \label{eq:non_expectation}
%     \log p_{\theta}(\mathcal{E})=\log p_{\theta}(e_{1},e_{2},...,e_{|\mathcal{E}|})=\sum_{i=1}^{|\mathcal{E}|}\log p_{\theta}(e_{i})
% \end{equation}
% where $e_{i}$ is the $i$-th link in $\mathcal{E}$ and $|\mathcal{E}|$ is the number of all observed links for signed directed graph $\mathcal{G}$.
where the true posterior $p_{\theta}(Z_{s},Z_{t}|\mathcal{E})$ in Eq.~\ref{eq:true_posters} is intractable because of the moderately complicated likelihood function of $p_{\psi}(\mathcal{E}|Z_{s},Z_{t})$ such as a neural network with non-linear layer~\cite{kingma2013auto,Zheng2018DegenerationIV,Zheng2019UnderstandingVI}. We thus introduce a tractable posterior $q_{\phi}(Z_{s},Z_{t}|\mathcal{E})$ to approximate $p_{\theta}(Z_{s},Z_{t}|\mathcal{E})$. In this case, the marginal log-likelihood $\log P(\mathcal{E})$ can be rewritten as:
\begin{equation}
    \log P(\mathcal{E})=D_{KL}[q_{\phi}(Z_{s},Z_{t}|\mathcal{E})||p_{\theta}(Z_{s},Z_{t}|\mathcal{E})]+\mathcal{L}
\label{eq:likelihood}
\end{equation}
where $D_{KL}$ means the Kullback-Liebler (KL) divergence and $\mathcal{L}$ is the (variational) \textit{evidence lower bound} (\textit{ELBO}) of $\log P(\mathcal{E})$. Since the KL divergence term is non-negative, we can maximize the log-likelihood $\log P(\mathcal{E})$ by maximizing $\mathcal{L}$.
Denoting the joint prior for $Z_{s}$ and $Z_{t}$ as $p_{\theta}(Z_{s},Z_{t})$, $\mathcal{L}$ is derived as:
\begin{equation}
    \mathcal{L}=-D_{KL}[q_{\phi}(Z_{s},Z_{t}|\mathcal{E})|p_{\theta}(Z_{s},Z_{t})]+\mathop{\mathbb{E}}_{q_{\phi}(Z_{s},Z_{t}|\mathcal{E})}[p_{\psi}(\mathcal{E}|Z_{s},Z_{t})]
\label{eq:ELBO_1}
\end{equation}
where $p_{\psi}(\mathcal{E}|Z_{s},Z_{t})$ indicates the probabilistic decoder parameterized by $\psi$.
In our method, we simplify the joint prior $p_{\theta}(Z_{s},Z_{t})$ by assuming $p_{\theta}(Z_{s},Z_{t})=p_{\theta}(Z_{s})p_{\theta}(Z_{s})$. Complex prior is a specific research topic in variational inference~\cite{tomczak2017vae,rezende2015variational,yin2018semi}. We do not explore more here since we mainly focus on the general variational auto-encoding idea for modeling both the \emph{first-order} and \emph{high-order} topology in signed directed networks. By decoupling the node embeddings into source node embeddings $Z_{s}$ and target node embeddings $Z_{t}$, we have the following proposition:

\begin{Proposition}
\label{proposition:1}
\textit{Given the observed links $\mathcal{E}$ in signed directed networks, the latent variable for source node embeddings $Z_{s}$ and the latent variable for target node embeddings $Z_{t}$ are conditional independent.}
\end{Proposition}

Thereby, with Proposition~\ref{proposition:1}, we rewrite the approximate posterior $q_{\phi}(Z_{s},Z_{t}|\mathcal{E})$ as:
\begin{equation}
\label{eq:approximate_poster_conditional}
    q_{\phi}(Z_{s},Z_{t}|\mathcal{E})=q_{\phi_{s}}(Z_{s}|\mathcal{E})q_{\phi_{t}}(Z_{t}|\mathcal{E})
\end{equation}
where $q_{\phi_{s}}(Z_{s}|\mathcal{E})$ and $q_{\phi_{t}}(Z_{t}|\mathcal{E})$ are the approximate posteriors parameterized by $\phi_{s}$ and $\phi_{t}$ respectively. If we denote $p_{\theta}(Z_{s})$ and $p_{\theta}(Z_{t})$ are the prior for $Z_{s}$ and $Z_{t}$ respectively, we rewrite the \emph{ELBO} in Eq.~\ref{eq:ELBO_1} as Eq.~\ref{eq:ELBO_2}.
\begin{equation}
\mathcal{L}=-D_{KL}[q_{\phi_{s}}(Z_{s}|\mathcal{E})||p_{\theta}(Z_{s})]-D_{KL}[q_{\phi_{t}}(Z_{t}|\mathcal{E})||p_{\theta}(Z_{t})]+\mathop{\mathbb{E}}_{\substack{q_{\phi_{s}}(Z_{s}|\mathcal{E})\\ q_{\phi_{t}}(Z_{t}|\mathcal{E})}} [\log p_{\psi}(\mathcal{E}|Z_{s},Z_{t})]
\label{eq:ELBO_2}
\end{equation}
More detailed derivation is provided in Appendix~\ref{append:detailed_ELBO}.
DVE tries to learn $Z_{s}$ and $Z_{t}$ via maximizing the above \textit{ELBO}.
To better understand DVE, we firstly introduce the two variational approximate posteriors $q_{\phi_{s}}(Z_{s}|\mathcal{E})$ and $q_{\phi_{t}}(Z_{t}|\mathcal{E})$. Modeling these two distributions also indicate the decoupled variational encoder in Figure~\ref{figure:model}. The conditional distribution $p_{\psi}(\mathcal{E}|Z_{s}, Z_{t})$ which indicates the structure decoder, will be discussed later.

\subsection{Decoupled Variational Encoder}
In this part, how the decoupled variational encoder in Figure~\ref{figure:model} works is introduced.
In our expectation, $Z_{s}$ and $Z_{t}$ are the representation for the source node and target node respectively. These two representations should capture the intrinsic local structures of nodes both involved in positive links and negative links. Take the source node representation $Z_{s}$ as an example, directly representing $Z_{s}$ through existing GCN methods is not appropriate, because this makes GCN do homophily effects on different signs. Instead, we decouple the signed directed graph into an undirected positive graph and an undirected negative graph, and consider that $Z_{s}$ could be generated by the node representation $Z_{s}^{p}$ involved in the undirected positive graph and node representation $Z_{s}^{n}$ involved in the undirected negative graph. In other words, $Z_{s}$ could be represented as $Z_{s}=f_{s}(Z_{s}^{p}, Z_{s}^{n})$, where $f_{s}$ is the generative function. A proper choice of $f_{s}$ can capture the interactions between positive and negative links.

In particular, in the learning process of $Z_{s}^{p}$ and $Z_{s}^{n}$, if we denote $A^{p}$ as the adjacent matrix of the undirected positive graph and $A^{n}$ as the adjacent matrix of the undirected negative graph, variational GCN is applied on $A^{p}$ and $A^{n}$. In other words, $Z_{s}\sim q_{\phi_{s}}(Z_{s}|\mathcal{E})$ is represented by the combination of $Z_{s}^{p}\sim q_{\phi_{s}^{p}}(Z_{s}^{p}|A^{p},X)$, $Z_{s}^{n}\sim q_{\phi_{s}^{n}}(Z_{s}^{n}|A^{n},X)$ and $f_{s}$, where $q_{\phi_{s}^{p}}(Z_{s}^{p}|A^{p},X)$ and $q_{\phi_{s}^{n}}(Z_{s}^{n}|A^{n},X)$ indicate the approximate posteriors for the true posteriors $p_{\theta}(Z_{s}^{p}|A^{p},X)$ and $p_{\theta}(Z_{s}^{n}|A^{n},X)$. And we set $f_{s}$ as concatenation operation here for simplicity. Note that the adjacent matrices of both the undirected positive graph and the undirected negative graph are composed of 0 and 1, where 1 means linked and 0 otherwise. The variational inference procedure for $q_{\phi_{s}}(Z_{s}|\mathcal{E})$ indicates the source node encoder shown in Figure~\ref{figure:model} and is introduced in the following part.

Let the node feature matrix be $X\in \mathbb{R}^{N\times F}$ where $N$ is the number of nodes and $F$ is the feature dimension\footnote{Since we do not have node features in our experiments, we simply set $X=I_{N}$, $I_{N}$ is a diagonal matrix with size $N$}. Let $Z_{s,i}^{p}\in \mathbb{R}^{1\times d}$ and $Z_{s,i}^{n}\in \mathbb{R}^{1\times d}$ be the source node embeddings of $i$-th node involved in the undirected positive graph and the undirected negative graph, respectively. If we denote $q_{\phi_{s}^{p}}(Z_{s}^{p}|A^{p},X)$ and $q_{\phi_{s}^{n}}(Z_{s}^{n}|A^{n},X)$ as the variational distribution for source node involved in the undirected positive graph and undirected negative graph respectively, we can have the following:
\begin{equation}
q_{\phi_{s}^{p}}(Z_{s}^{p}|A^{p},X)=\prod_{i=1}^N q_{\phi_{s}^{p}}(Z_{s,i}^{p}|A^{p},X)
\label{equation:q_s_pos}
\end{equation}
\begin{equation}
q_{\phi_{s}^{n}}(Z_{s}^{n}|A^{n},X)=\prod_{i=1}^N q_{\phi_{s}^{n}}(Z_{s,i}^{n}|A^{n},X)
\label{equation:q_s_neg}
\end{equation}
Inspired by the idea that different semantics can come from the same family of functions (e.g. Gaussian) since these semantics are modeled by different parameters and are in different spaces~\cite{kingma2013auto,pu2016variational,kusner2017grammar}.
We assume that both $q_{\phi_{s}^{p}}(Z_{s}^{p}|A^{p},X)$ and $q_{\phi_{s}^{n}}(Z_{s}^{n}|A^{n},X)$ follow Gaussian distribution, then the reparametrization Gaussian parameters $\mu_{s}^{p,l}\in \mathbb{R}^{N\times d}, \sigma_{s}^{p,l}\in \mathbb{R}^{N\times d}, \mu_{s}^{n,l}\in \mathbb{R}^{N\times d}, \sigma_{s}^{n,l}\in \mathbb{R}^{N\times d}$ in $l$-th layer are defined as\footnote{A quick note: $s$ means the source node, $\mu,\sigma$ denote the mean value and standard deviation parameter of Gaussian distribution, $p$ means undirected positive graph, $n$ indicates undirected negative graph and $l$ means the $(l+1)$-th layer.}:
\begin{equation}
\begin{cases}
\mu_{s}^{p,l+1} = \widetilde{A}^{p}H_{s,\mu}^{p,l}W_{s,\mu}^{p,l},~H_{s,\mu}^{p,l}=h(\mu_{s}^{p,l}),~H_{s,\mu}^{p,0}=X
\vspace{0.1cm} \\
\log\sigma_{s}^{p,l+1} =
\widetilde{A}^{p}H_{s,\sigma}^{p,l}W_{s,\sigma}^{p,l},~H_{s,\sigma}^{p,l}=h(\log \sigma_{s}^{p,l}),~H_{s,\sigma}^{p,0}=X
\end{cases}
\label{eq:mu_sigma_sp}
\end{equation}

\begin{equation}
\begin{cases}
\mu_{s}^{n,l+1} = \widetilde{A}^{n}H_{s,\mu}^{n,l}W_{s,\mu}^{n,l},~H_{s,\mu}^{n,l}=h(\mu_{s}^{n,l}),~H_{s,\mu}^{n,0}=X
\vspace{0.1cm} \\
\log\sigma_{s}^{n,l+1} =
\widetilde{A}^{n}H_{s,\sigma}^{n,l}W_{s,\sigma}^{n,l},~H_{s,\sigma}^{n,l}=h(\log \sigma_{s}^{n,l}),~H_{s,\sigma}^{n,0}=X
\end{cases}
\label{eq:mu_sigma_sn}
\end{equation}
where $\widetilde{A}^{p}=[\widehat{D}^{p}]^{-\frac{1}{2}}(A^{p}+I_N)[\widehat{D}^{p}]^{-\frac{1}{2}}$ and $\widetilde{A}^{n}=[\widehat{D}^{n}]^{-\frac{1}{2}}(A^{n}+I_N)[\widehat{D}^{n}]^{-\frac{1}{2}}$ are the propagation matrices. $\widehat{D}^{p}$ and $\widehat{D}^{n}$ are the degree matrices of $A^{p}+I_{N}$ and $A^{n}+I_{N}$, respectively.
$h(\cdot)$ denotes the non-linear $ReLu$ function. $W_{s,\mu}^{p,l}\in \mathbb{R}^{F\times d}$ and $W_{s,\sigma}^{p,l}\in \mathbb{R}^{F\times d}$ denote the $l$-layer reparametrization parameters for $Z_{s}^{p}$. Similarly, $W_{s,\mu}^{n,l}\in \mathbb{R}^{F\times d}$ and $W_{s,\sigma}^{n,l}\in \mathbb{R}^{F\times d}$ are the $l$-layer reparametrization parameters for $Z_{s}^{n}$. 
Accordingly, if we denote $p_{\theta}(Z_{s}^{p})$ and $p_{\theta}(Z_{s}^{n})$ are prior distributions for $Z_{s}^{p}$ and $Z_{s}^{n}$ respectively, 
the prior regularization loss on $q_{\phi_{s}^{p}}(Z_{s}^{p}|A^{p},X)$ and $q_{\phi_{s}^{n}}(Z_{s}^{n}|A^{n},X)$ are written as:
\begin{equation}
    \min_{\phi_{s}} L_{KL}^{s}=D_{KL}[q_{\phi_{s}^{p}}(Z_{s}^{p}|A^{p},X)||p_{\theta}(Z_{s}^{p})]+D_{KL}[q_{\phi_{s}^{n}}(Z_{s}^{n}|A^{n},X)||p_{\theta}(Z_{s}^{n})]
\end{equation}
% where $p_{\theta}(Z_{s}^{p})$ and $p_{\theta}(Z_{s}^{n})$ are prior distributions for $Z_{s}^{p}$ and $Z_{s}^{n}$, respectively.
where $\phi_s=\{\phi_{s}^{p},\phi_{s}^{n}\}=\{W_{s,\mu/\sigma}^{p,l_{0}/l_{1}},W_{s,\mu/\sigma}^{n,l_{0}/l_{1}}\}$ is the parameter of the source node encoder.
Therefore, representation for the source node can be obtained by $Z_{s}=Z_{s}^{p}\oplus Z_{s}^{n}$ ($\oplus$ means concatenation), where $Z_{s}^{p}\sim q_{\phi_{s}^{p}}(Z_{s}^{p}|A^{p},X)$ and $Z_{s}^{n}\sim q_{\phi_{s}^{n}}(Z_{s}^{n}|A^{n},X)$. The target node representation $Z_{t}$ can be obtained in similar procedure. 
% The source node encoder and the target node encoder are shown in Figure~\ref{figure:model}.

It is worthwhile to highlight that GCN working on both $A^{p}$ and $A^{n}$ with different parameters models the distinctive effects of different link signs. Conducting GCN on $A^{p}$ and $A^{n}$ is reasonable since GCN does not specify positive or negative meaning of links in graphs. Instead, GCN emphasizes the correlation that links two nodes. How to leverage the information from $A^{p}$ and $A^{n}$ in the subsequent modules determines the positive or negative semantics.
In our case, we use GCN to summarize the correlation pattern among nodes, and then ask the following module to employ the positive semantics in $A^{p}$ and negative semantics in $A^{n}$. By this way, the signed local structures can be captured in a decoupled manner. 

\subsection{Structure Decoder}
In auto-encoding theory, decoder is an essential module and our structure decoder is introduced here.
The structure decoder is expected to reconstruct the signed directed links and guide the encoder learning. This requires that the structure decoder should preserve the structural characteristics in signed directed networks. Note that the Extended Structural Balance Theory~\cite{qian2013extended} states the closeness of users in signed networks. The essential insight of this theory is that for four users $i,j,k,r$, if the link signs are $\epsilon_{ij}=1, \epsilon_{ik}=0, \epsilon_{ir}=-1$, the closeness among them follows Eq.~\ref{equation:balance_theory}.
\begin{equation}
g(i,j)<g(i,k)<g(i,r)
\label{equation:balance_theory}
\end{equation}
where $g(i,j)$ denotes distance between user $i$ and $j$. For example, if a positive link means trust and a negative link means distrust in social networks, user $i$ prefers to trust $j$ than $k$ and trusts $k$ more than $r$. Actually, this theory states the \emph{first-order} topology that indicates the closeness relationships among nodes. By combining this theory with Bayesian Personalized Ranking~\cite{rendle2009bpr}, we naturally develop a novel Balance Pair-wise Ranking (BPWR) loss to guide the whole model learning. 
To clarify, we denote the distance in Eq.~\ref{equation:balance_theory} as the existence score of positive links. The higher score is, the more probably the positive link exists. Thus, the Extended Structural Balance Theory can be interpreted as Eq.~\ref{equation:balance_theory_existence}:
\begin{equation}
f(i\rightarrow j)>f(i\rightarrow k)>f(i\rightarrow r)
\label{equation:balance_theory_existence}
\end{equation}
where $f(i\rightarrow j)$ indicates the existence score of positive links from source node $i$ to target node $j$. If $j>_{i} k$ indicates the relation score of $i\rightarrow j$ is larger than that of $i\rightarrow k$ with node $i$ as the reference object,
for samples $(i,j,k,r)$ with $\epsilon_{i\rightarrow j}=1,\epsilon_{i\rightarrow k}=0,\epsilon_{i\rightarrow r}=-1$, the maximum posteriors satisfy:
\begin{equation}
\begin{cases}
\max \limits_{\phi_s,\phi_t} \prod \limits_{(i,j,k)}p(\phi_s,\phi_t|j>_ik)\propto \prod \limits_{(i,j,k)}p(j>_ik|\phi_s,\phi_t)p(\phi_s,\phi_t)
\vspace{0.1cm}\\
\max \limits_{\phi_s,\phi_t} \prod \limits_{(i,k,r)}p(\phi_s,\phi_t|k>_i r)\propto \prod \limits_{(i,k,r)}p(k>_i r|\phi_s,\phi_t)p(\phi_s,\phi_t)
\end{cases}
\label{equation:max_likelihood1_likelihood2}
\end{equation}
% where $j>_{i} k$ indicates the relation score of $i\rightarrow j$ is larger than that of $i\rightarrow k$ with node $i$ as the reference object. 
where $\phi_s$ and $\phi_t$ are the parameters of the decoupled variational encoder to obtain $Z_{s}$ and $Z_{t}$. 
% Sampling source node embeddings $Z_{s}$ from $p(Z_{s})$ and target node embeddings $Z_{t}$ from $p(Z_{t})$ and assuming $f(i,j)=Z_{s,i}Z_{t,j}^{T}$ like in~\citet{rendle2009bpr}, 
$p(j>_i k|\phi_s,\phi_t)$ and $p(k>_i r|\phi_s,\phi_t)$ indicate the likelihood functions which are written as:
\begin{equation}
\begin{cases}
p(j>_ik|\phi_s,\phi_t) = \sigma(f(i\rightarrow j)-f(i\rightarrow k))
\vspace{0.1cm}\\
p(k>_i r|\phi_s,\phi_t) = \sigma(f(i\rightarrow k)-f(i\rightarrow r))
\end{cases}
\label{equation:probability_theta1_theta2}
\end{equation}
where $f(i\rightarrow j)$ is calculated by the inner product of the source node embeddings $Z_{s,i}$ of node $i$ and target node embeddings $Z_{t,j}$ of node $j$. $f(i\rightarrow k)$ and $f(i\rightarrow r)$ can be obtained in similar way. $\sigma$ is the sigmoid function. Therefore, following~\cite{rendle2009bpr},
the Balance Pair-wise Ranking (BPWR) loss of our structure decoder can be written as:
\begin{equation}
\begin{split}
    \min_{\phi_s,\phi_t}~~ L_{BPWR} =& -\mathbb{E}_{(i,j,k)\sim P(\mathcal{E})} \ln\sigma(Z_{s,i}^{T} Z_{t,j}-Z_{s,i}^{T} Z_{t,k}) \\ &-\mathbb{E}_{(i,k,r)\sim P(\mathcal{E})}\ln\sigma(Z_{s,i}^{T} Z_{t,k}-Z_{s,i}^{T} Z_{t,r})
\end{split}
\label{equation:BPWR_loss}
\end{equation}
where $i,j,k,r$ are the node indexes that satisfy $e_{i\rightarrow j}\in \mathcal{E}^{p}$ and $e_{i\rightarrow r}\in \mathcal{E}^{n}$ and $e_{i\rightarrow k}$ is the sampled non-existent link . 
When $Z_{s}$ and $Z_{t}$ are not learned from the decoupled variational encoder, $Z_{s}$ and $Z_{t}$ can be initialized trainable embedding matrices. In other words, BPWR can be an independent model to learn node embeddings in signed directed networks.

\subsection{Model Learning}
Putting the encoder and decoder together, we can write the objective function of DVE as follows\footnote{Note that we take the expectation formula here for scaled loss values instead of summarization.}:
\begin{equation}
\begin{split}
\min_{\phi_{s},\phi_{t}}~~ L_{DVE}=&-\mathbb{E} _{(i,j,k)\sim P(\mathcal{E})}\ln\sigma(Z_{s,i}^{T} Z_{t,j}-Z_{s,i}^{T} Z_{t,k}) \\
&-\mathbb{E}_{(i,k,r)\sim P(\mathcal{E})}\ln\sigma(Z_{s,i}^{T} Z_{t,k}-Z_{s,i}^{T} Z_{t,r}) \\
&+\frac{1}{N}\sum_{i=1}^{N}\{D_{KL}[q_{\phi_{s}^{p}}(Z_{s,i}^{p}|A^{p},X)||p_{\theta}(Z_{s}^{p})]+D_{KL}[q_{\phi_{s}^{n}}(Z_{s,i}^{n}|A^{n},X)||p_{\theta}(Z_{s}^{n})]\} \\
&+\frac{1}{N}\sum_{i=1}^{N}\{D_{KL}[q_{\phi_{t}^{p}}(Z_{t,i}^{p}|A^{p},X)||p_{\theta}(Z_{t}^{p})]+D_{KL}[q_{\phi_{t}^{n}}(Z_{t,i}^{n}|A^{n},X)||p_{\theta}(Z_{t}^{n})]\}
\end{split} 
\label{eq:DVE_loss}
\end{equation}
where $\phi_s=\{\phi_{s}^{p},\phi_{s}^{n}\}=\{W_{s,\mu/\sigma}^{p,l_{0}/l_{1}},W_{s,\mu/\sigma}^{n,l_{0}/l_{1}}\}$ is the parameter of the source node encoder and $\phi_t=\{\phi_{t}^{p},\phi_{t}^{n}\}=\{W_{t,\mu/\sigma}^{p,l_{0}/l_{1}},W_{t,\mu/\sigma}^{n,l_{0}/l_{1}}\}$ denotes the parameter of the target node encoder. The source node embeddings and target node embeddings are respectively denoted as $Z_{s}=Z_{s}^{p}\oplus Z_{s}^{n}$, $Z_{t}=Z_{t}^{p}\oplus Z_{t}^{n}$. All priors $p_{\theta}(Z_{s}^{p})$,$p_{\theta}(Z_{s}^{n})$,$p_{\theta}(Z_{t}^{p})$ and $p_{\theta}(Z_{t}^{n})$ are standard Gaussian distributions.
In DVE, the matrix multiplication is conducted between a sparse adjacent matrix and a dense matrix, e.g. Eq.~\ref{eq:mu_sigma_sp},\ref{eq:mu_sigma_sn}, which can be implemented with high efficiency in recent deep learning programming frameworks.

For each positive link $e_{i\rightarrow j}$ and negative link $e_{i\rightarrow r}$, we randomly sample $n_{noise}$ non-linked nodes to play as $k$ and construct the training triplets $(i,j,k)$ and $(i,k,r)$. We adopt Dropout technique for regularization rather than $L_{2}$ norm. Many widely used optimization algorithms such as RMSProp can be applied for model learning.

\subsection{Comparison Between DVE and Existing Methods}
There are differences and connections between DVE and existing methods. A key difference is that DVE integrally captures both the \emph{first-order} and \emph{high-order} topology for signed directed networks. However, most existing methods~\cite{hsieh2012low,wang2017signed,wang2017attributed} mainly focus on modeling the \emph{first-order} topology.
There are some works~\cite{yuan2017sne,kim2018side} based on random walks, but they fail to capture the \emph{high-order} topology since they apply homophily effects on different link signs.

Meanwhile, there are connections between DVE and existing methods in terms of the \emph{first-order} topology modeling. Both DVE and existing methods perform the \emph{first-order} topology modeling regarding signed directed links. Existing methods works with restrictive distance metrics and usually ignore the mediator function of non-existent links. Instead, BPWR of DVE working in a personalized ranking scheme has more potential to capture the closeness relationships of nodes. Besides, it is worthwhile to point out that both the margin ranking (MR) loss in SNEA~\cite{wang2017attributed} and BPWR loss in DVE have similar target that are developed from Extended Structural Balance Theory. However, MR is a deterministic non-smooth metric and BPWR derived from maximizing the posterior of the observations is smooth and easy to be optimized by gradient based techniques~\cite{rendle2009bpr}. The superior performance of BPWR over SNEA-MR is also verified in Section~\ref{section:comparison}.

\subsection{Time Complexity Analysis}
Stochastic training of DNN methods involves two steps, the forward and backward computations.
DVE supports the mini-batch training and the time cost lies in the decoupled variational encoder and structure decoder. We thus decompose the time complexity of DVE into two parts, namely the time complexity of decoupled variational encoder and structure decoder. In each batch of DVE, the decoupled variational encoder learns node embeddings for all nodes. Thus, following the analysis of GCN in~\citet{wu2019comprehensive}, the time complexity of decoupled variational encoder is $\mathcal{O}(2|\mathcal{E}^{p}|+2|\mathcal{E}^{n}|)=\mathcal{O}(2|\mathcal{E}|)$, where $|\mathcal{E}^{p}|$, $|\mathcal{E}^{p}|$ and $|\mathcal{E}|$ denote the number of edges of undirected positive graph, undirected negative graph and signed directed graph, respectively. Note that for this decoupled variational encoder, the source node encoder and target node encoder can be parallelly conducted. In this case, the time complexity for decoupled variational encoder can be reduced to $\mathcal{O}(|\mathcal{E}^{p}|+|\mathcal{E}^{n}|)=\mathcal{O}(|\mathcal{E}|)$. 
Furthermore, the graph convolution on $A^{p}$ and $A^{n}$ to learn $Z_{s}$ and $Z_{t}$
 could also be paralleled, which leads to the time complexity as $\mathcal{O}(\max\{|\mathcal{E}^{p}|,|\mathcal{E}^{n}|\})$. 
As for the structure decoder in each batch, we compute the BPWR loss with non-existent link sampling. If we denote the sampling size as $n_{noise}$ and the batch size of training positive/negative links as $B$, the time complexity is $\mathcal{O}(n_{noise}B)$. 

In summary, the time complexity of the non-parallel DVE in each batch is $\mathcal{O}(2|\mathcal{E}|+n_{noise}B)$, the half-parallel counterpart is $\mathcal{O}(|\mathcal{E}|+n_{noise}B)$ and the quarter-parallel counterpart is $\mathcal{O}(\max\{|\mathcal{E}^{p}|,|\mathcal{E}^{n}|\}+n_{noise}B)$. Generally, the main time cost lies in the decoupled variational encoder, since we usually have $2|\mathcal{E}|>|\mathcal{E}|>\max\{|\mathcal{E}^{p}|,|\mathcal{E}^{n}|\}>>n_{noise}B$. Actually, the time complexity of decoupled variational encoder is related to the specific graph convolutional network. For datasets with too large $\max\{|\mathcal{E}^{p}|,|\mathcal{E}^{n}|\}$, the $\mathcal{O}(\max\{|\mathcal{E}^{p}|,|\mathcal{E}^{n}|\})$ in each batch may still be time-consuming. This can be solved by using other kinds of GCN~\cite{hamilton2017inductive,chen2018fastgcn,chen2017stochastic} that reduce $\mathcal{O}(\max\{|\mathcal{E}^{p}|,|\mathcal{E}^{n}|\})$ to the scale of the training batch size $B$. This makes DVE scalable to much larger datasets. We do not explore more here since we mainly focus on the general idea of variational auto-encoding to capture both the \emph{first-order} and \emph{high-order} topological information for signed directed networks.

\section{Experimental Results and Analysis}
In this section, we conduct experiments on three widely used datasets. Performance on both link sign prediction task and node recommendation task are implemented to verify the effectiveness of DVE. Further ablation study and qualitative analysis are investigated to provide deep understanding about DVE.

\subsection{Experimental Setups}
\subsubsection{\textbf{Dataset Description}} We conduct the experiments on three widely used real-world datasets. Epinions\footnote{https://snap.stanford.edu/data/soc-sign-epinions.html}: Epinions is one popular product review site in which users can create both trust (positive) and distrust (negative) links to others. Slashdot\footnote{https://snap.stanford.edu/data/soc-sign-Slashdot090216.html} is a technology news platform where users can create friend (positive) and foe (negative) links to others. Wiki\footnote{https://snap.stanford.edu/data/wiki-RfA.html} is a dataset collected from the Wikipedia site, where users vote for or against other users in order to determine administration promotion. For each dataset, we randomly sample a subset links as our experimental dataset. We also filter out users who have no link with others. The statistics of processed data are shown in Table~\ref{table:data_statistics}. From the table, it is obvious that both the undirected positive graph and the undirected negative graph are very sparse.
\begin{table}[]
\caption{The statistics of Epinions, Slashdot and Wiki utilized in our experiments.}
\label{table:data_statistics}
\begin{tabular}{cccc}
\hline
Dataset            & Epinions  & Slashdot  & Wiki      \\ \hline
\#nodes            & 22,503    & 17,496    & 6,836     \\
\#edges            & 84,102    & 54,920    & 89,365    \\
\#positive edges        & 60,044    & 46,189    & 70,075    \\
\#negative edges        & 24,058    & 8,731     & 19,290    \\
undirected positive graph density & 0.0119\%  & 0.0151\%  & 0.1499\%  \\
undirected negative graph density & 4.75e-3\% & 2.85e-3\% & 4.12e-4\% \\ \hline
\end{tabular}
\end{table}

\subsubsection{\textbf{Baselines}} We compare DVE with 9 competitive baselines.
\begin{itemize}
\item LINE~\cite{tang2015line}: LINE defines loss functions to preserve the first-order
or second-order proximity between nodes in a graph. Here, we only perform LINE on the positive links since it does not work on signed graphs. Since LINE's first order proximity usually presents better performance than the second order one, we report the performance of LINE's first-order proximity here. 
\item MF~\cite{hsieh2012low}: Matrix factorization is one popular technique for network embedding. We perform MF with the same noise sampling method as DVE here to learn low-dimensional node embeddings for signed directed networks. 
\item SNE~\cite{yuan2017sne}: This method develops the log-bilinear model with random walk to learn low-dimensional node embeddings for signed networks. On signed directed networks, we apply directed random walk for SNE here.
\item SiNE~\cite{wang2017signed}: SiNE is a deep neural network method that makes a distinction between positively linked nodes and negatively linked nodes. It is capable of capturing the non-linear pattern in signed directed networks. 
\item SIDE~\cite{kim2018side}: SIDE is a random walk based method, which formulates the social balance theory into a likelihood for signed directed networks.
\item SNEA-MR~\cite{wang2017attributed}: SNEA is a method for attributed signed social networks. Considering that the margin ranking (MR) loss~\footnote{It refers to Eq.5 in the original paper.} in SNEA is also based on Extended Structural Balance Theory, we thus extend it here as a baseline to make comparison with BPWR.
\item BPWR (Ours): As the structure decoder of DVE method, this Balance Pair-wise Ranking loss could be an independent model and be a comparison to the loss in SiNE and SNEA-MR.
\item SLVE (Ours): SLVE substitutes the decoupled variational encoder in DVE with a non-decoupled one by leveraging the signed Laplacian matrix~\cite{gallier2016spectral}. 
\item DE (Ours): DE is the non-variational variant of DVE method.
\end{itemize}

\subsubsection{\textbf{Parameter Settings.}}
For each baseline, we follow the parameter settings in their papers or codes. Batch training size is 1000 for all methods. For our model, we set the training epoch size as 200 and the number of GCN layers as $l=2$. Dropout probability is 0.2. Learning rate is taken as 0.01. RMSProp optimizer~\cite{tieleman2014rmsprop} is adopted to optimize our objective function. According to the training loss, the size of randomly sampled noise ($e_{i\rightarrow k}$=0) is set as 5 for Epinions and 20 for Slashdot and Wiki.
Embedding size is $d_1=128$ and $d=64$ on all three datasets. 
We randomly split each dataset into 80\% train data and 20\% test data. For every model, we conduct 10 times and report the averaged best performance on test set as the model performance.
\begin{table*}[]
\caption{Link sign prediction performance. Names with * refer to our methods. Compared to SiNE, the absolute improvement percentage of DE and DVE are given. Compared to DVE, the t-test results of other baselines are shown in this table. $\ddagger$ means p-value<0.01, $\dagger$ indicates p-value<0.05 and $-$ means p-value>0.05.}
\label{table:link_sign_prediction_task}
\begin{tabular}{c|cc|cc|cc}
\hline
Dataset & \multicolumn{2}{c|}{Epinions}                  & \multicolumn{2}{c|}{Slashdot}                 & \multicolumn{2}{c}{Wiki}                     \\ \hline
Method  & AUC                    & F1                    & AUC                   & F1                    & AUC                   & F1                    \\ \hline
LINE    & 0.906$^\ddagger$                  & 0.902$^\ddagger$                 & 0.855$^\ddagger$                 & 0.920$^\ddagger$                 & 0.782$^\ddagger$                 & 0.889$^\ddagger$                 \\
MF      & 0.934$^\ddagger$                  & 0.927$^\ddagger$                 & 0.801$^\ddagger$                 & 0.918$^\ddagger$                 & 0.595$^\ddagger$                 & 0.884$^\ddagger$                 \\
SNE     & 0.952$^\ddagger$                  & 0.933$^\ddagger$                 & 0.869$^\ddagger$                 & 0.928$^\ddagger$                 & 0.848$^\ddagger$                 & 0.902$^\ddagger$                 \\
SiNE    & 0.929$^\ddagger$                  & 0.919$^\ddagger$                 & 0.870$^\ddagger$                 & 0.916$^\ddagger$                 & 0.864$^\ddagger$                 & 0.904$^\ddagger$                 \\
SIDE    & 0.807$^\ddagger$                  & 0.861$^\ddagger$                 & 0.798$^\ddagger$                 & 0.917$^\ddagger$                 & 0.647$^\ddagger$                 & 0.884$^\ddagger$                 \\
SNEA-MR    & 0.864$^\ddagger$                  & 0.891$^\ddagger$                 & 0.753$^\ddagger$                 & 0.913$^\ddagger$                 & 0.732$^\ddagger$                 & 0.889$^\ddagger$                 \\ \hline
BPWR$^*$    & 0.933$^\ddagger$                  & 0.926$^\ddagger$                 & 0.891$^\ddagger$                 & 0.929$^\ddagger$                 & 0.881$^\ddagger$                 & 0.911$^\ddagger$                 \\
SLVE$^*$    & 0.924$^\ddagger$                  & 0.918$^\ddagger$                 & 0.871$^\ddagger$                 & 0.922$^\ddagger$                 & 0.874$^\ddagger$                 & 0.905$^\ddagger$                 \\
DE$^*$      & 0.958(+2.9\%)$^-$ & 0.939(2.0\%)$^-$ & 0.899(2.9\%)$^\ddagger$          & 0.930(1.4\%)$^\ddagger$          & 0.885(2.1\%)$^\ddagger$          & 0.909(0.5\%)$^\ddagger$          \\
DVE$^*$     & \textbf{0.960(3.1\%)}           & \textbf{0.940(2.1\%)}          & \textbf{0.905(3.5\%)} & \textbf{0.934(1.8\%)} & \textbf{0.889(2.5\%)} & \textbf{0.911(0.7\%)} \\ \hline
\end{tabular}
\end{table*}

\subsection{Performance Comparison}
\label{section:comparison}
\subsubsection{\textbf{Link Sign Prediction}}
We first compare the model performance on link sign prediction task. Link sign prediction aims to predict the unobserved signs of existing links. 
Following the evaluation protocols of existing works~\cite{yuan2017sne,wang2017signed}, we train a binary classifier which is a two-layer MLP with $Relu$ as the non-linear function. Then, we use signed links in the model training stage as the train data for the binary classifier and predict signs for the test links. More specifically, we concatenate two node embeddings as the link representation and take the link representation as input for the binary classifier. Due to the unbalanced signs in test links, AUC and F1 are adopted to assess the performance. We consider sign +1 as the positive class.
The results are shown in Table~\ref{table:link_sign_prediction_task}. From this table, we summarize that:
\begin{itemize}
    \item The proposed DVE outperforms recent competitive methods and reaches the best performance. For example, regarding AUC on Slashdot, DVE obtains a 3.5\% improvement compared to SiNE and a 3.6\% improvement compared to SNE. This verifies that DVE learns more representative node embeddings in signed directed networks.
    \item Comparing BPWR with other baselines (SiNE, SNE, MF), BPWR outperforms them on AUC and F1 on all three datasets. 
    This exposes the deficiencies of the baselines in mining the \emph{first-order} topology. Developed from Extended Structural Balance Theory, BPWR working in a personalized pair-wise ranking manner is more capable of mining the closeness relationships among nodes. 
    It is worthwhile to point out that although SNEA-MR and BPWR are both developed from the Extended Structural Balance Theory and they have a similar training goal. However, the objective of SNEA-MR is not smooth while BPWR based on maximizing the posterior of signed directed links is smooth and easy to be optimized by gradient based optimization methods. The comparison results of SNEA-MR and BPWR as well verify the effectiveness of BPWR.
    \item Modeling the \emph{high-order} topology facilitates to learn more representative node embeddings in signed directed networks. Comparing DVE, DE with BPWR, DVE and DE in an auto-encoder formula are able to model both the \emph{first-order} and the \emph{high-order} topology in signed directed networks. However, BPWR works as an independent model can only model the \emph{first-order} topology.
    The gap between DVE,DE and BPWR is more obvious in the following node recommendation task.  
    \item It is obvious that DVE always outperforms SLVE, which indicates the importance of our decoupling idea. Particularly, SLVE is the non-decoupled variant of DVE by applying signed Laplacian matrix~\cite{gallier2016spectral} in GCN. Thus the only difference between SLVE and DVE is the encoder part. From the comparison between SLVE and BPWR, we can see that SLVE even damages its own decoder's (BPWR) performance. This highlights the necessity of applying distinctive effects on different types of links in signed directed networks.
    \item Regarding DVE and DE, DVE models the uncertainty of node embeddings in signed directed networks. DVE performs better on Slashdot and Wiki compared to the non-variational DE. A more informative prior matching the complex data rather than standard Gaussian will be better for variational inference. Thus, if provided with a more proper prior, the advantages of modeling uncertainty are expected to be more obvious.
\end{itemize}

\subsubsection{\textbf{Node Recommendation}}
Another practical application of network embedding in signed directed networks is node recommendation. It matches a fact that friend recommendation in social media. We thus conduct the node recommendation task here to investigate the quality of learned node embeddings.
\begin{table*}[]
\caption{Node recommendation performance on Epinions. Names with $*$ refer to our methods. The metrics for this task are $Recall@k$ and $Precision@k$. We pick k=10,20,50 here. Compared to SiNE, the absolute improvement percentage of DVE is given in the table. Compared to DVE, the t-test results of other baselines are as well shown in the table. $\ddagger$ means p-value<0.01, $\dagger$ indicates p-value<0.05 and $-$ means p-value>0.05.}
\label{table:node_recommendation_task_epinions}
\begin{tabular}{c|cccccc}
\hline
Dataset & \multicolumn{6}{c}{Epinions}                                                                                                                       \\ \hline
Methods & R@10                   & R@20                   & R@50                   & P@10                   & P@20                  & P@50                   \\ \hline
LINE      & 0.004$^\ddagger$                  & 0.011$^\ddagger$                  & 0.025$^\ddagger$                  & 0.004$^\ddagger$                  & 0.004$^\ddagger$                 & 0.005$^\ddagger$                  \\
MF      & 0.022$^\ddagger$                  & 0.036$^\ddagger$                  & 0.069$^\ddagger$                  & 0.029$^\ddagger$                  & 0.025$^\ddagger$                 & 0.021$^\ddagger$                  \\
SNE     & 0.002$^\ddagger$                  & 0.003$^\ddagger$                  & 0.008$^\ddagger$                  & 0.001$^\ddagger$                  & 9.5e-4$^\ddagger$                  & 9.0e-4$^\ddagger$                   \\
SiNE    & 0.027$^\ddagger$                  & 0.039$^\ddagger$                  & 0.074$^\ddagger$                  & 0.031$^\ddagger$                  & 0.026$^\ddagger$                 & 0.021$^\ddagger$                  \\
SIDE    & 5.5e-4$^\ddagger$                   & 8.2e-4$^\ddagger$                   & 0.002$^\ddagger$                  & 8.3e-4$^\ddagger$                   & 7.1e-4$^\ddagger$                  & 5.9e-4$^\ddagger$                   \\
SNEA-MR    & 0.024$^\ddagger$                   & 0.037$^\ddagger$                   & 0.065$^\ddagger$                  & 0.024$^\ddagger$                   & 0.020$^\ddagger$                  & 0.016$^\ddagger$                   \\ \hline
BPWR$^*$    & 0.031$^\ddagger$                  & 0.053$^\ddagger$                  & 0.088$^\ddagger$                  & 0.026$^\ddagger$                  & 0.024$^\ddagger$                 & 0.021$^\ddagger$                  \\
SLVE$^*$    & 0.0181$^\ddagger$                 & 0.029$^\ddagger$                  & 0.070$^\ddagger$                  & 0.017$^\ddagger$                  & 0.015$^\ddagger$                 & 0.012$^\ddagger$                  \\
DE$^*$      & 0.030$^\ddagger$                  & 0.045$^\ddagger$                  & 0.082$^\ddagger$                  & 0.022$^\ddagger$                  & 0.020$^\ddagger$                 & 0.017$^\ddagger$                  \\
DVE$^*$     & \textbf{0.035(0.8\%)} & \textbf{0.053(1.4\%)} & \textbf{0.100(2.6\%)} & \textbf{0.035(0.4\%)} & \textbf{0.030(0.4\%)} & \textbf{0.024(0.3\%)} \\ \hline
\end{tabular}
\end{table*}

\begin{table*}[]
\caption{Node recommendation performance on Slashdot. Names with $*$ refer to our methods. The metrics for this task are $Recall@k$ and $Precision@k$. We pick k=10,20,50 here. Compared to SiNE, the absolute improvement percentage of DVE is given in the table. Compared to DVE, the t-test results of other baselines are as well shown in the table. $\ddagger$ means p-value<0.01, $\dagger$ indicates p-value<0.05 and $-$ means p-value>0.05.}
\label{table:node_recommendation_task_slashdot}
\begin{tabular}{c|cccccc}
\hline
Dataset & \multicolumn{6}{c}{Slashdot}                                                                                                                  \\ \hline
Methods & R@10                  & R@20                  & R@50                  & P@10                  & P@20                  & P@50                  \\ \hline
LINE      & 0.011$^\ddagger$                 & 0.017$^\ddagger$                 & 0.033$^\ddagger$                 & 0.006$^\ddagger$                 & 0.006$^\ddagger$                 & 0.006$^\ddagger$                 \\
MF      & 0.015$^\ddagger$                 & 0.026$^\ddagger$                 & 0.053$^\ddagger$                 & 0.025$^\ddagger$                 & 0.022$^\ddagger$                 & 0.018$^\ddagger$                 \\
SNE     & 0.002$^\ddagger$                 & 0.006$^\ddagger$                 & 0.011$^\ddagger$                 & 0.002$^\ddagger$                 & 0.002$^\ddagger$                 & 0.002$^\ddagger$                 \\
SiNE    & 0.052$^\ddagger$                 & 0.068$^\ddagger$                 & 0.107$^\ddagger$                 & 0.027$^\ddagger$                 & 0.025$^\ddagger$                 & 0.021$^\ddagger$                 \\
SIDE    & 8.2e-4$^\ddagger$                & 0.001$^\ddagger$                 & 0.005$^\ddagger$                 & 6.5e-4$^\ddagger$                & 6.5e-4$^\ddagger$                & 7.6e-4$^\ddagger$                \\
SNEA-MR & 0.005$^\ddagger$                 & 0.007$^\ddagger$                 & 0.014$^\ddagger$                 & 0.005$^\ddagger$                & 0.004$^\ddagger$                & 0.003$^\ddagger$                \\ \hline
BPWR$^*$    & 0.058$^\ddagger$                 & 0.073$^\ddagger$                 & 0.111$^\ddagger$                 & 0.028$^\ddagger$                 & 0.023$^\ddagger$                 & 0.020$^\ddagger$                 \\
SLVE$^*$    & 0.037$^\ddagger$                 & 0.049$^\ddagger$                 & 0.089$^\ddagger$                 & 0.018$^\ddagger$                 & 0.017$^\ddagger$                 & 0.016$^\ddagger$                 \\
DE$^*$      & 0.039$^\ddagger$                 & 0.057$^\ddagger$                 & 0.101$^\ddagger$                 & 0.025$^\ddagger$                 & 0.022$^\ddagger$                 & 0.019$^\ddagger$                 \\
DVE$^*$     & \textbf{0.060(0.8\%)} & \textbf{0.086(1.8\%)} & \textbf{0.134(2.7\%)} & \textbf{0.036(0.9\%)} & \textbf{0.031(0.6\%)} & \textbf{0.024(0.3\%)} \\ \hline
\end{tabular}
\end{table*}

\begin{table*}[]
\caption{Node recommendation performance on Wiki. Names with $*$ refer to our methods. The metrics for this task are $Recall@k$ and $Precision@k$. We pick k=10,20,50 here. Compared to SiNE, the absolute improvement percentage of DVE is given in the table. Compared to DVE, the t-test results of other baselines are as well shown in the table. $\ddagger$ means p-value<0.01, $\dagger$ indicates p-value<0.05 and $-$ means p-value>0.05.}
\label{table:node_recommendation_task_wiki}
\begin{tabular}{c|cccccc}
\hline
Dataset & \multicolumn{6}{c}{Wiki}                                                                                                                  \\ \hline
Methods & R@10                  & R@20                  & R@50                  & P@10                  & P@20                  & P@50                  \\ \hline
LINE      & 0.037$^\ddagger$                 & 0.054$^\ddagger$                 & 0.112$^\ddagger$                 & 0.014$^\ddagger$                 & 0.010$^\ddagger$                 & 0.009$^\ddagger$                 \\
MF      & 0.011$^\ddagger$                 & 0.024$^\ddagger$                 & 0.048$^\ddagger$                 & 0.005$^\ddagger$                 & 0.005$^\ddagger$                 & 0.004$^\ddagger$                 \\
SNE     & 0.002$^\ddagger$                 & 0.005$^\ddagger$                 & 0.011$^\ddagger$                 & 8.4e-4$^\ddagger$                & 9.9e-4$^\ddagger$                & 9.5e-4$^\ddagger$                \\
SiNE    & 0.033$^\ddagger$                 & 0.055$^\ddagger$                 & 0.111$^\ddagger$                 & 0.012$^\ddagger$                 & 0.010$^\ddagger$                 & 0.009$^\ddagger$                 \\
SIDE    & 0.001$^\ddagger$                 & 0.002$^\ddagger$                 & 0.009$^\ddagger$                 & 5.8e-4$^\ddagger$                & 5.2e-4$^\ddagger$                & 7.6e-4$^\ddagger$                \\ 
SNEA-MR    & 0.002$^\ddagger$                 & 0.004$^\ddagger$                 & 0.014$^\ddagger$                 & 9.4e-4$^\ddagger$                & 7.6e-4$^\ddagger$                & 0.011$^\ddagger$                \\ \hline
BPWR$^*$    & 0.049$^\ddagger$                 & 0.087$^\ddagger$                 & 0.174$^\ddagger$                 & 0.018$^\ddagger$                 & 0.016$^\ddagger$                 & 0.014$^\ddagger$                 \\
SLVE$^*$    & 0.013$^\ddagger$                 & 0.037$^\ddagger$                 & 0.101$^\ddagger$                 & 0.004$^\ddagger$                 & 0.006$^\ddagger$                 & 0.007$^\ddagger$                 \\
DE$^*$      & 0.043$^\ddagger$                 & 0.073$^\ddagger$                 & 0.152$^\ddagger$                 & 0.018$^\ddagger$                 & 0.016$^\ddagger$                 & 0.014$^\ddagger$                 \\
DVE$^*$     & \textbf{0.050(1.7\%)} & \textbf{0.092(3.7\%)} & \textbf{0.179(6.8\%)} & \textbf{0.020(0.8\%)} & \textbf{0.018(0.8\%)} & \textbf{0.016(0.7\%)} \\ \hline
\end{tabular}
\end{table*}
In particular, for a specific node, we recommend nodes that have high-probabilities to build positive directed links. For example, denote $i$ as the source node, we want to recommend a target node list $\mathcal{J}_{i}=[j_{1},j_{2},...,j_{k}]$ which is ranked according to the prediction scores to build positive links. $k$ means the cut off number. Specifically, we use the learned embeddings and calculate the prediction scores by the trained model on the test nodes. We take $Recall@k$ and $Precision@k$ as the evaluation metrics here. The results are shown in Table~\ref{table:node_recommendation_task_epinions},\ref{table:node_recommendation_task_slashdot},\ref{table:node_recommendation_task_wiki}, from which we have the following observations:
\begin{itemize}
    \item DVE outperforms other baselines of $Recall@k$ and $Precision@k$ on all datasets. Compared to SiNE on $Recall@50$, DVE even reaches a 2.6\% improvement on Epinions and a 2.7\% improvement on Slashdot and a 6.8\% increase on Wiki. Compared to the baseline methods that ignores the \emph{high-order} topology, DVE integrally extracts both the \emph{first-order} and \emph{high-order} topology, and learns more representative node embeddings for signed directed networks.
    \item Compared with other baselines (SiNE, SNE, MF), BPWR has better ability to mine the relative closeness relationships among nodes. For example, SiNE defines a limited distance metric by considering only signed links. In contrary, BPWR mines the mediator value of non-existent links. Furthermore, compared to SNEA-MR, BPWR in personalized ranking formulation is a smooth objective function and can be easily optimized by gradient based algorithms.
\end{itemize}
In order to investigate whether there is a statistical improvement of our method, we further conduct t-test experiments with 10 times for each setting. Results are shown in Table~\ref{table:link_sign_prediction_task},\ref{table:node_recommendation_task_epinions},\ref{table:node_recommendation_task_slashdot},\ref{table:node_recommendation_task_wiki}. 

From these tables, it is clear that the improvement of DVE is statistical significant with p-value<0.01 in comparison with baseline methods. The statistical improvement of DVE over BPWR indicates the importance of \emph{high-order} topological information extracted by decoupled variational encoder. Note that on link sign prediction task for Epinions dataset, p-value of DVE and DE is larger than 0.05, which indicates the improvement is not statistical significant. While for other two datasets, we get the opposite figures and conclusion. This is because DE is the non-variational variant of DVE and whether the variational one presents a better performance relies on the data distribution and prior distribution, which does not serve as a conflict of our main idea.

\subsubsection{\textbf{The Effect of Sparse Training Data}}
\begin{figure*}[h!]
\centering
\begin{minipage}[t]{0.32\textwidth}
\centering
\includegraphics[width=\textwidth]{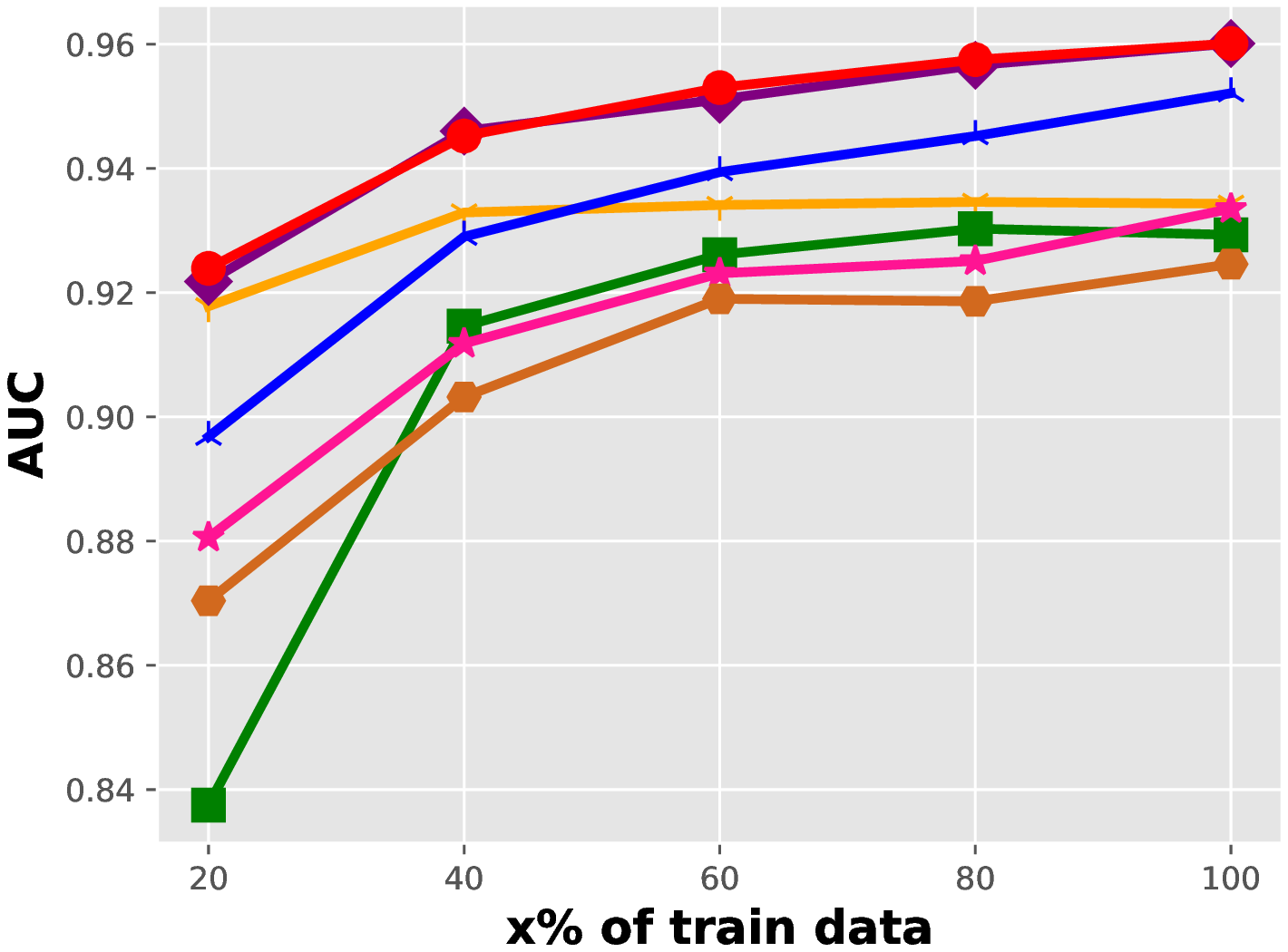}
\vspace{-17pt}
\caption*{(a) Epinions - AUC}
\end{minipage}
\begin{minipage}[t]{0.32\textwidth}
\centering
\includegraphics[width=\textwidth]{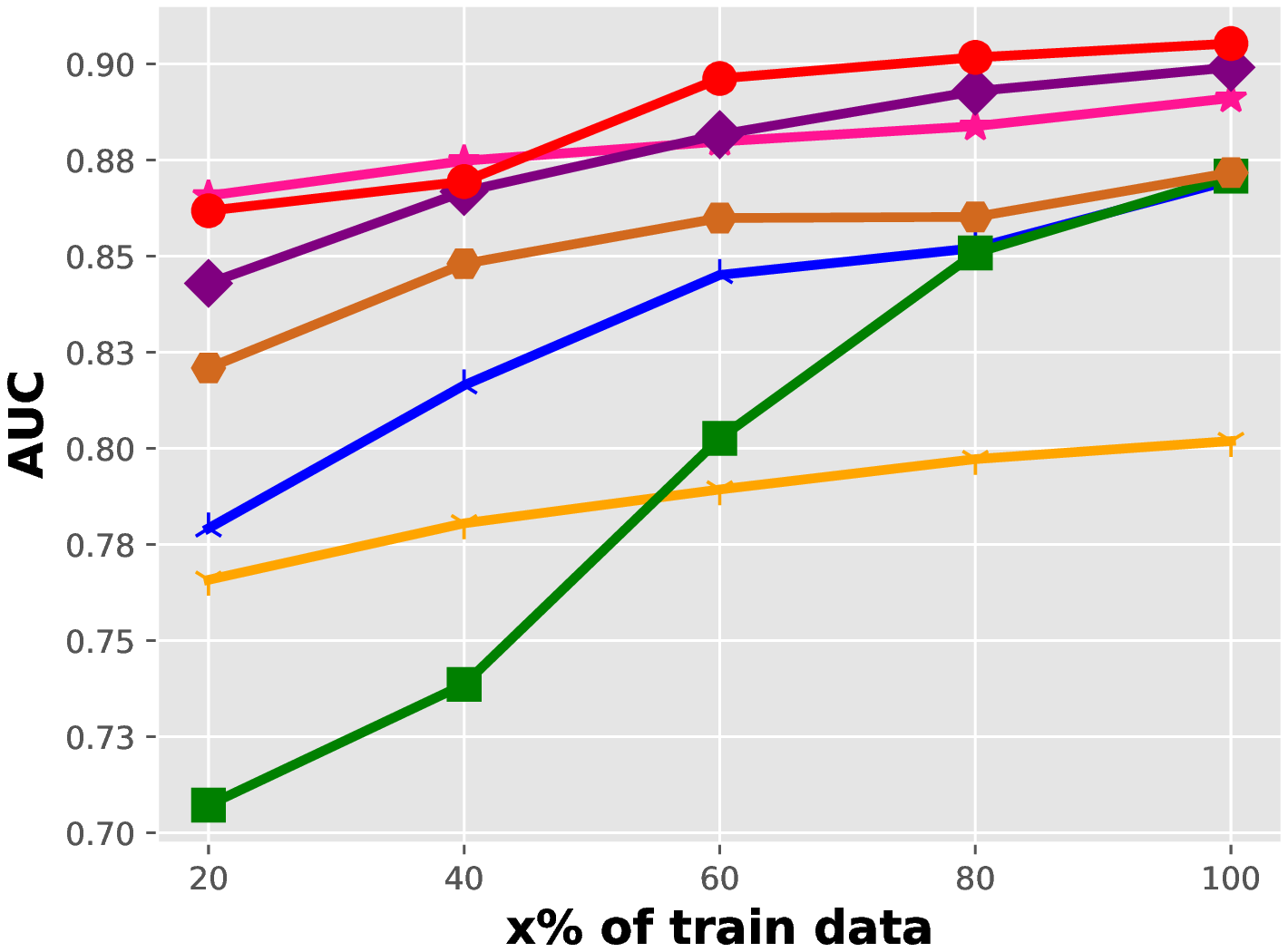}
\vspace{-17pt}
\caption*{(b) Slashdot - AUC}
\end{minipage} 
\begin{minipage}[t]{0.32\textwidth}
\centering
\includegraphics[width=\textwidth]{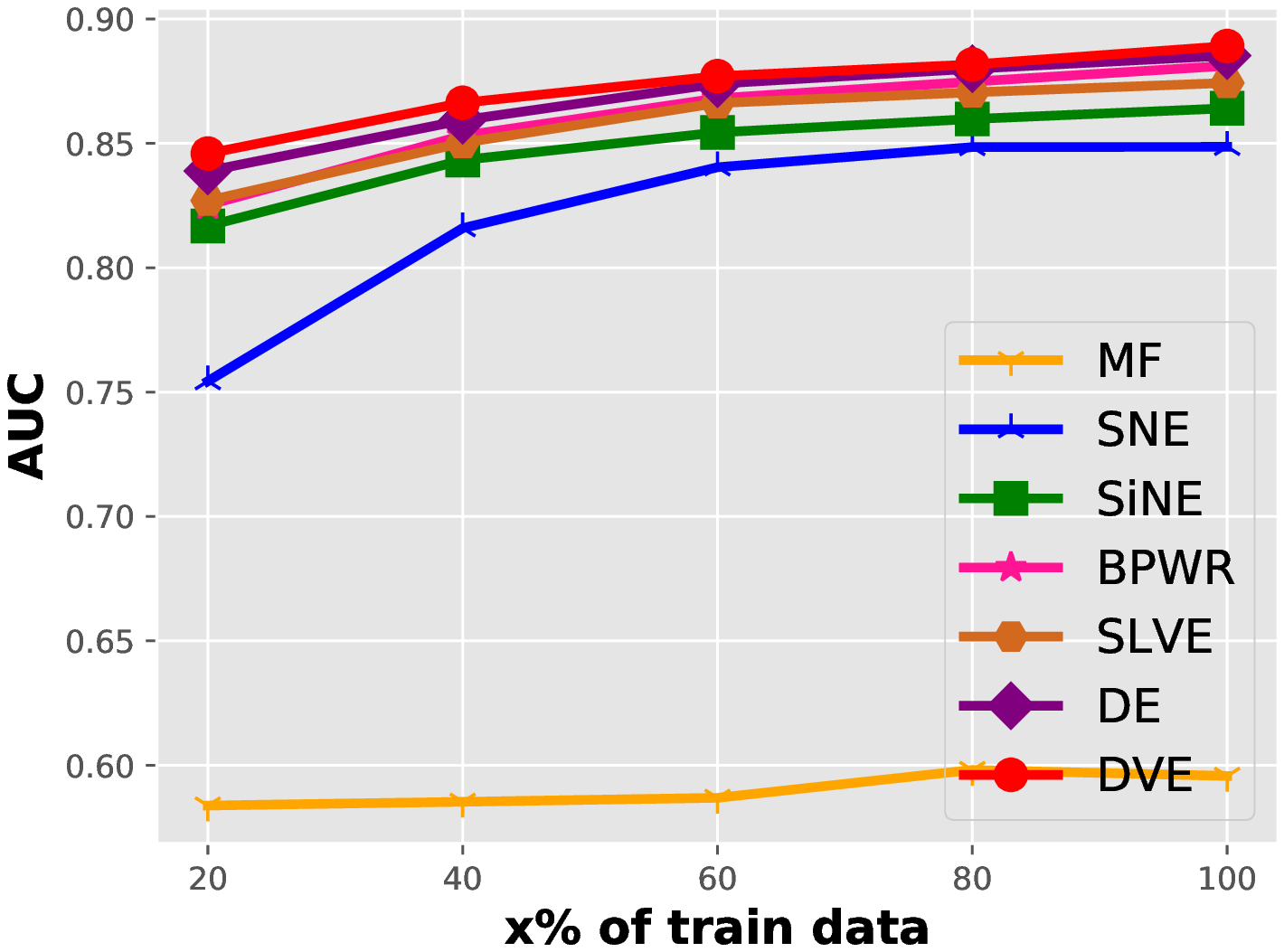}
\vspace{-17pt}
\caption*{(c) Wiki - AUC}
\end{minipage}  \\
\begin{minipage}[t]{0.32\textwidth}
\centering
\includegraphics[width=\textwidth]{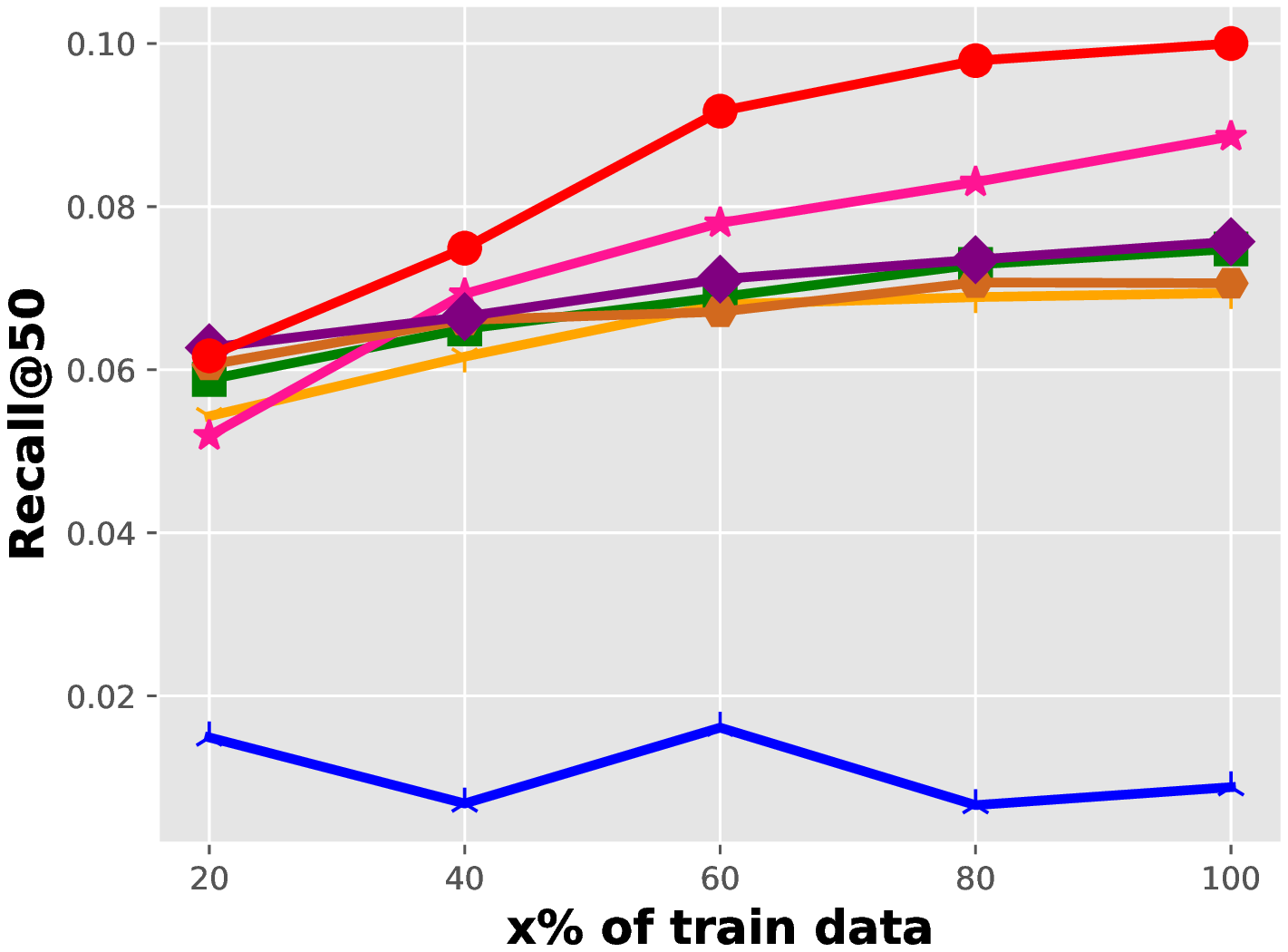}
\vspace{-17pt}
\caption*{(d) Epinions - Recall@50}
\end{minipage} 
\begin{minipage}[t]{0.32\textwidth}
\centering
\includegraphics[width=\textwidth]{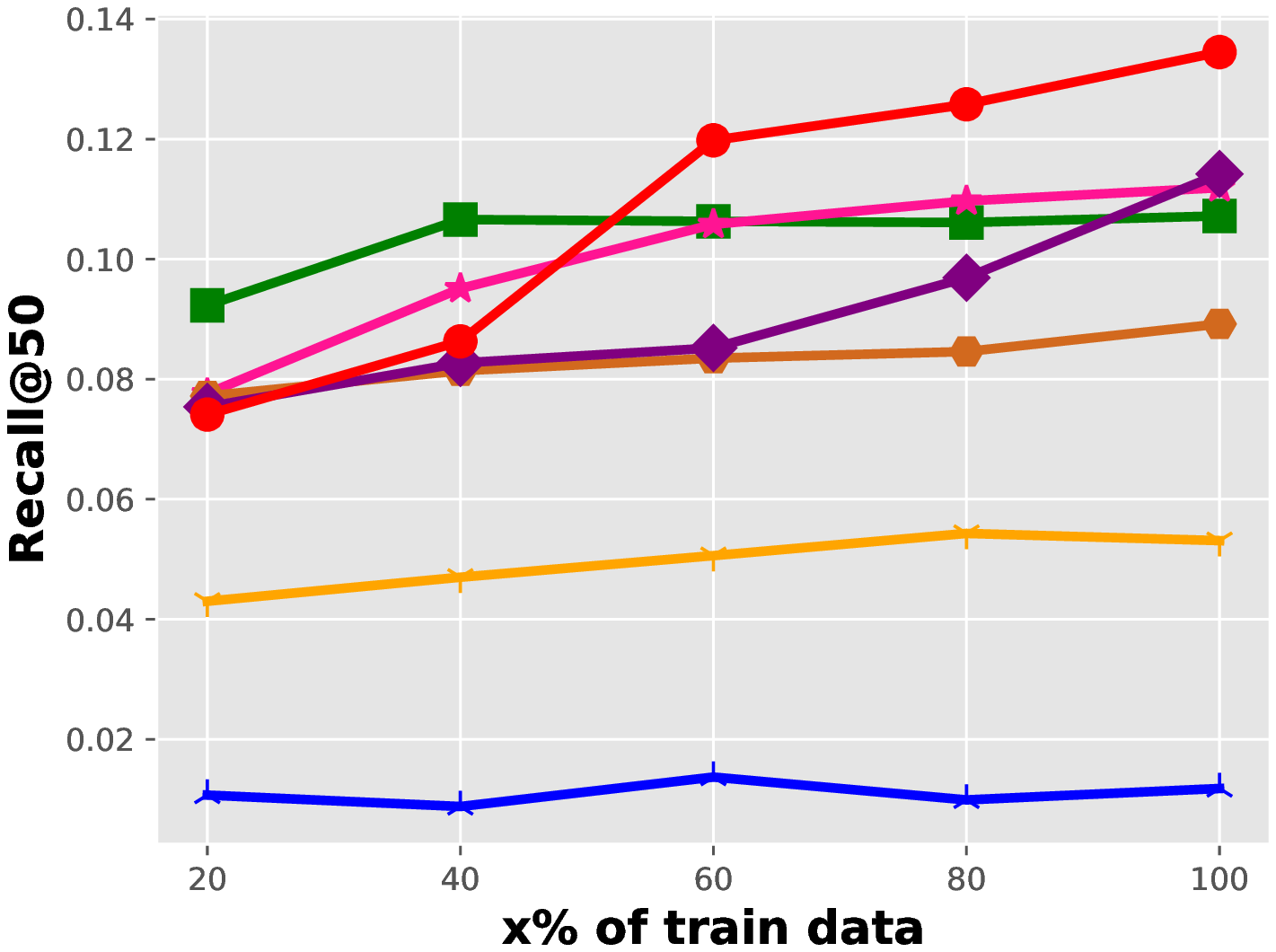}
\vspace{-17pt}
\caption*{(e) Slashdot - Recall@50}
\end{minipage}
\begin{minipage}[t]{0.32\textwidth}
\centering
\includegraphics[width=\textwidth]{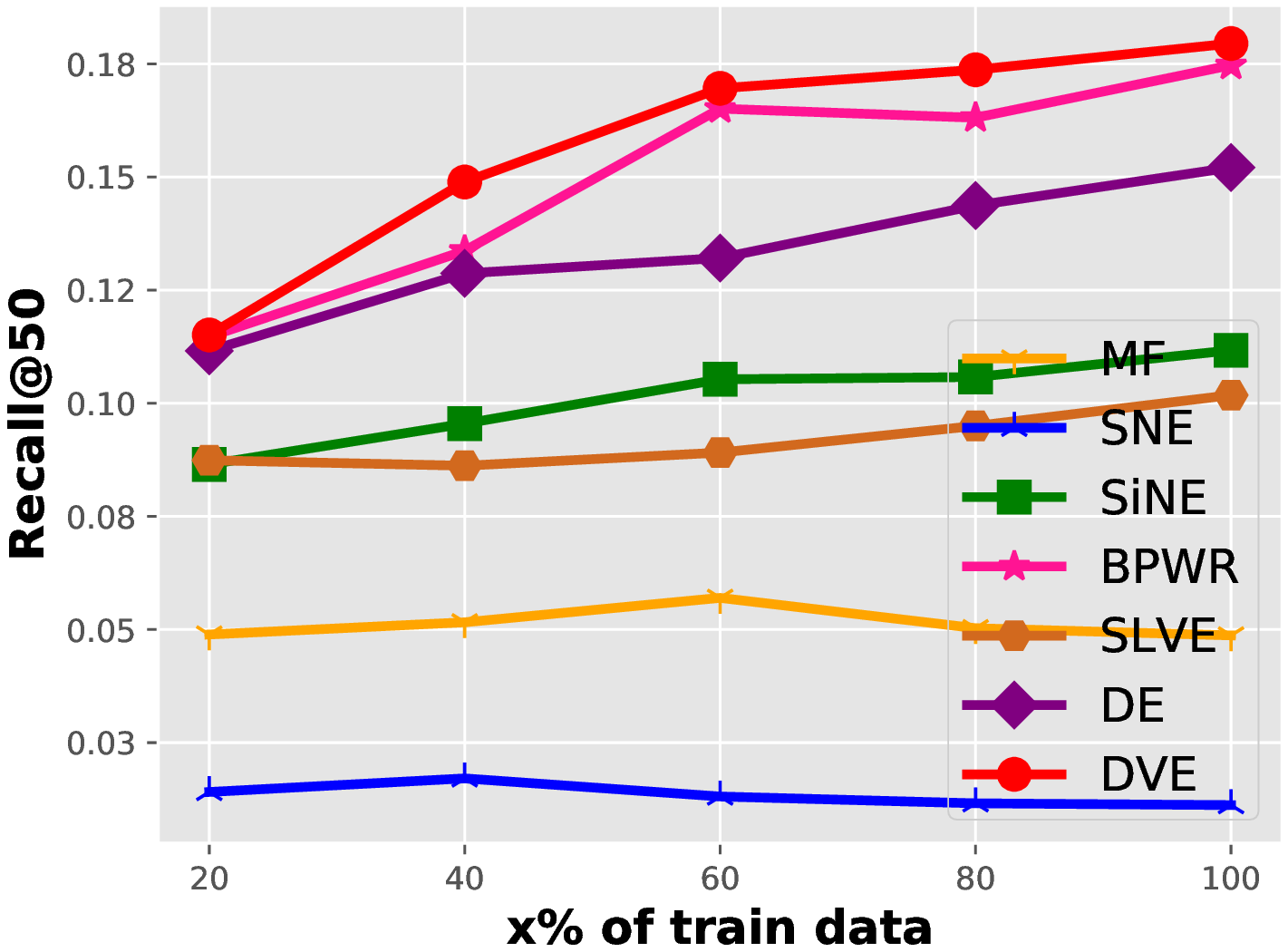}
\vspace{-17pt}
\caption*{(f) Wiki - Recall@50}
\end{minipage}
\caption{Comparison of methods with different training data on Epinions, Slashdot and Wiki for two tasks. AUC in (a)(b)(c) is the metric for link sign prediction task and Recall@50 in (d)(e)(f) is the metric for node recommendation task.}
\label{figure:ratio_train_data}
\end{figure*}

We investigate the effect of sparse training data on model performance. In particular, we vary the ratio of the 80\% training data as new train data and keep the 20\% test data fixed. Results are shown in Figure~\ref{figure:ratio_train_data}. From this figure, we see that:
\begin{itemize}
    \item Generally, the performance of each model decreases with the decline of training data, which indicates that sparse data causes deterioration to the model performance. Meanwhile, DVE consistently reaches the best performance in most sparse cases, which shows DVE has better adaptability in comparison with the baseline methods.  
    \item For node recommendation task in Figure~\ref{figure:ratio_train_data}~(d)(e)(f), it is obvious that both MF and SNE perform worse than other methods (e.g. SiNE, BPWR) in most cases, because MF and SNE are not ranking based loss. In contrary, SiNE and BPWR are both based on ranking loss which is more advantageous in node recommendation task.
\end{itemize} 
 
\begin{figure*}[h!]
\centering
\begin{minipage}[t]{0.45\textwidth}
\centering
\includegraphics[width=\textwidth]{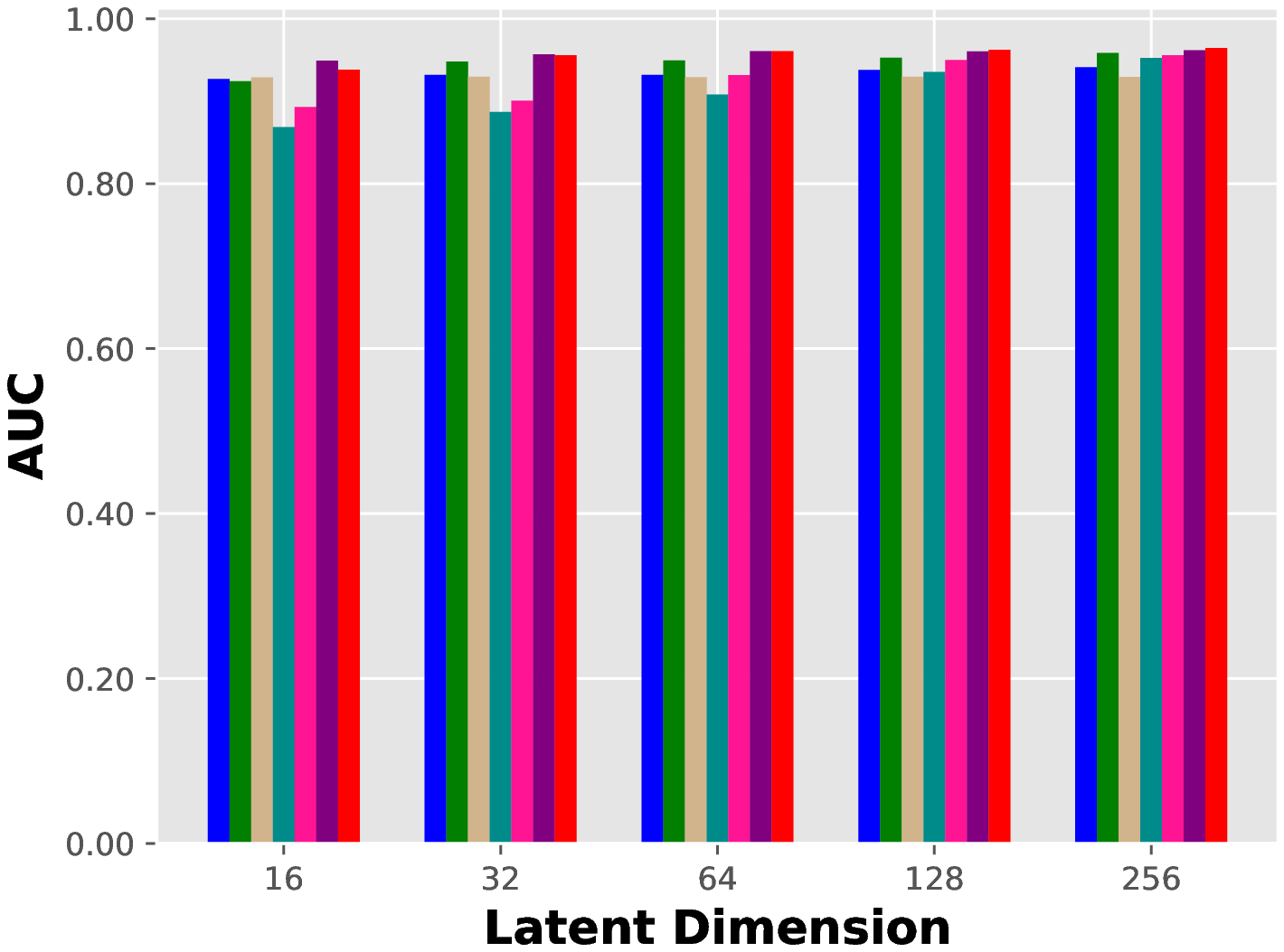}
\vspace{-17pt}
\caption*{(a) Epinions - AUC}
\end{minipage}
\begin{minipage}[t]{0.45\textwidth}
\centering
\includegraphics[width=\textwidth]{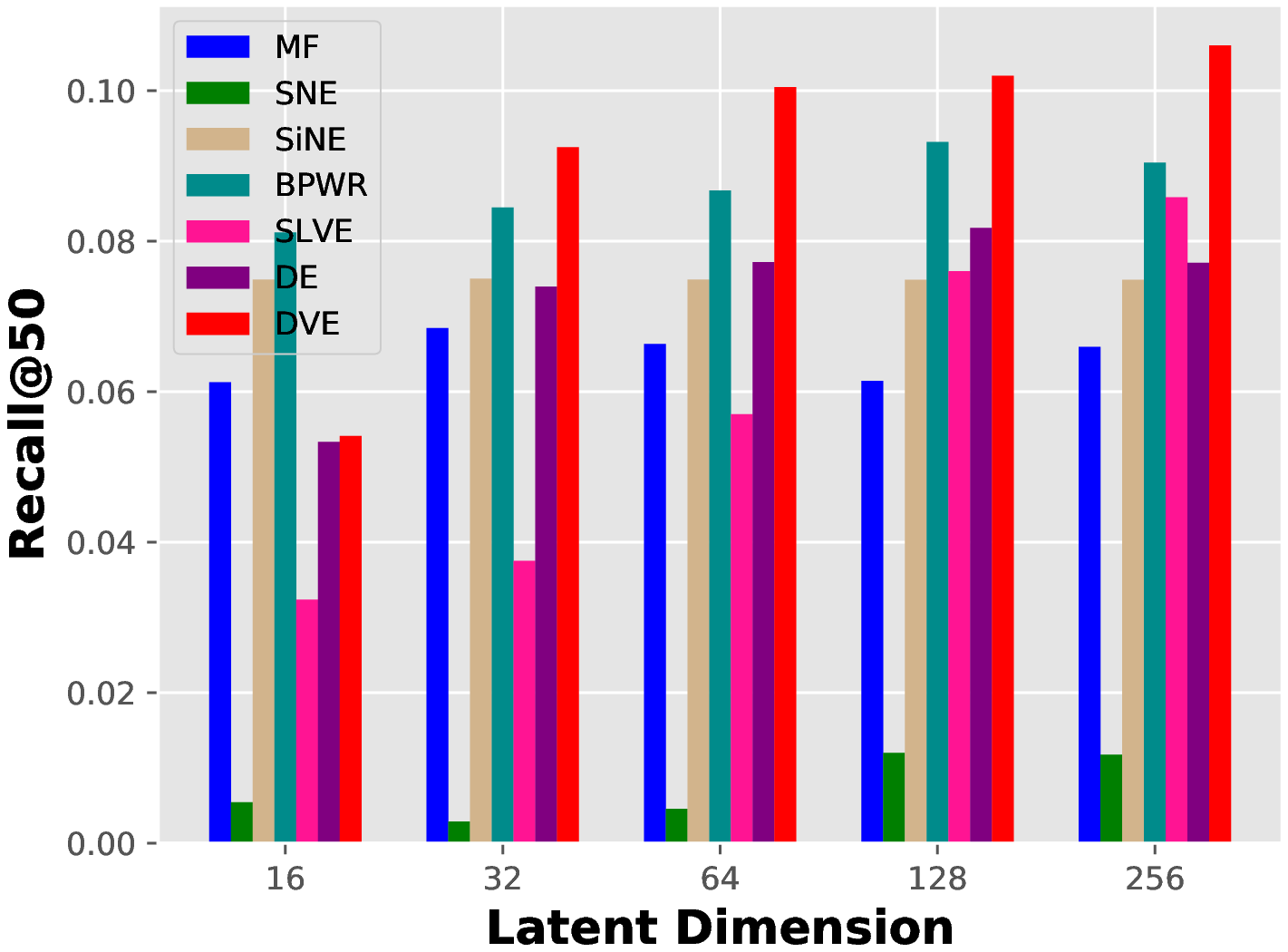}
\vspace{-17pt}
\caption*{(b) Slashdot - Recal@50} 
\end{minipage}  \\
\begin{minipage}[t]{0.45\textwidth}
\centering
\includegraphics[width=\textwidth]{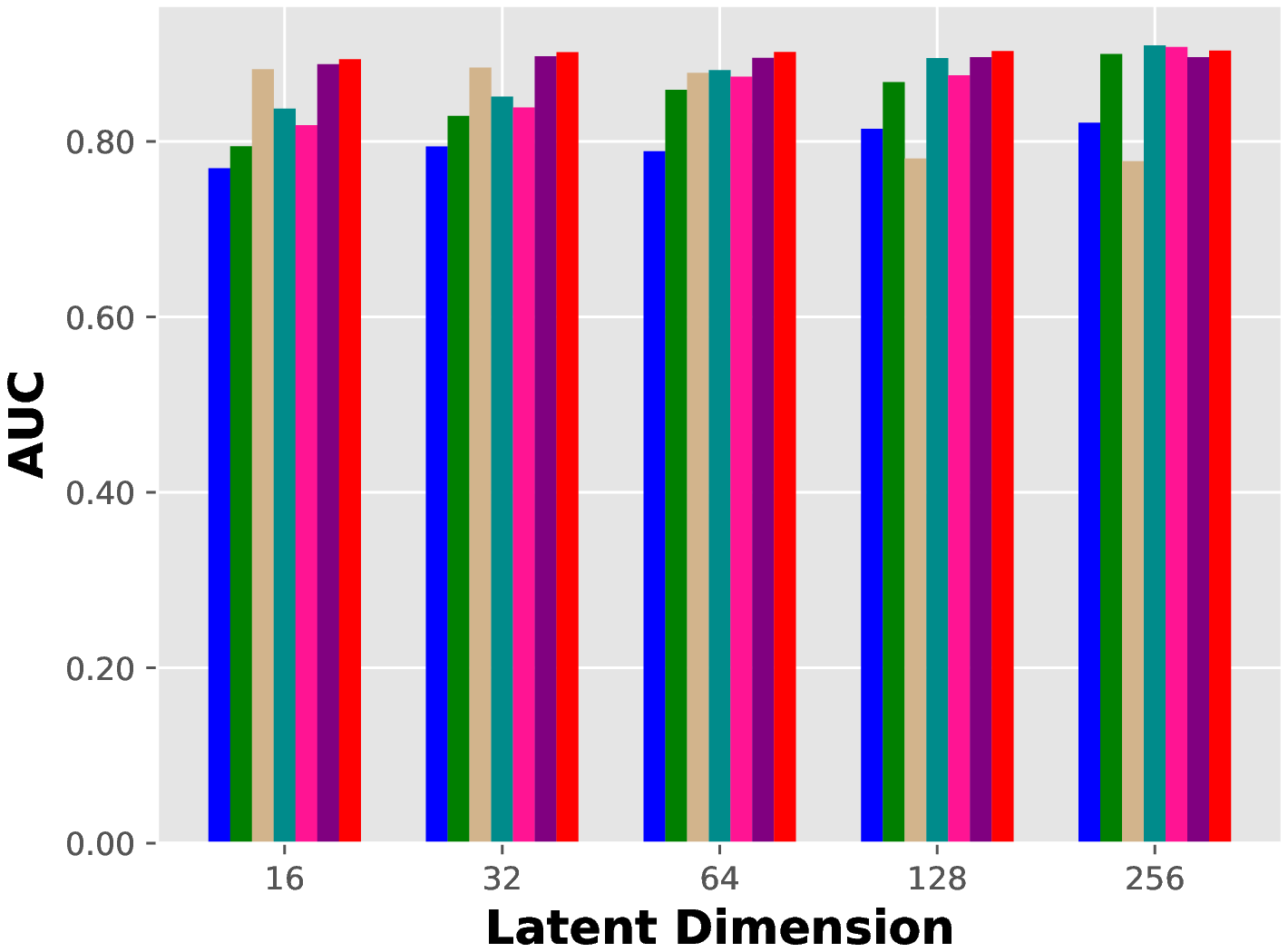}
\vspace{-17pt}
\caption*{(c) Wiki - AUC}
\end{minipage}  
\begin{minipage}[t]{0.45\textwidth}
\centering
\includegraphics[width=\textwidth]{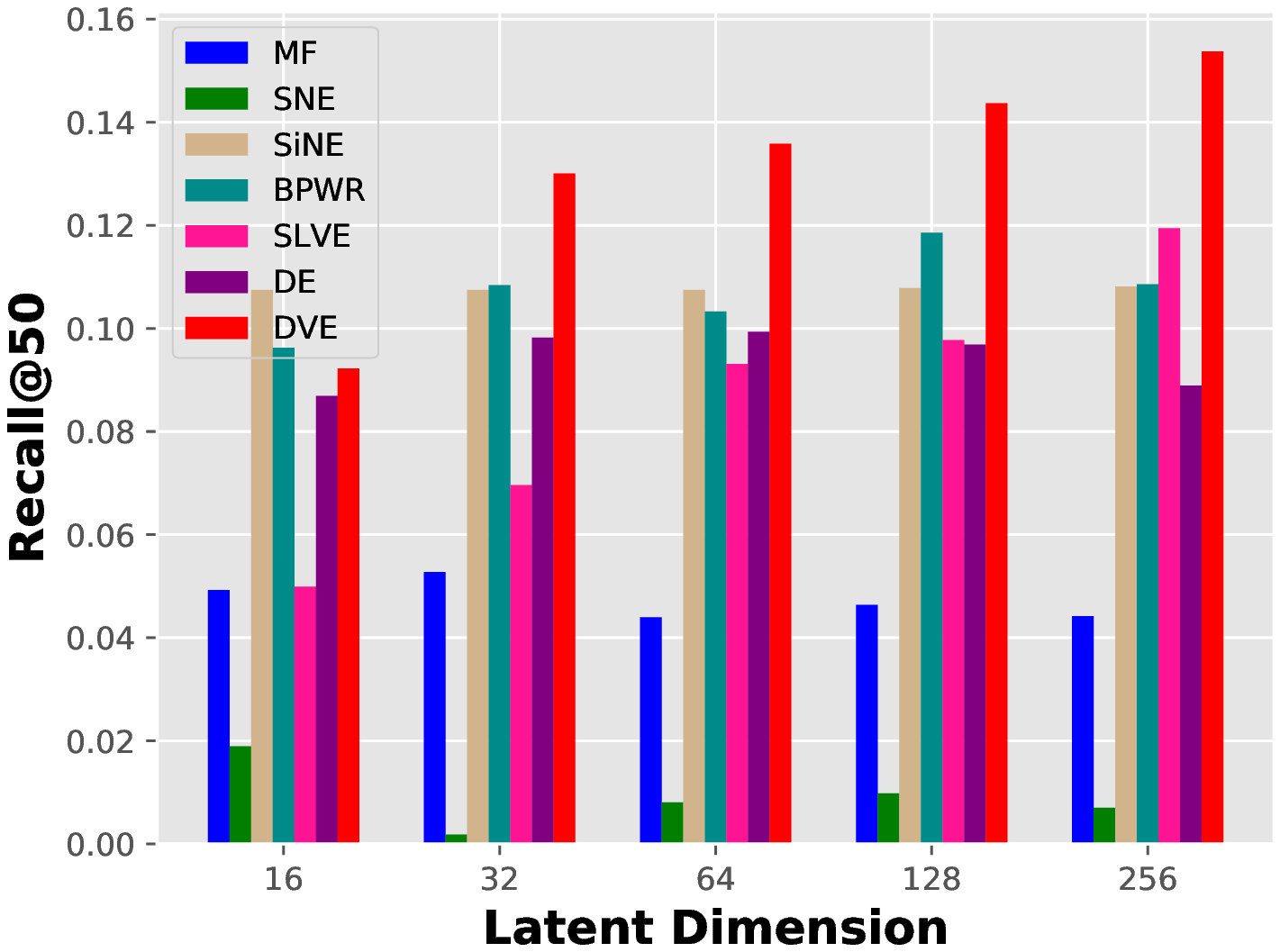}
\vspace{-17pt}
\caption*{(d) Epinions - Recall@50}
\end{minipage}  \\
\begin{minipage}[t]{0.45\textwidth}
\centering
\includegraphics[width=\textwidth]{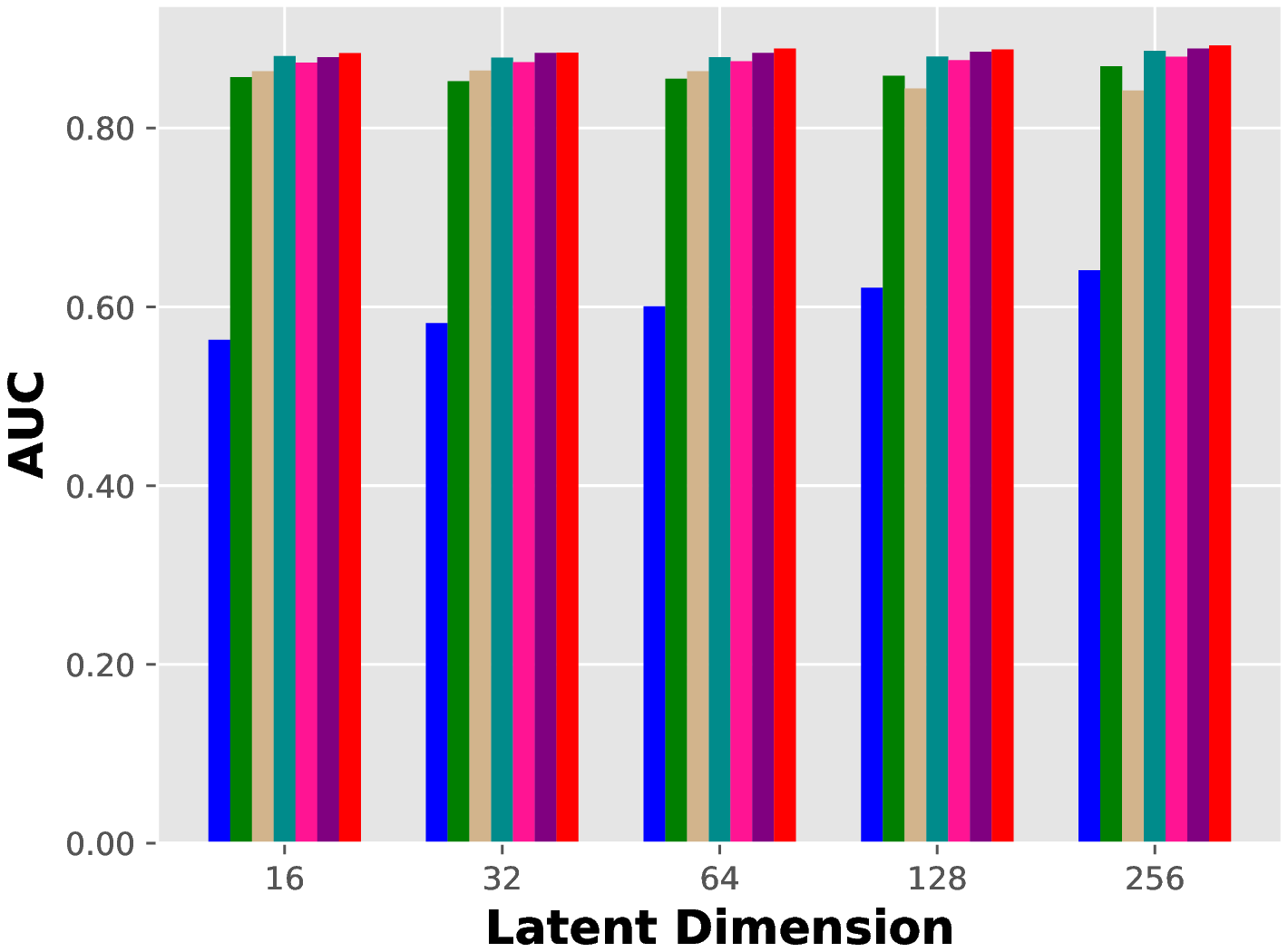}
\vspace{-17pt}
\caption*{(e) Slashdot - AUC}
\end{minipage}
\begin{minipage}[t]{0.45\textwidth}
\centering
\includegraphics[width=\textwidth]{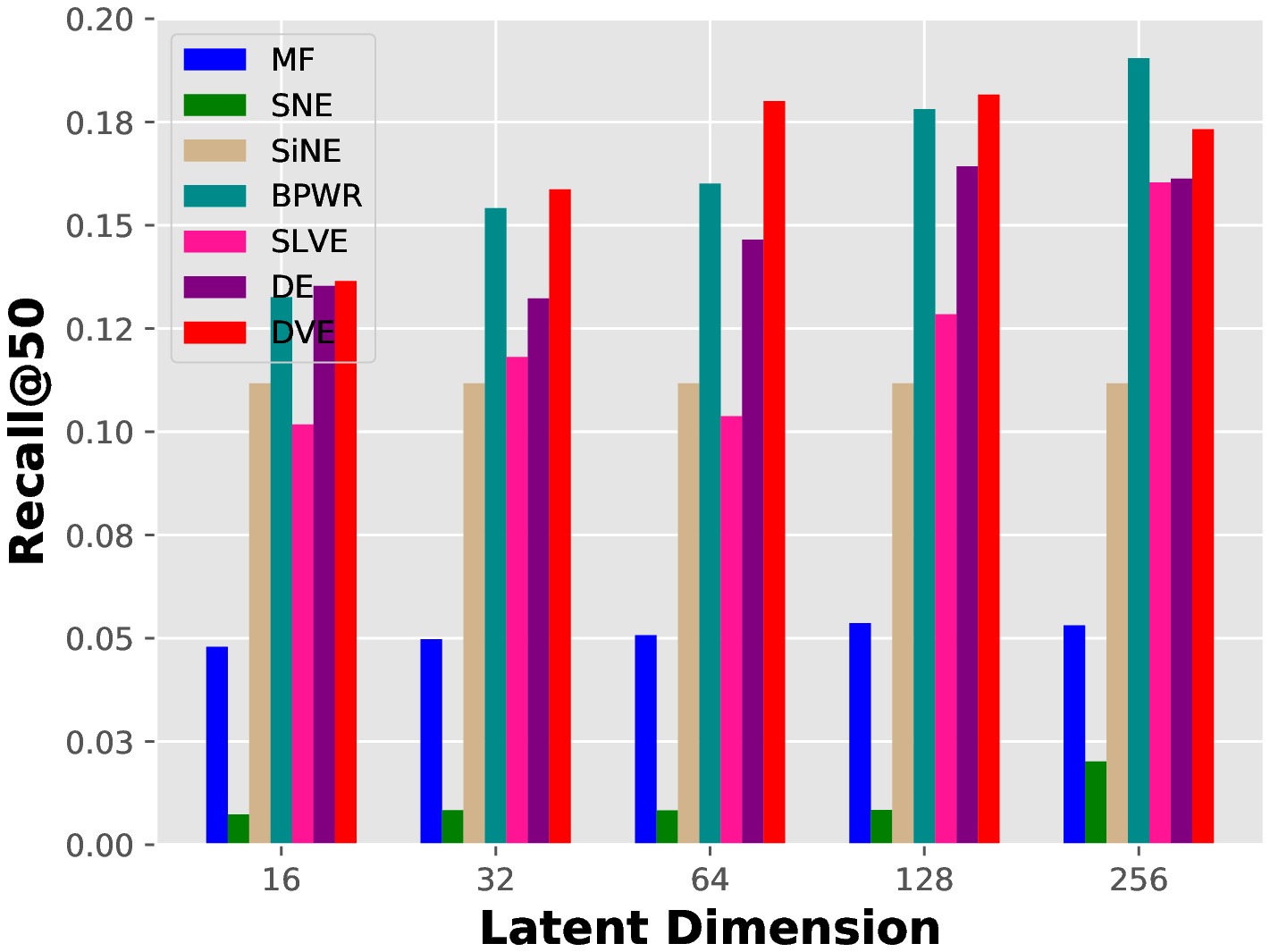}
\vspace{-17pt}
\caption*{(f) Wiki - Recall@50}
\end{minipage}
% \vspace{-5pt}
\caption{Comparison of methods with different latent dimension on Epinions, Slashdot and Wiki for two tasks. AUC in (a)(c)(e) is the metric for link sign prediction task and Recall@50 in (b)(d)(f) is the metric for node recommendation task.}
\label{figure:latent_d}
\end{figure*}

\subsubsection{\textbf{The Effect of Different Latent Dimensions}}
The latent dimension of embeddings is an important factor that accounts for the model performance in network embedding. We thus conduct an experiment to investigate the effect of different latent dimensions varying in $[16,32,64,128,256]$. The results are shown in Figure~\ref{figure:latent_d}. From this figure, we have the following observations:
\begin{itemize}
    \item The proposed methods (BPWR, DE and DVE) consistently outperform other baselines with different latent dimensions. DVE achieves the best performance in most cases, because DVE is the only method that simultaneously captures both the \emph{first-order} and \emph{high-order} topological information in signed directed networks.
    \item It is worthwhile to notice that DVE tends to reach a better performance at a higher dimension in comparison with SiNE. DVE considering both the \emph{high-order} and the \emph{first-order} topology requires a high dimension to encode the additional information. As for the baselines that only considers the \emph{first-order} topology, when the dimension is high, the information in learned embeddings tends to be redundant and yields unsatisfying performance on the test set.
\end{itemize}

\begin{figure}[h!]
\centering
\begin{minipage}[t]{0.32\textwidth}
\centering
\includegraphics[width=\textwidth]{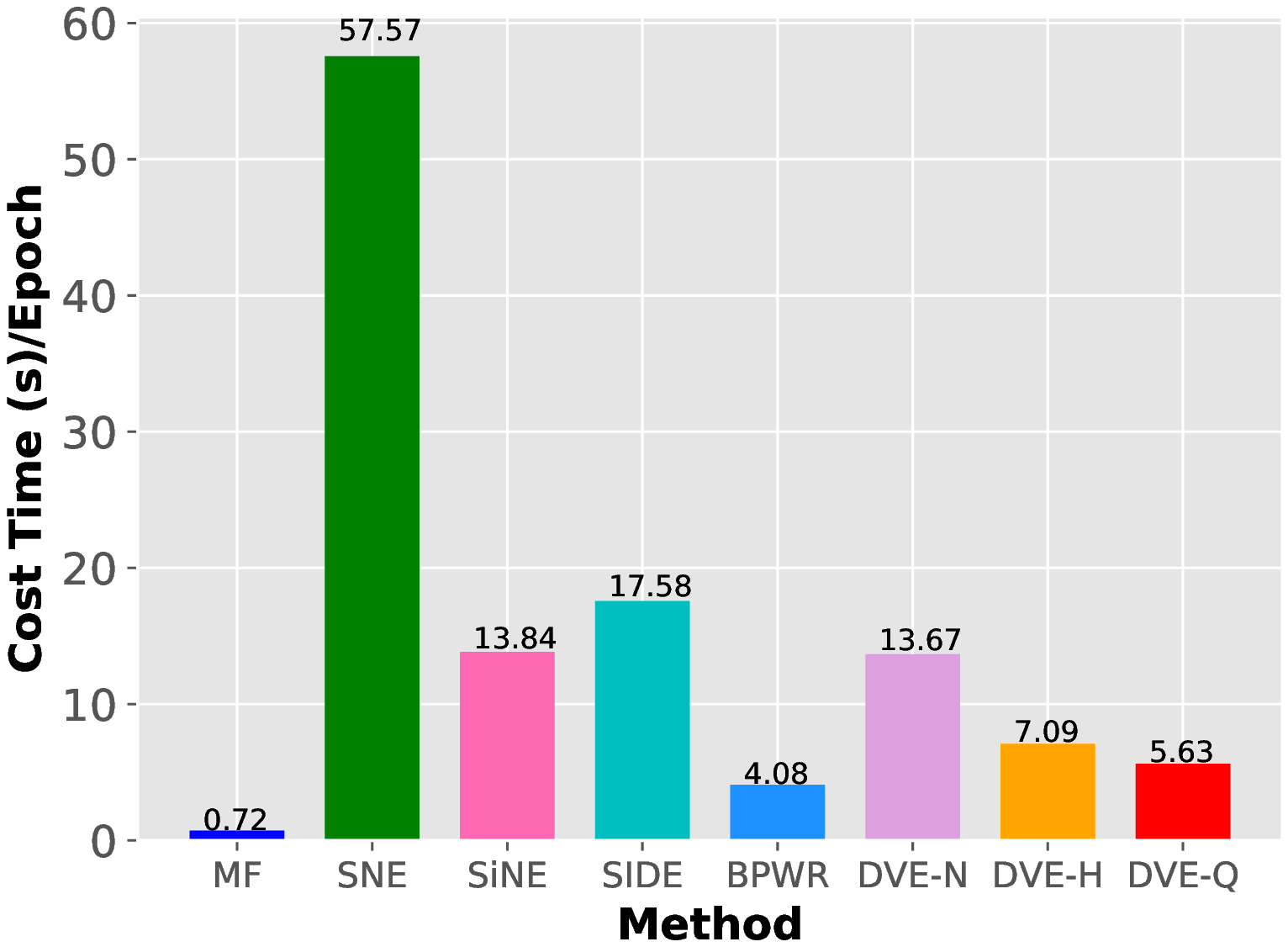}
\vspace{-17pt}
\caption*{(a) Epinions}
\end{minipage}
\begin{minipage}[t]{0.32\textwidth}
\centering
\includegraphics[width=\textwidth]{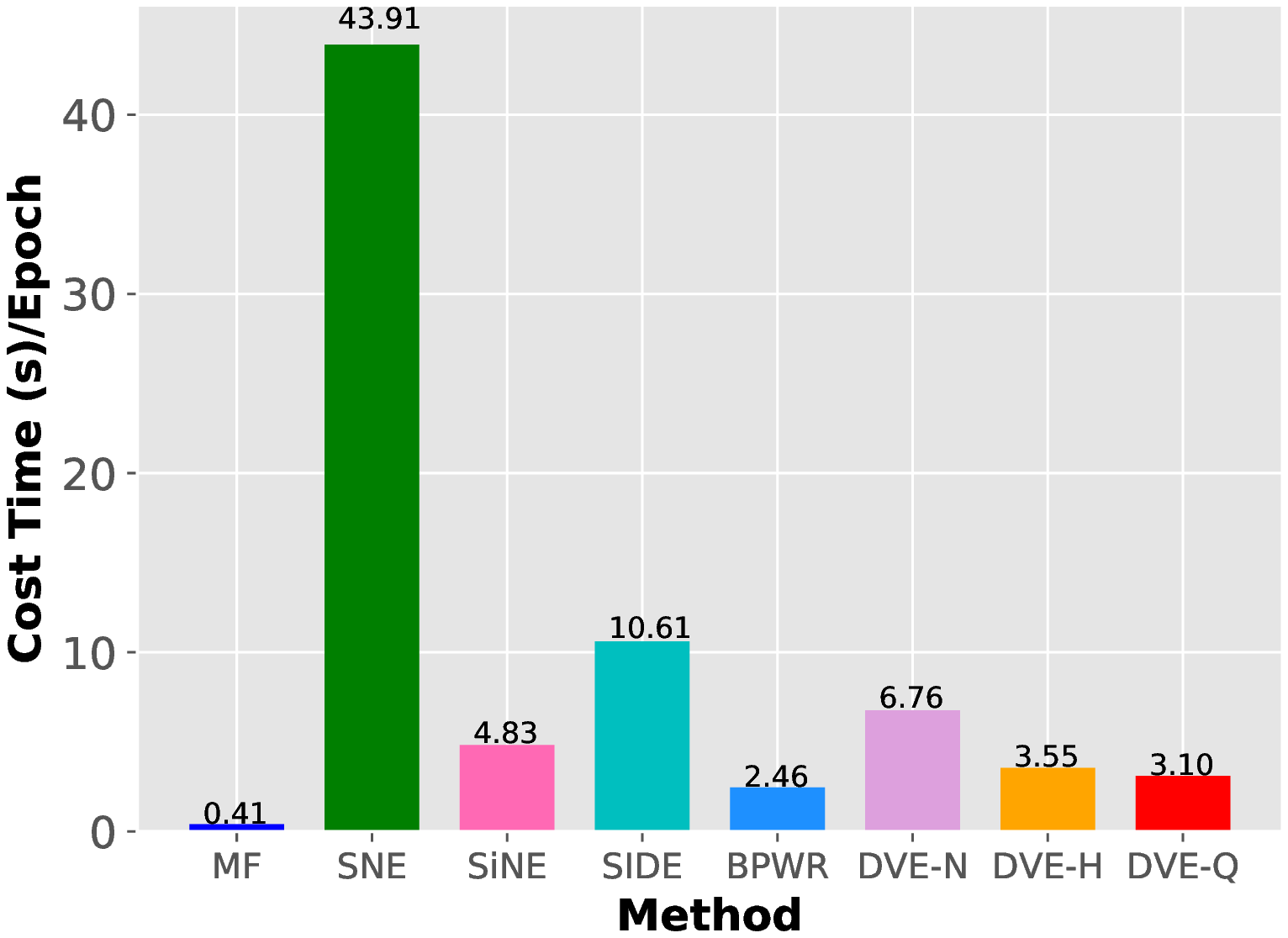}
\vspace{-17pt}
\caption*{(b) Slashdot}
\end{minipage}
\begin{minipage}[t]{0.32\textwidth}
\centering
\includegraphics[width=\textwidth]{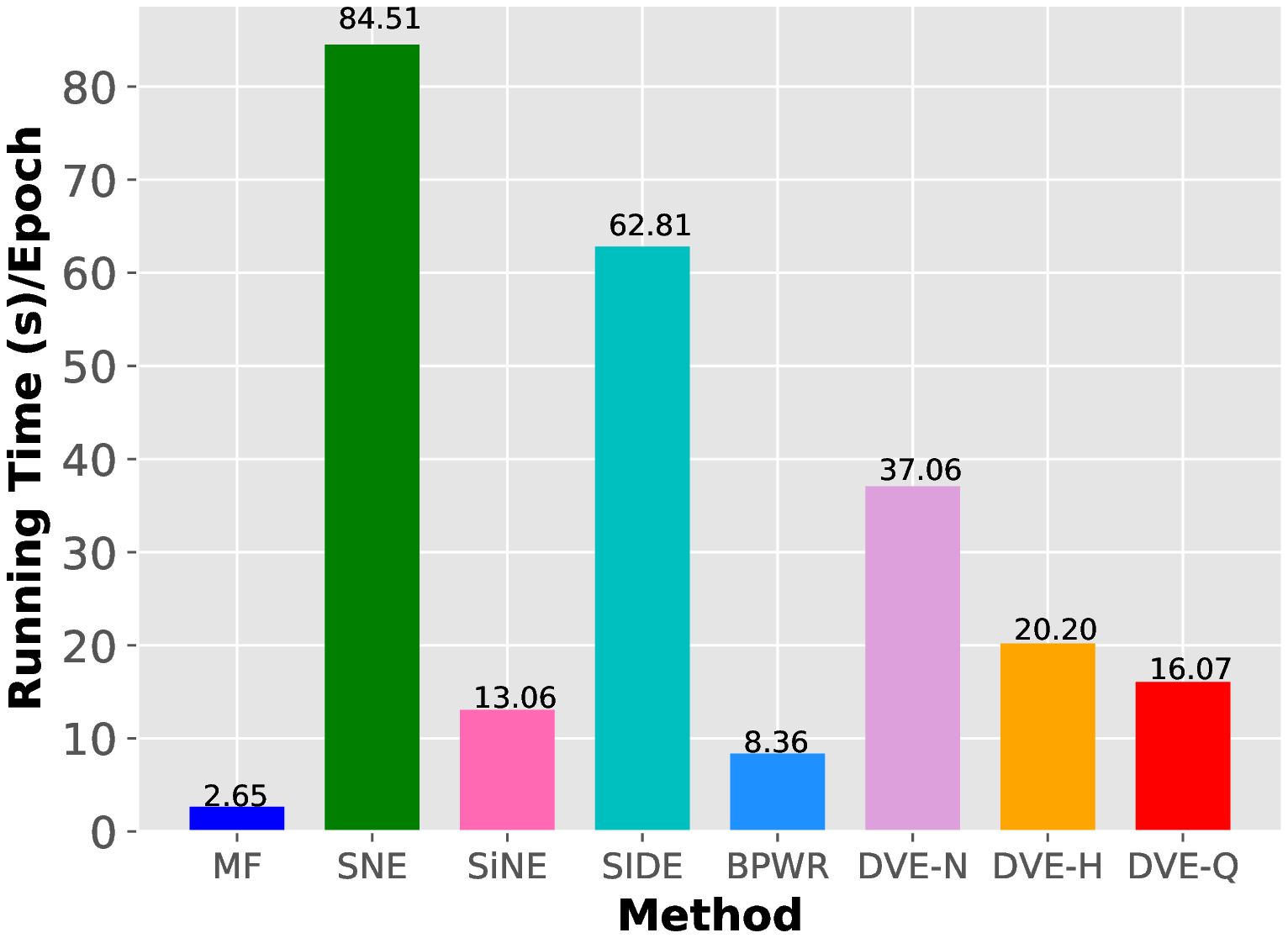}
\vspace{-17pt}
\caption*{(c) Wiki}
\end{minipage} 
\caption{The empirical running time in each epoch of different methods. In this figure, DVE-N indicates the non-parallel DVE, DVE-H means the half-parallel one and the DVE-Q denotes the quarter-parallel one.}
\label{figure:running_time}
\end{figure}
\subsubsection{\textbf{Empirical Running Time Analysis.}} To investigate the time complexity, we conduct an experiment to compare the empirical running time in each epoch of different methods. 
In particular, we set the training batch number is 1,000 for all methods. For the baselines (MF, SNE, SiNE, SIDE), we follow the hyper-parameter settings in the source codes provided by the authors. All these methods are implemented with deep learning programming frameworks such as Pytorch, Tensorflow or Theano. For DVE, we implement it with Tensorflow and the sampling size $n_{noise}$ of non-existent links is 5,5,20 on Epinions, Slashdot and Wiki, respectively. Since DVE has parallel versions, we thus denote DVE-N as the non-parallel one, DVE-H as the half-parallel one and DVE-Q as the quarter-parallel one. We conduct the experiments 10 times on the same machine with one Nvidia-1080 GPU. The mean value of running time per epoch is reported in Figure~\ref{figure:running_time}. We see that:
\begin{itemize}
    \item MF costs the least time because of its simple scheme. SNE and SIDE involving the softmax operation consume much time than other methods. The proposed DVE in non-parallel version (DVE-N) generally costs more time than MF and SiNE, because DVE is more complex to capture both the \emph{high-order} and \emph{first-order} topological information. 
    \item DVE-H and DVE-Q take much less time than DVE-N. They are even faster than SiNE in some cases. Meanwhile, DVE provides better performance in comparison with SiNE. In summary, the decoupling idea has advantages to accelerate the training process as well as learns more representative node embeddings for signed directed networks.  
\end{itemize}

\subsection{Ablation Study}
\begin{figure}[h!]
\centering
\begin{minipage}[t]{0.32\textwidth}
\centering
\includegraphics[width=\textwidth]{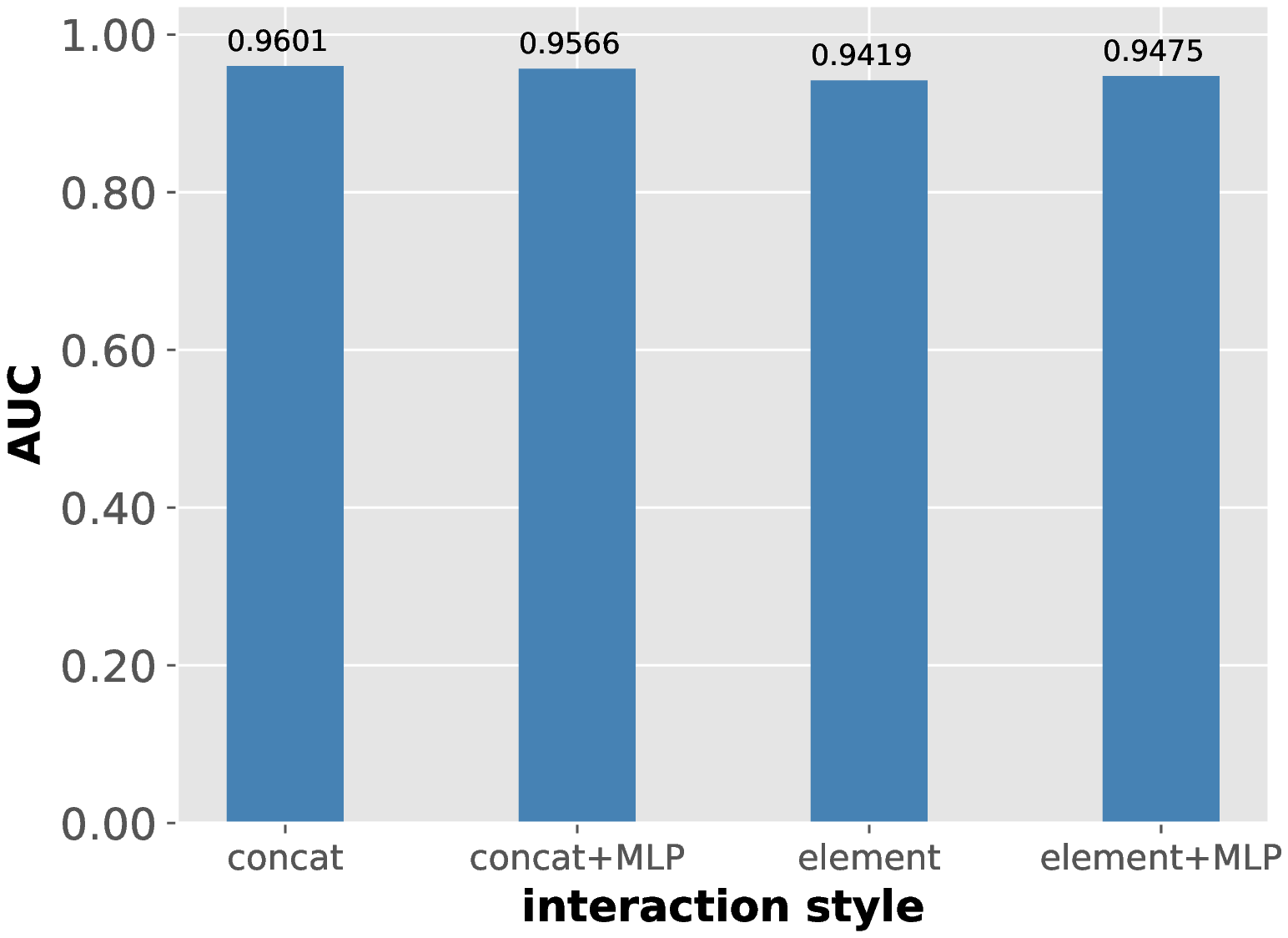}
\vspace{-17pt}
\caption*{(a) Epinions - AUC}
\end{minipage}
\begin{minipage}[t]{0.32\textwidth}
\centering
\includegraphics[width=\textwidth]{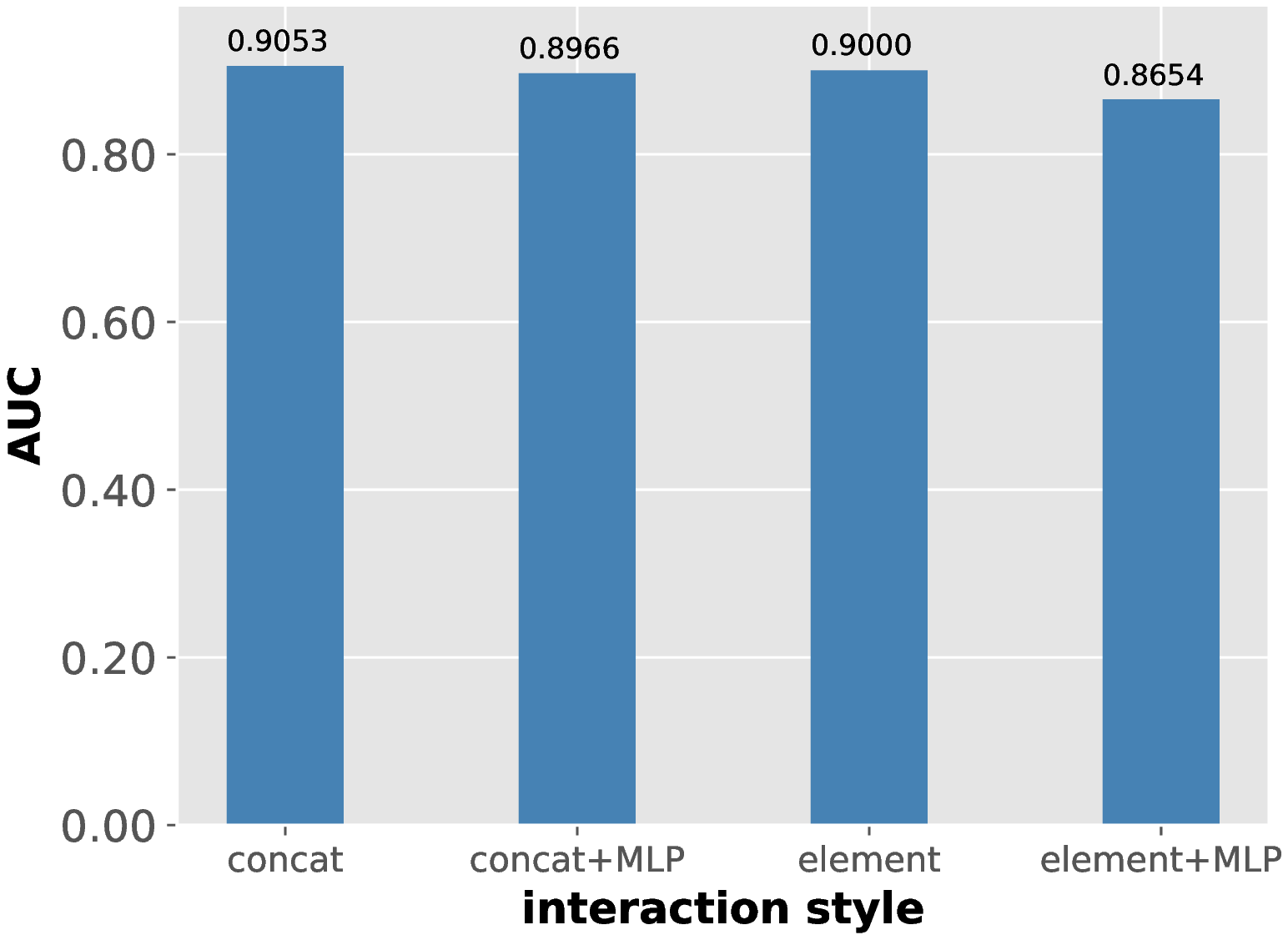}
\vspace{-17pt}
\caption*{(b) Slashdot - AUC}
\end{minipage}
\begin{minipage}[t]{0.32\textwidth}
\centering
\includegraphics[width=\textwidth]{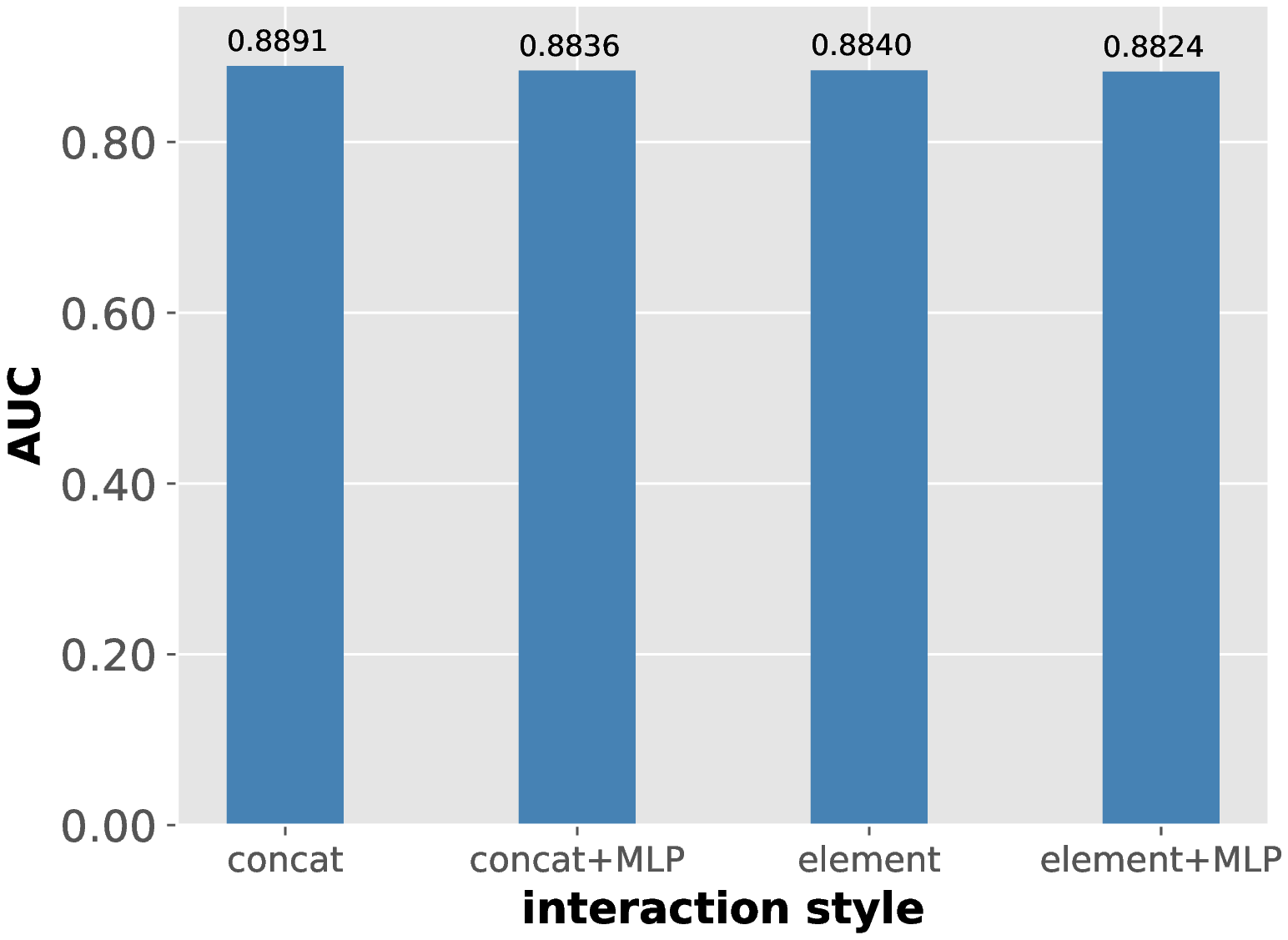}
\vspace{-17pt}
\caption*{(c) Wiki - AUC}
\end{minipage} \\
\begin{minipage}[t]{0.32\textwidth}
\centering
\includegraphics[width=\textwidth]{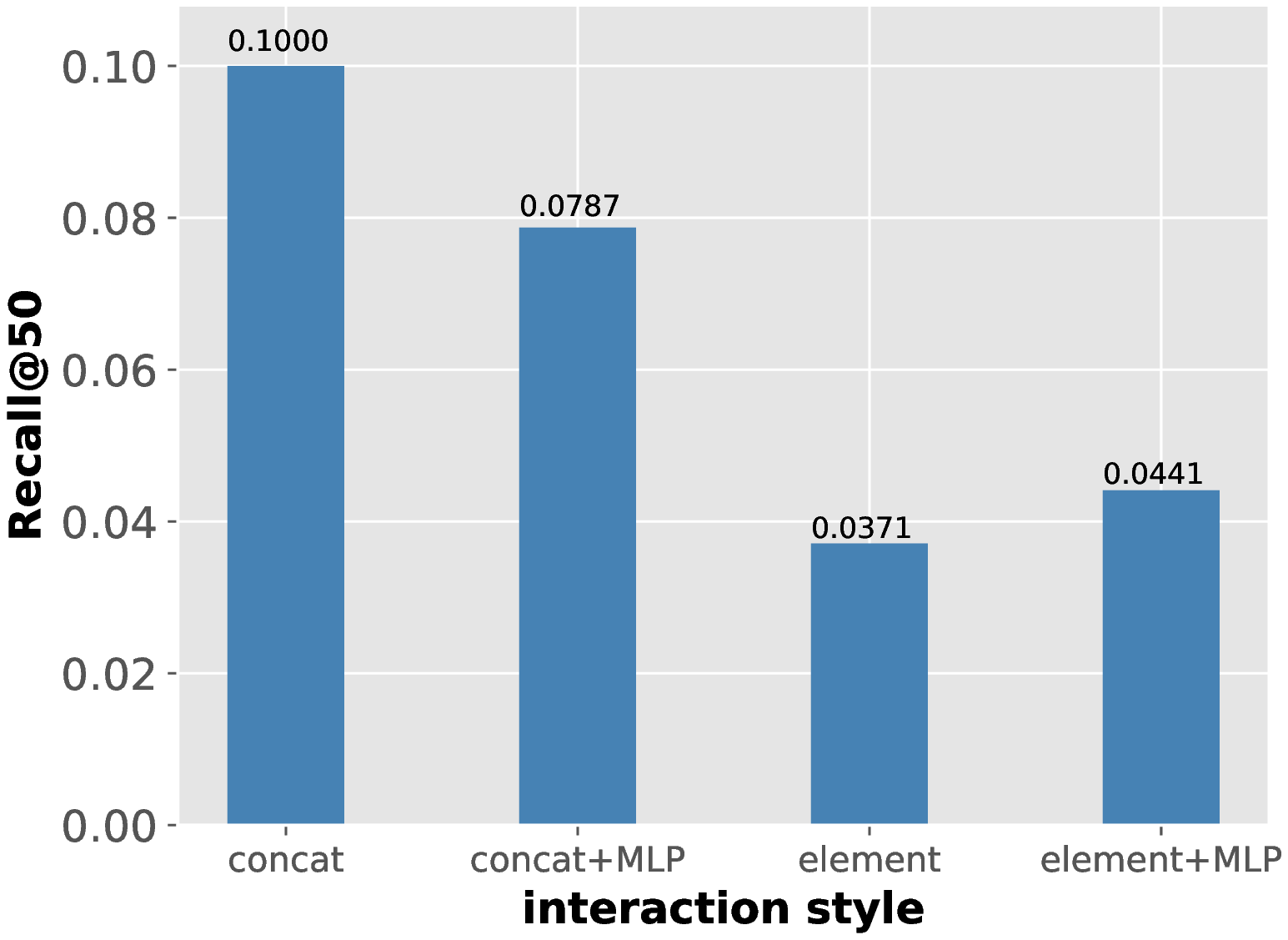}
\vspace{-17pt}
\caption*{(d) Epinions - Recall@50}
\end{minipage}
\begin{minipage}[t]{0.32\textwidth}
\centering
\includegraphics[width=\textwidth]{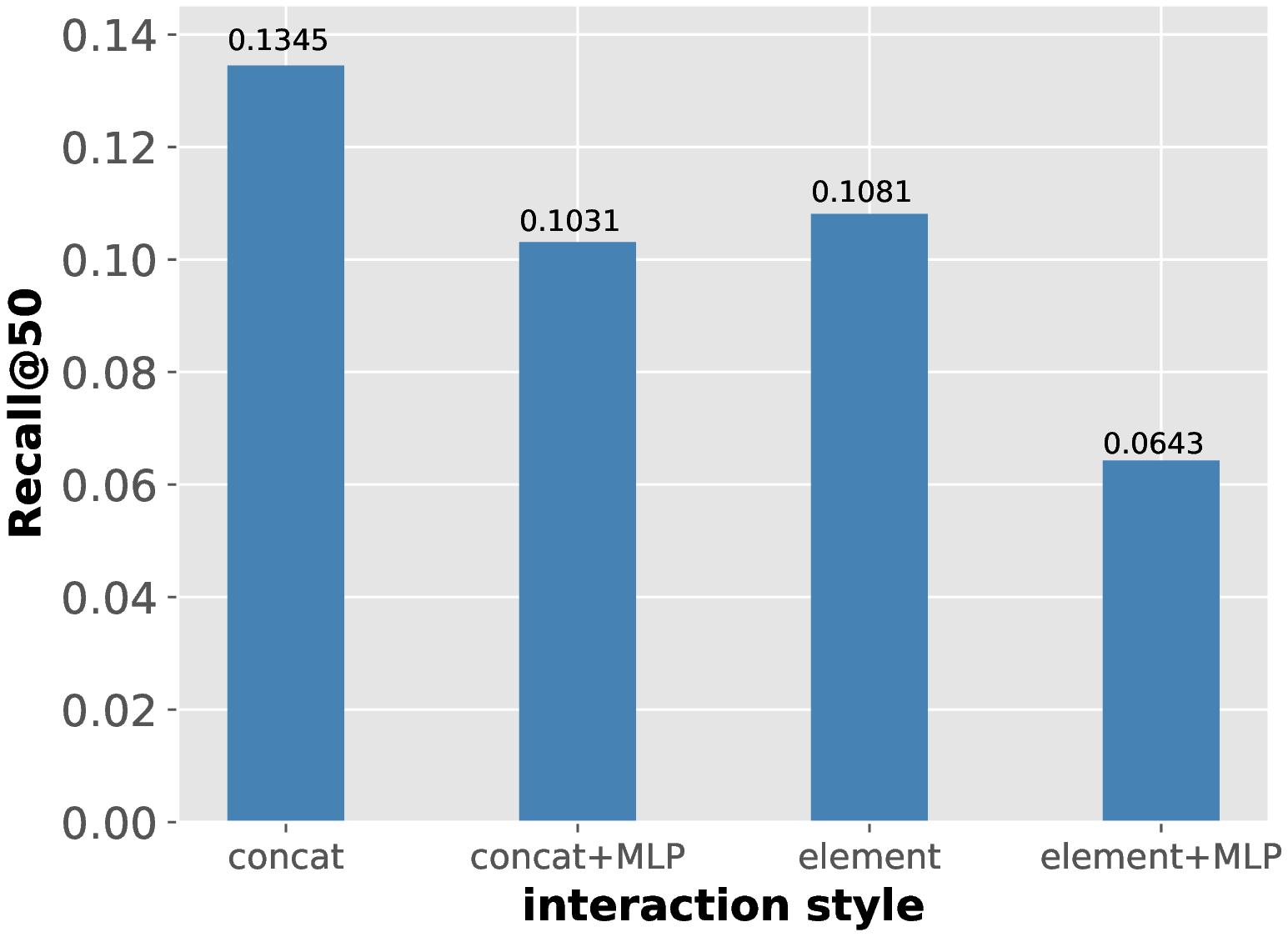}
\vspace{-17pt}
\caption*{(e) Slashdot - Recall@50}
\end{minipage}
\begin{minipage}[t]{0.32\textwidth}
\centering
\includegraphics[width=\textwidth]{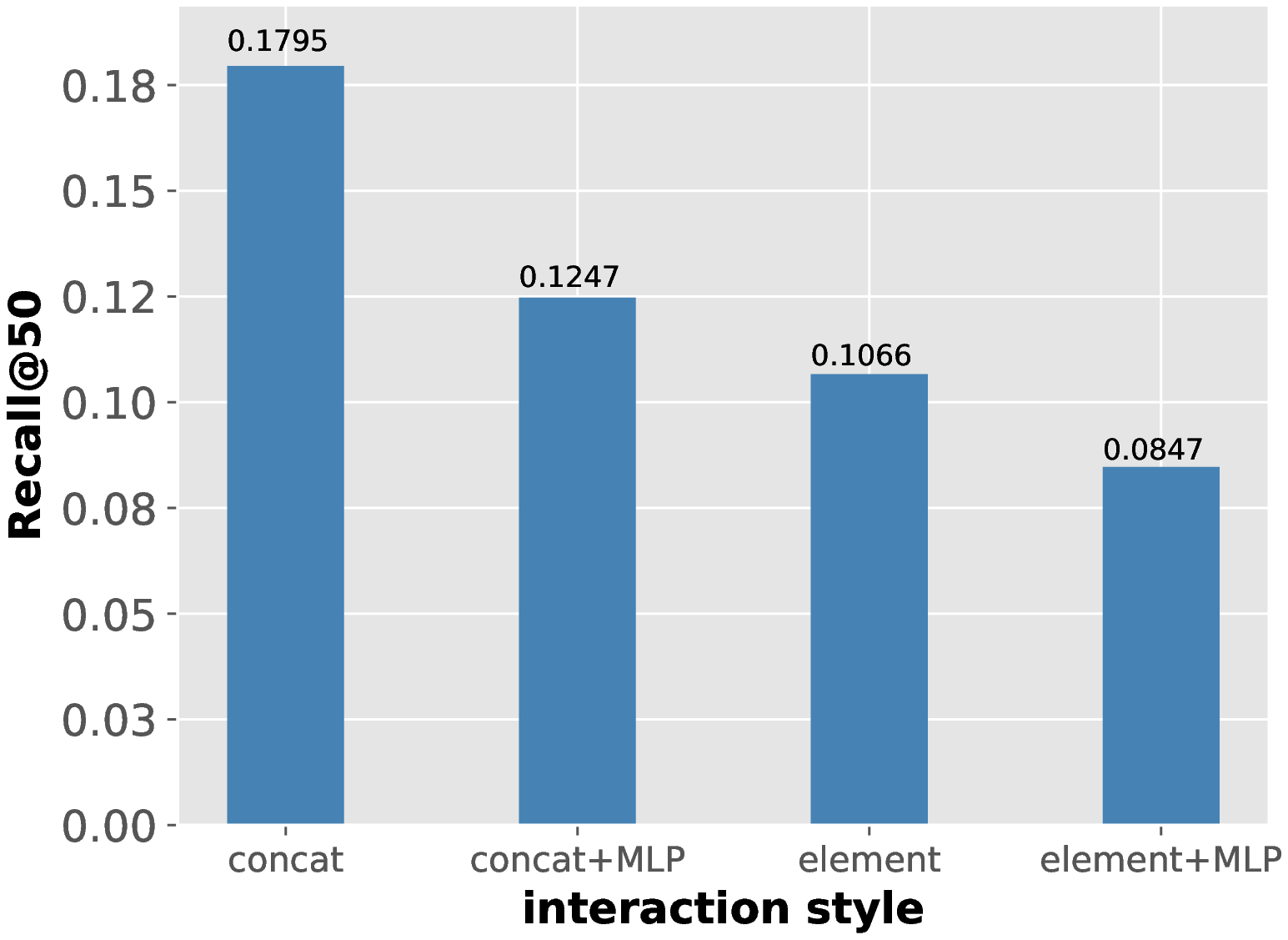}
\vspace{-17pt}
\caption*{(f) Wiki - Recall@50}
\end{minipage} 
\caption{Performance of DVE with different generative functions. AUC in (a)(b)(c) refers to the metric for link sign prediction task. Recall@50 in (d)(e)(f) is the metric for node recommendation task.}
\label{figure:generate_func_sensitivity}
\end{figure}
\subsubsection{\textbf{The Effect of Different Generative Functions}}
Remind that we assume the source node embeddings $Z_{s}$ could be generated through $Z_{s}=f_{s}(Z_{s}^{p},Z_{s}^{n})$, where $f_{s}$ is the generative function. In order to explore the influence of different generative functions, we conduct an experiment with various functions that are defined in Table~\ref{table:different_functions}. Note that we only use $f_{s}$ as an example to illustrate the experiment setting here. The target node representation $Z_{t}$ has similar formulation, while the notations are $Z_{t}^{p},Z_{t}^{n}$ and $f_{t}$. The results are shown in Figure~\ref{figure:generate_func_sensitivity}.
\begin{table}[]
\caption{Different generative functions for $f_{s}$. In this table, $[\cdot,\cdot]$ means the concatenation operation and $W_{C}\in \mathbb{R}^{2d\times 2d}$ is the weight of MLP for concatenation. $\odot$ indicates the element-wise product operation and $W_{E}\in \mathbb{R}^{d\times d}$ is the weight of MLP for element-wise product. }
\label{table:different_functions}
\begin{tabular}{c|c}
\hline
type                     & formula \\ \hline
concat                   & $Z_{s}=[Z_{s}^{p}, Z_{s}^{n}]$       \\
concat+MLP               & $Z_{s}=([Z_{s}^{p}, Z_{s}^{n}])W_{C}$       \\
element-wise product     & $Z_{s}=Z_{s}^{p}\odot Z_{s}^{n}$       \\
element-wise product+MLP & $Z_{s}=(Z_{s}^{p}\odot Z_{s}^{n})W_{E}$       \\ \hline
\end{tabular}
\end{table}

From Figure~\ref{figure:generate_func_sensitivity}, we observe two interesting phenomenons: 1) concatenation does better than concatenation+MLP and inner product does better than element-wise product+MLP; 2) concatenation performs better than element-wise product and concatenation+MLP performs better than element-wise product+MLP. The main reason for the first phenomenon may be that additional trainable parameters lead to over-fitting on the sparse graph data. For the second phenomenon, it is because both $Z_{s}^{p}$ and $Z_{s}^{n}$ are learned with distinctive deep neural networks, which indicates they are in different latent spaces. The aligned element-wise product may lead to information loss to represent source node embeddings $Z_{s}$. Therefore, concatenation operation tends to be the most suitable choice among them in terms of both efficiency and easy implementation.

\begin{figure}[h!]
\centering
\begin{minipage}[t]{0.32\textwidth}
\centering
\includegraphics[width=\textwidth]{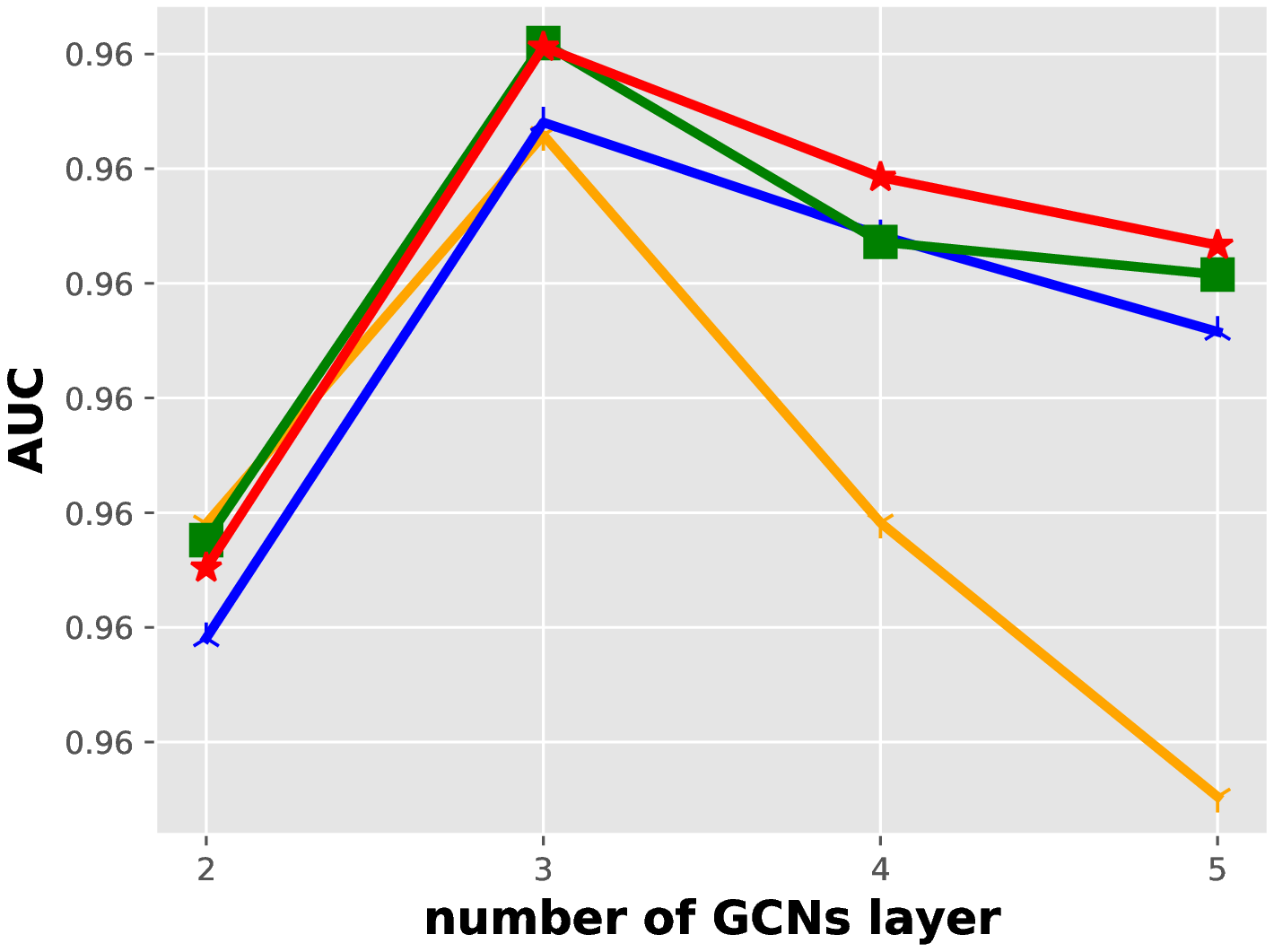}
\vspace{-17pt}
\caption*{(a) Epinions - AUC}
\end{minipage}
\begin{minipage}[t]{0.32\textwidth}
\centering
\includegraphics[width=\textwidth]{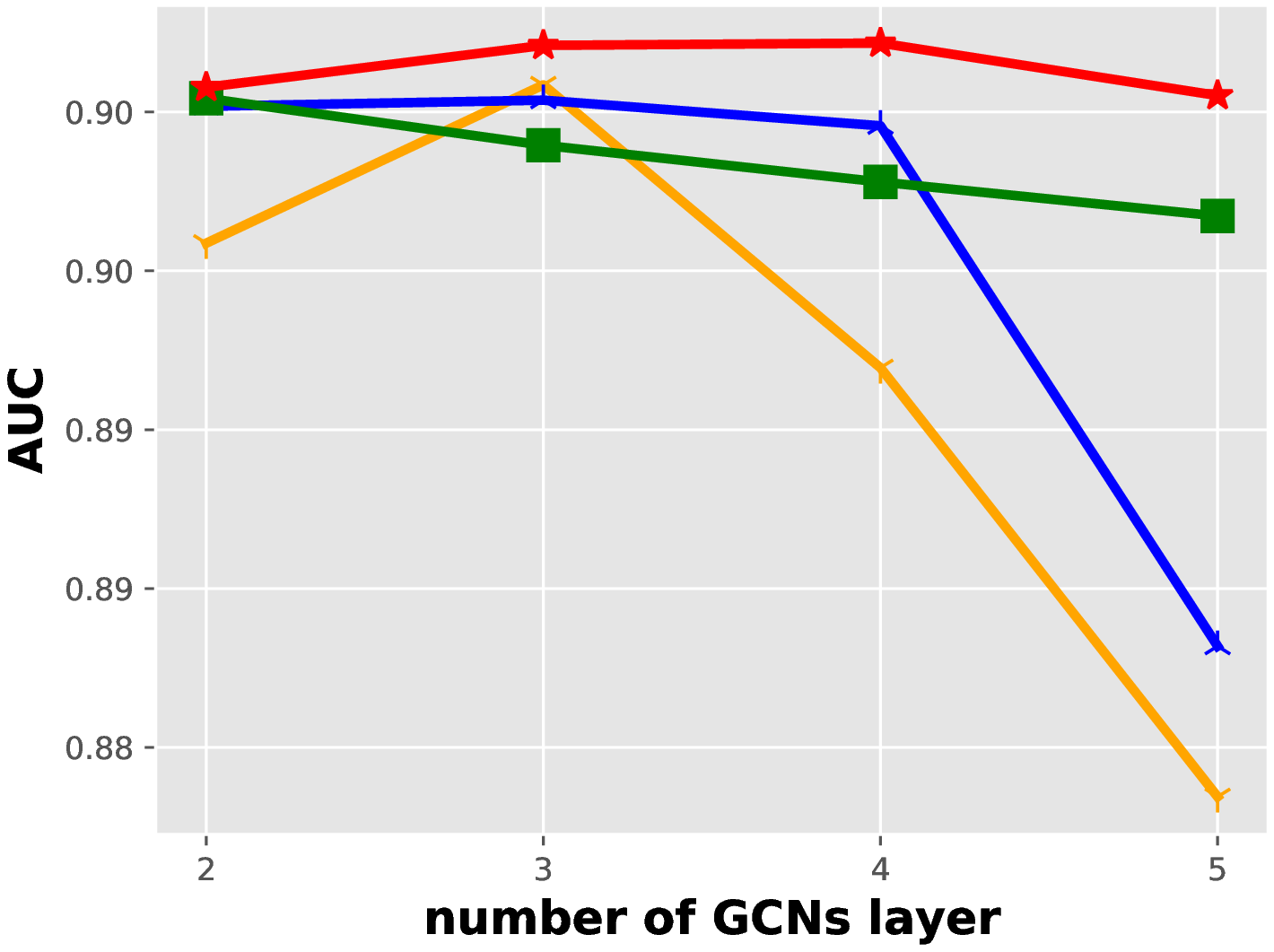}
\vspace{-17pt}
\caption*{(b) Slashdot - AUC}
\end{minipage}
\begin{minipage}[t]{0.32\textwidth}
\centering
\includegraphics[width=\textwidth]{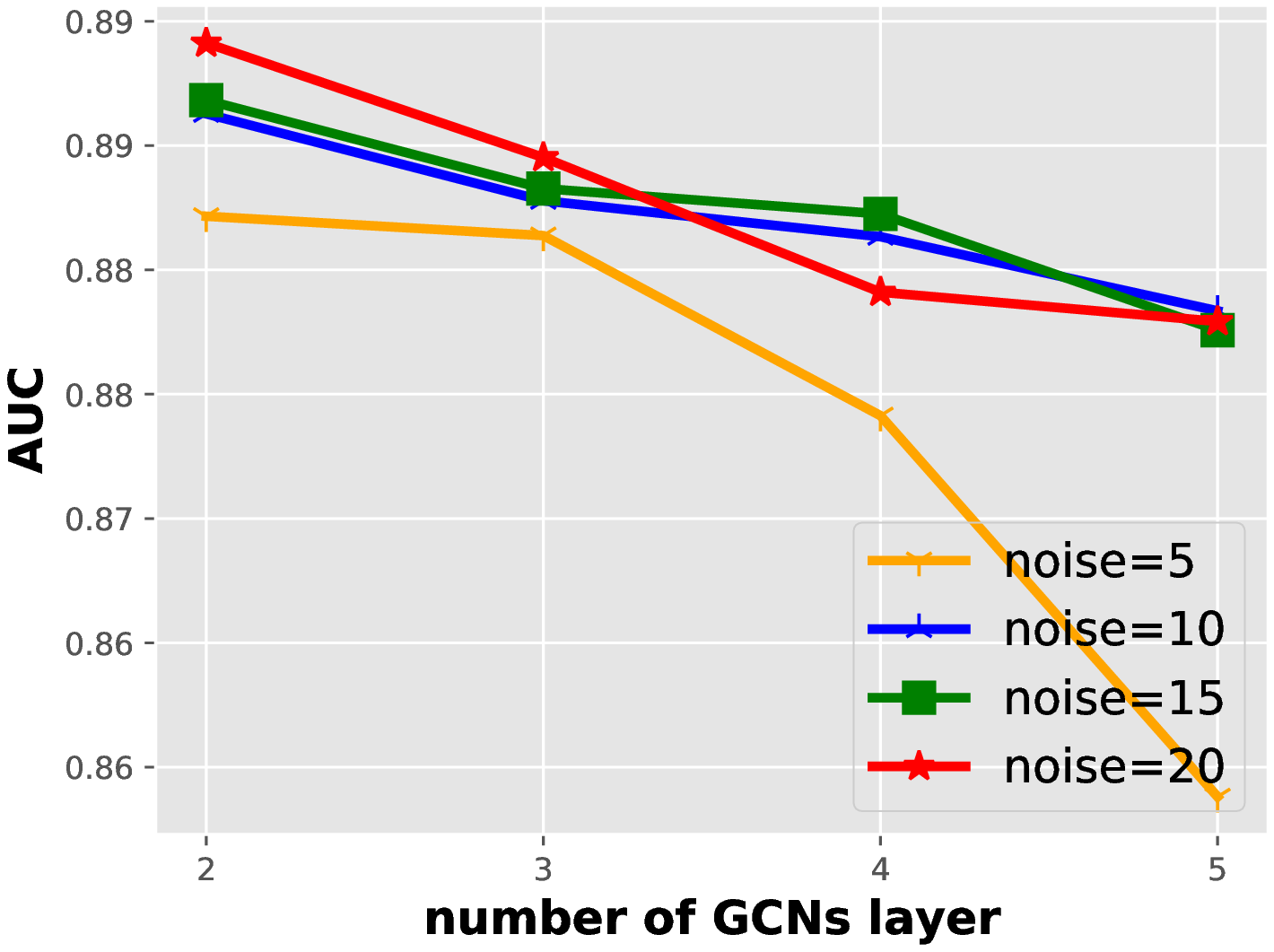}
\vspace{-17pt}
\caption*{(c) Wiki - AUC}
\end{minipage} \\ 
\begin{minipage}[t]{0.32\textwidth}
\centering
\includegraphics[width=\textwidth]{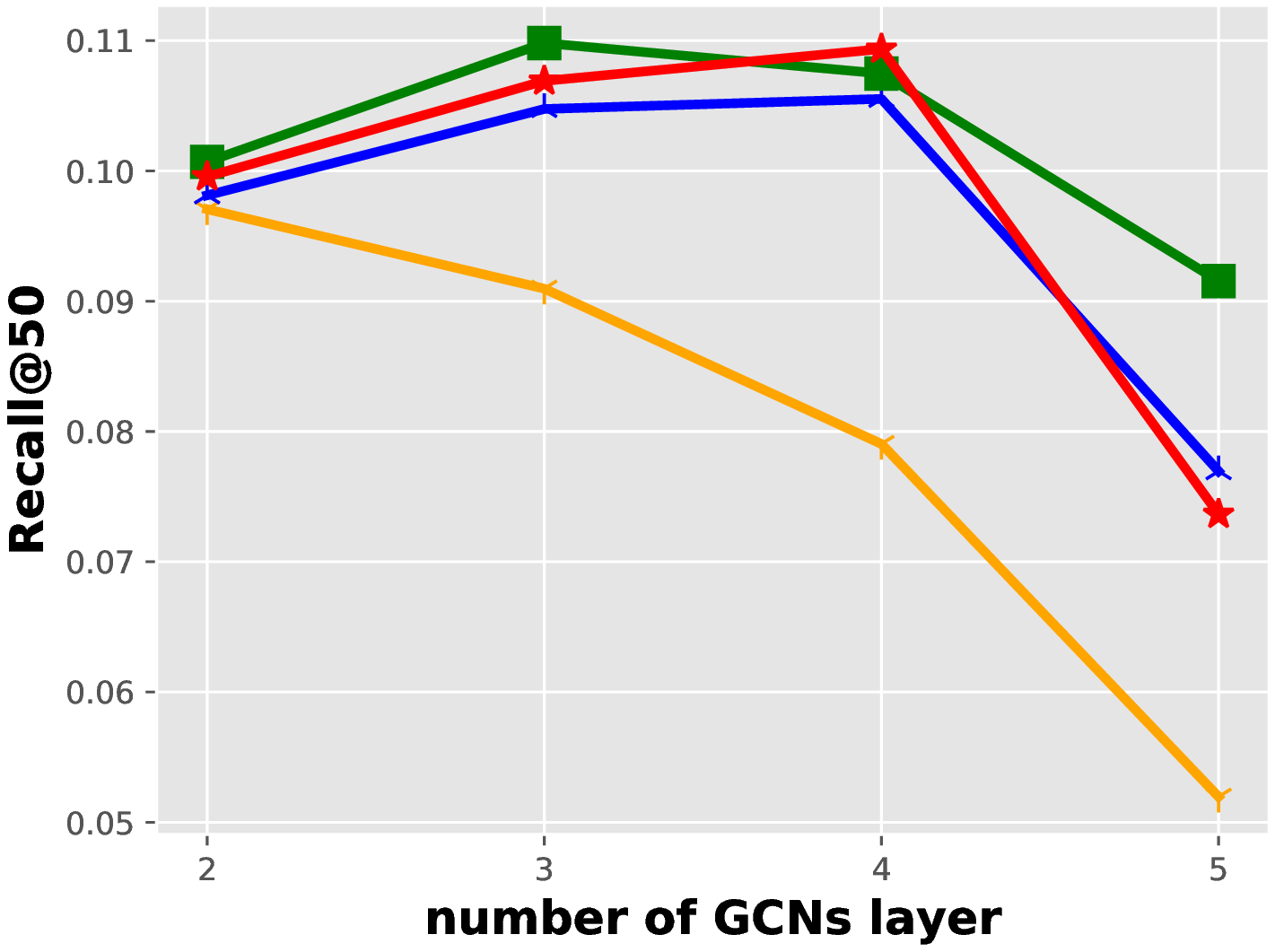}
\vspace{-17pt}
\caption*{(d) Epinions - Recall@50}
\end{minipage} 
\begin{minipage}[t]{0.32\textwidth}
\centering
\includegraphics[width=\textwidth]{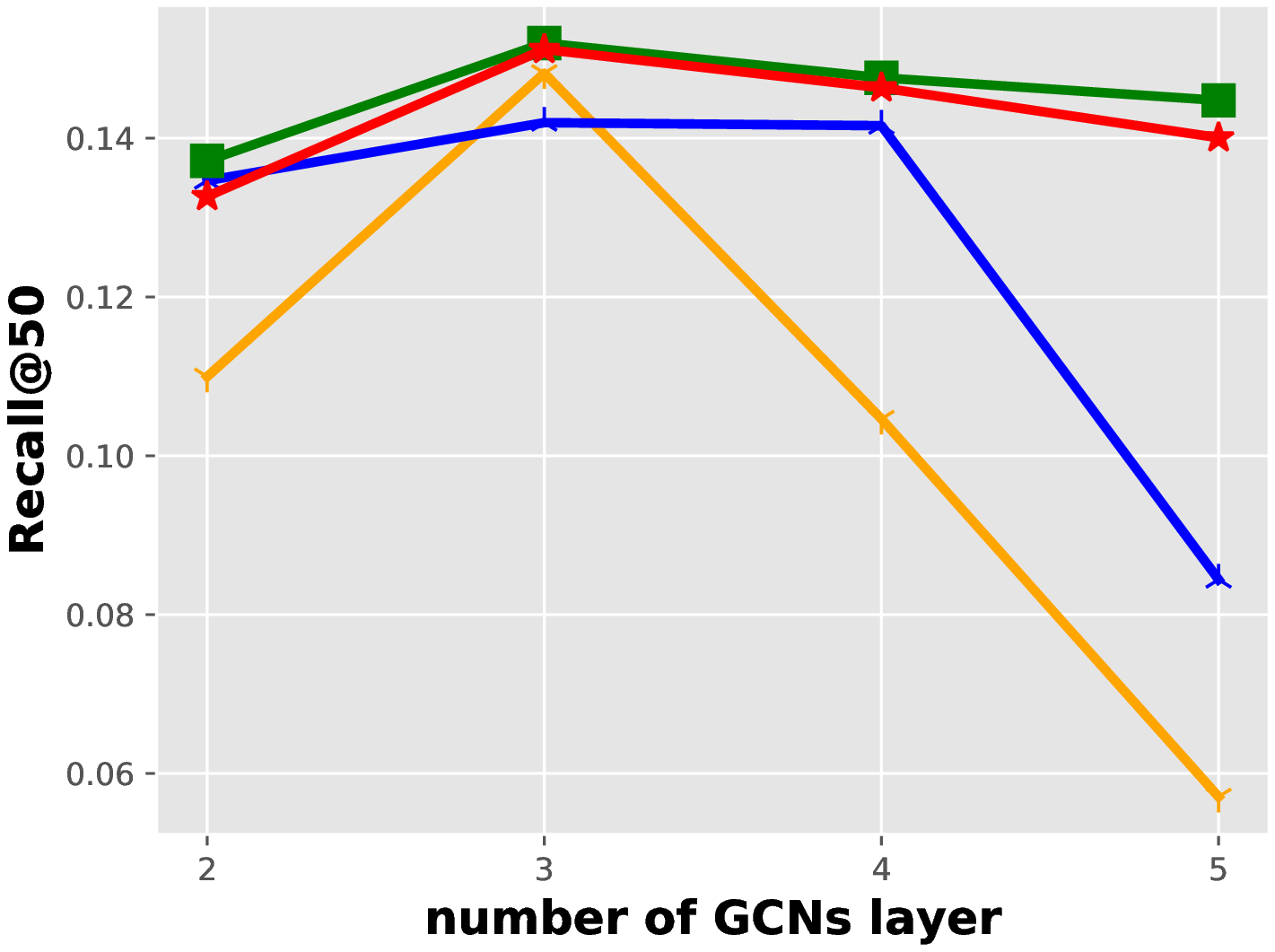}
\vspace{-17pt}
\caption*{(e) Slashdot - Recall@50}
\end{minipage}
\begin{minipage}[t]{0.32\textwidth}
\centering
\includegraphics[width=\textwidth]{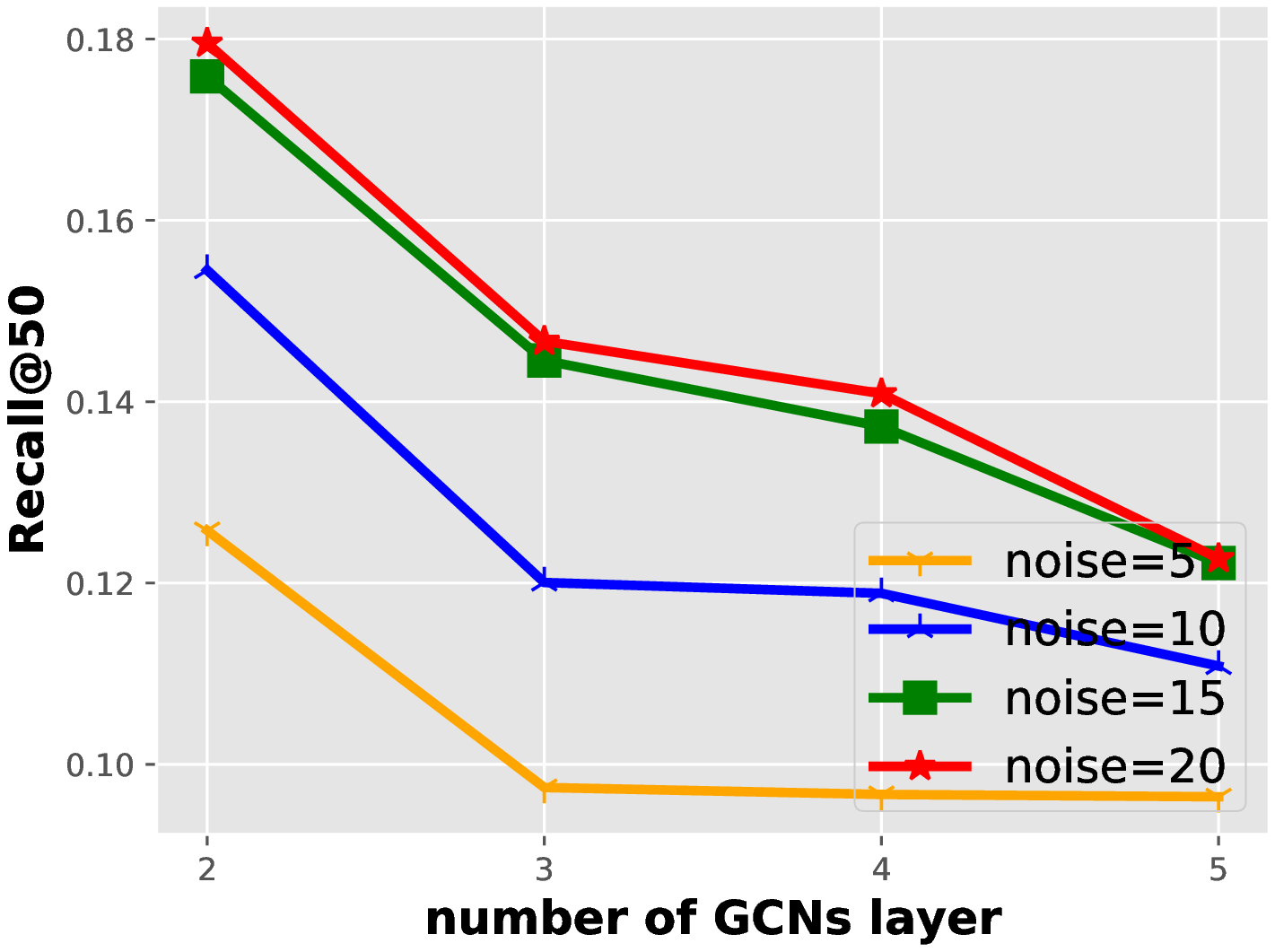}
\vspace{-17pt}
\caption*{(f) Wiki - Recall@50}
\end{minipage}
\caption{Performance of DVE with different parameter settings. AUC in (a)(b)(c) is the metric for link sign prediction task and Recall@50 in (d)(e)(f) is the metric for node recommendation task.}
\label{figure:parameter_sensitivity}
\end{figure}
\subsubsection{\textbf{Hyper-Parameter Sensitivity}}
In DVE, the two hyper-parameters are the number of GCN layer $n_{GCN}$ and the noise sampling size $n_{noise}$. $n_{GCN}$ controls the order of a node's local structures and $n_{noise}$ influence the sampling size of non-existent links. We here investigate the model sensitivity on these two hyper-parameters. The results are show in Figure~\ref{figure:parameter_sensitivity}, from which we have the following observations:
\begin{itemize}
    \item DVE achieves its best performance on different datasets when $n_{GCN}=2,3$ and $n_{noise}=5,20$. The slight change of $n_{GCN}$ indicates that most useful topological information is within low-order neighborhoods. While the noise sampling size $n_{noise}$ varies a lot on different datasets, which means $n_{noise}$ is better chosen according to the datasets. 
    \item From Figure~\ref{figure:parameter_sensitivity} (a)(b)(c) for link sign prediction task, we see that the performance of link sign prediction does not change a lot (0.958$\sim$ 0.964 on Epinions and 0.88$\sim$ 0.90 on Slashdot and 0.86$\sim$ 0.88 on Wiki). This slight change indicates that the link sign prediction performance is robust to $n_{GCN}$ and $n_{noise}$. However, as shown in Figure~\ref{figure:parameter_sensitivity}~(d)(e)(f), the node recommendation performance changes obviously (0.10$\sim$ 0.06 on Epinions and 0.14$\sim$ 0.06 on Slashdot and 0.09$\sim$ 0.17 on Wiki). This is because the node recommendation is instinctively measured from model training, while a binary classifier is additionally trained for link sign prediction. The binary classifier reduces model sensitivity on $n_{GCN}$ and $n_{noise}$. 
\end{itemize}

\subsubsection{\textbf{Dropout Regularization}}
\begin{figure}[h!]
\centering
\begin{minipage}[t]{0.32\textwidth}
\centering
\includegraphics[width=\textwidth]{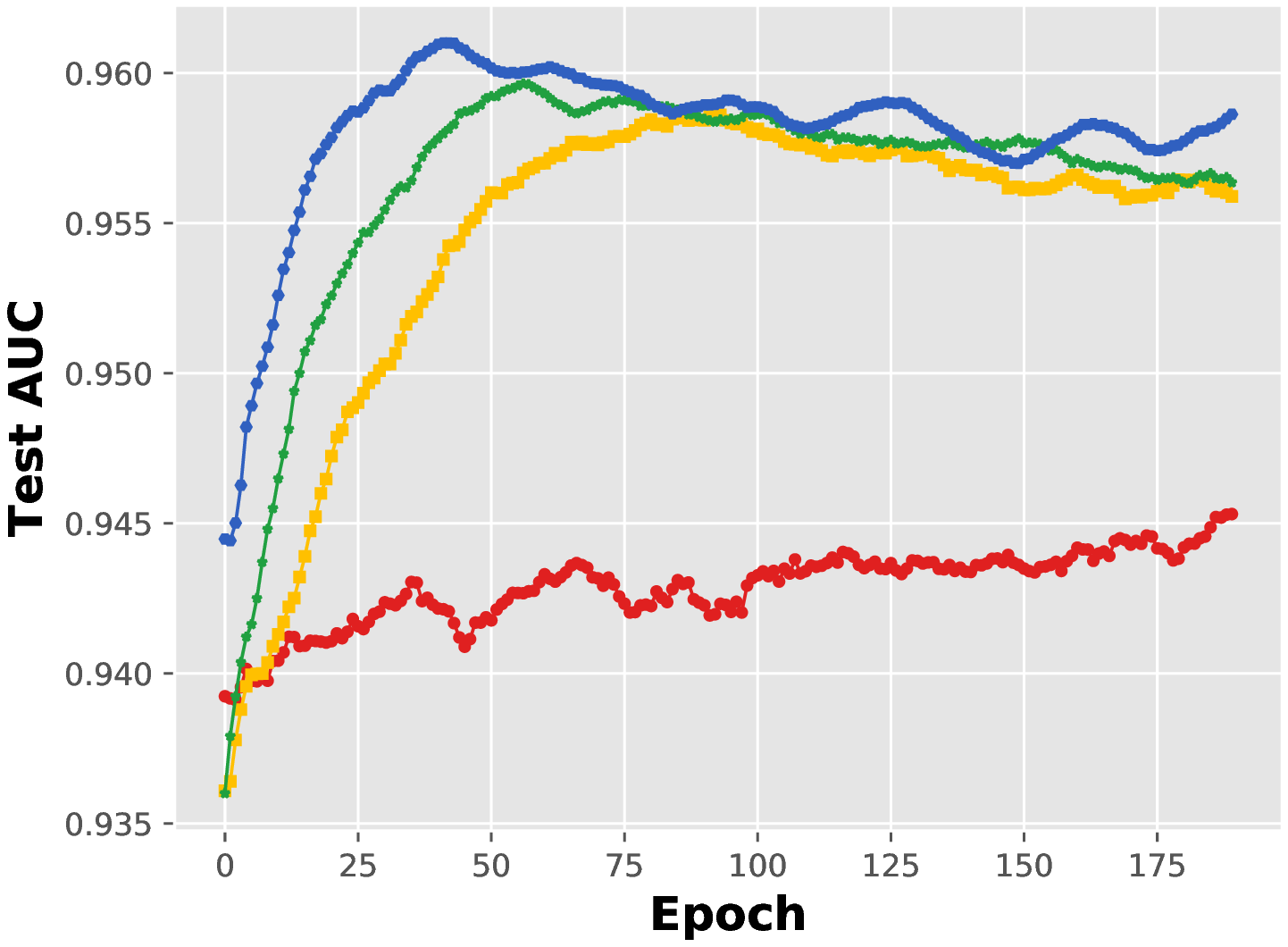}
\vspace{-17pt}
\caption*{(a) Epinions - AUC}
\end{minipage}
\begin{minipage}[t]{0.32\textwidth}
\centering
\includegraphics[width=\textwidth]{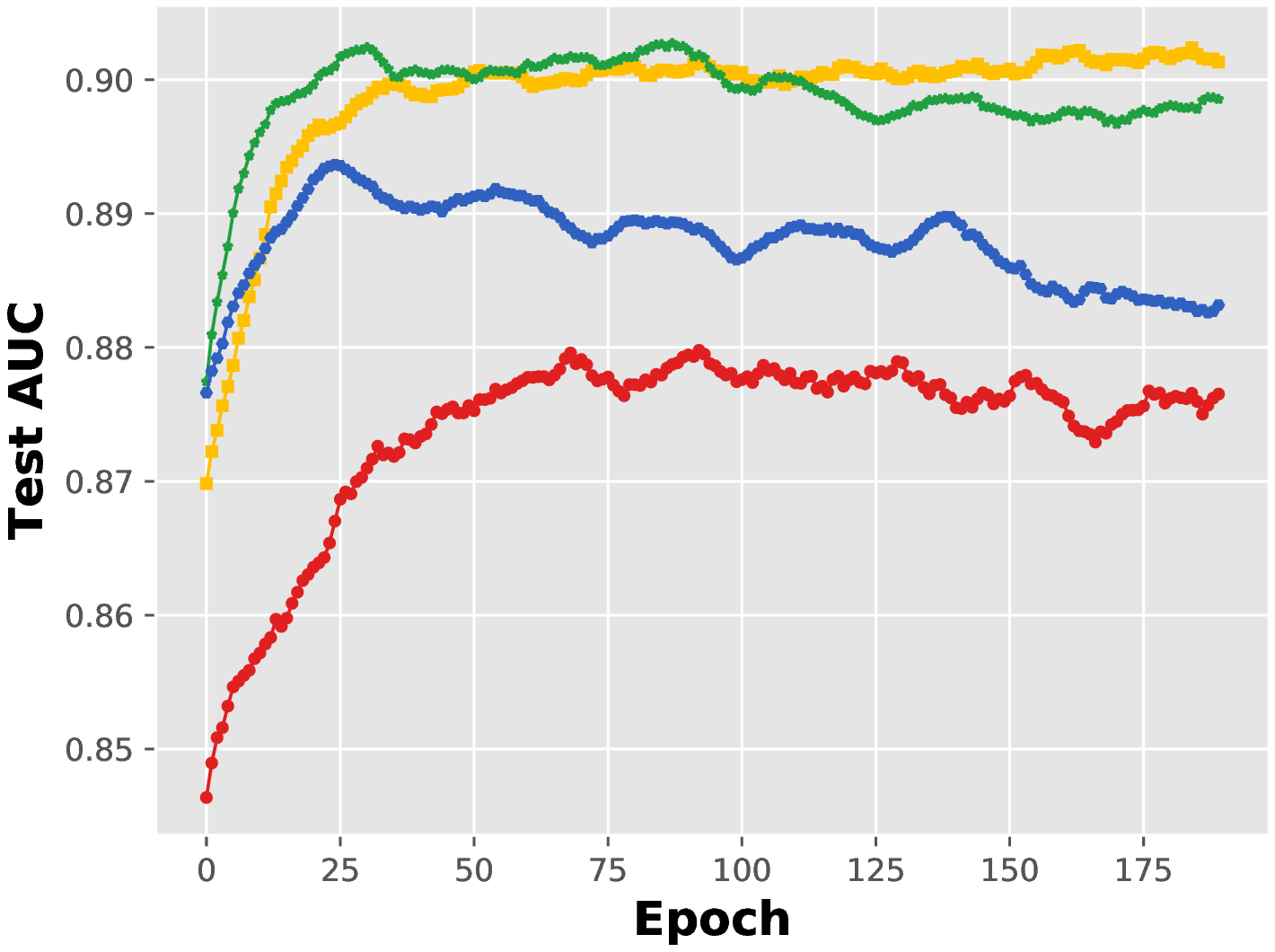}
\vspace{-17pt}
\caption*{(b) Slashdot - AUC}
\end{minipage}
\begin{minipage}[t]{0.32\textwidth}
\centering
\includegraphics[width=\textwidth]{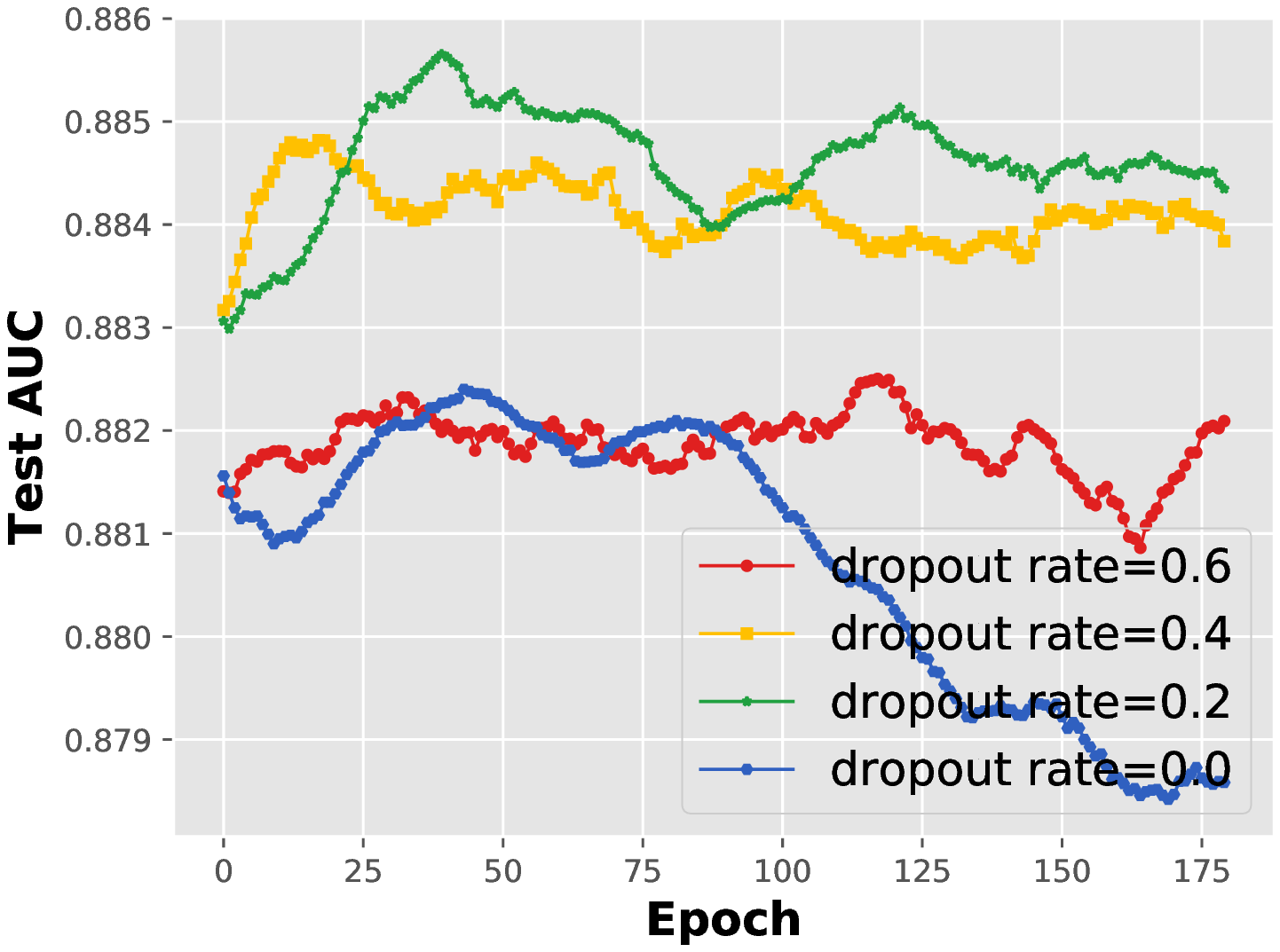}
\vspace{-17pt}
\caption*{(c) Wiki - AUC}
\end{minipage} \\ 
\begin{minipage}[t]{0.32\textwidth}
\centering
\includegraphics[width=\textwidth]{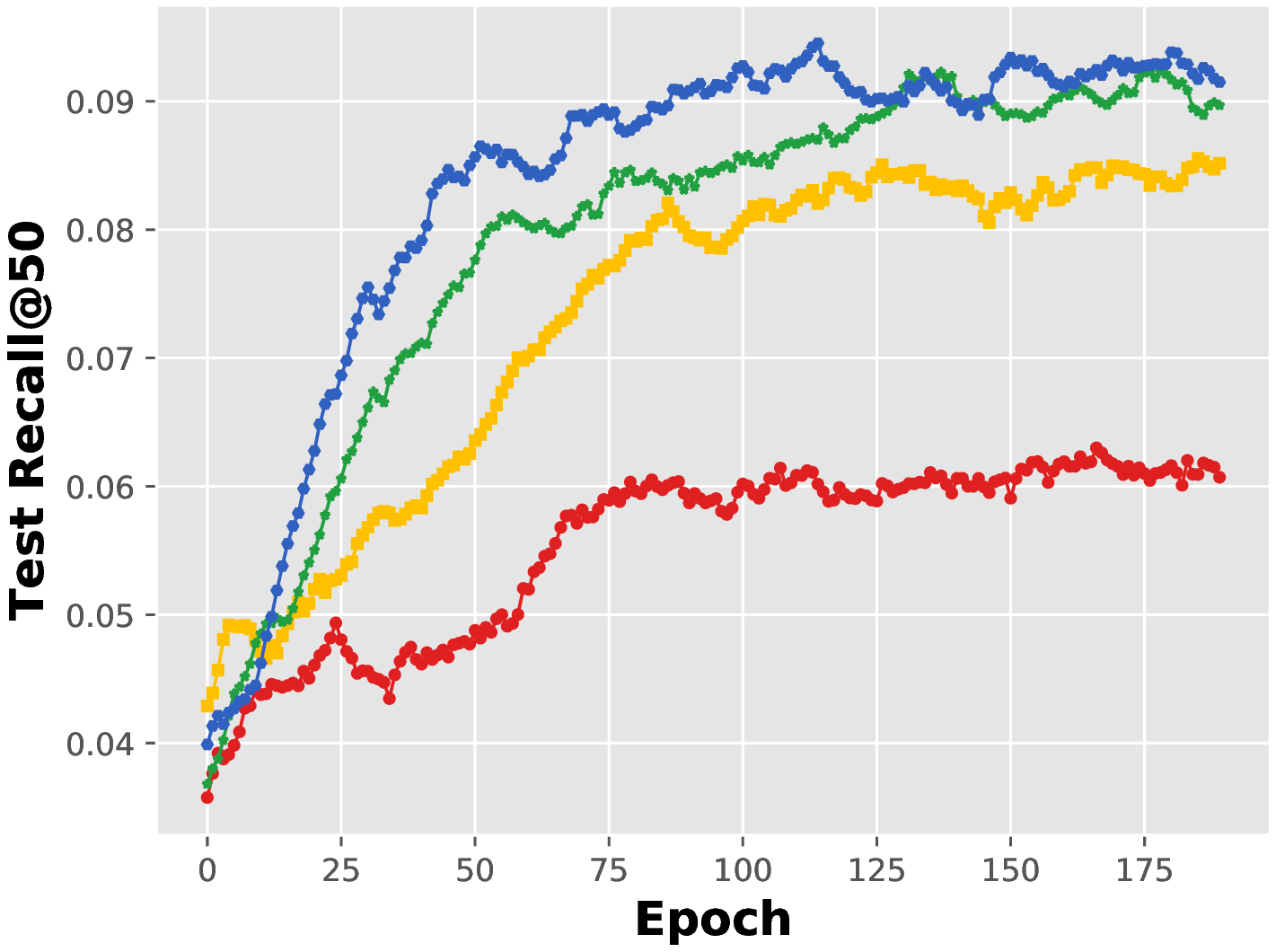}
\vspace{-17pt}
\caption*{(d) Epinions - Recall@50}
\end{minipage} 
\begin{minipage}[t]{0.32\textwidth}
\centering
\includegraphics[width=\textwidth]{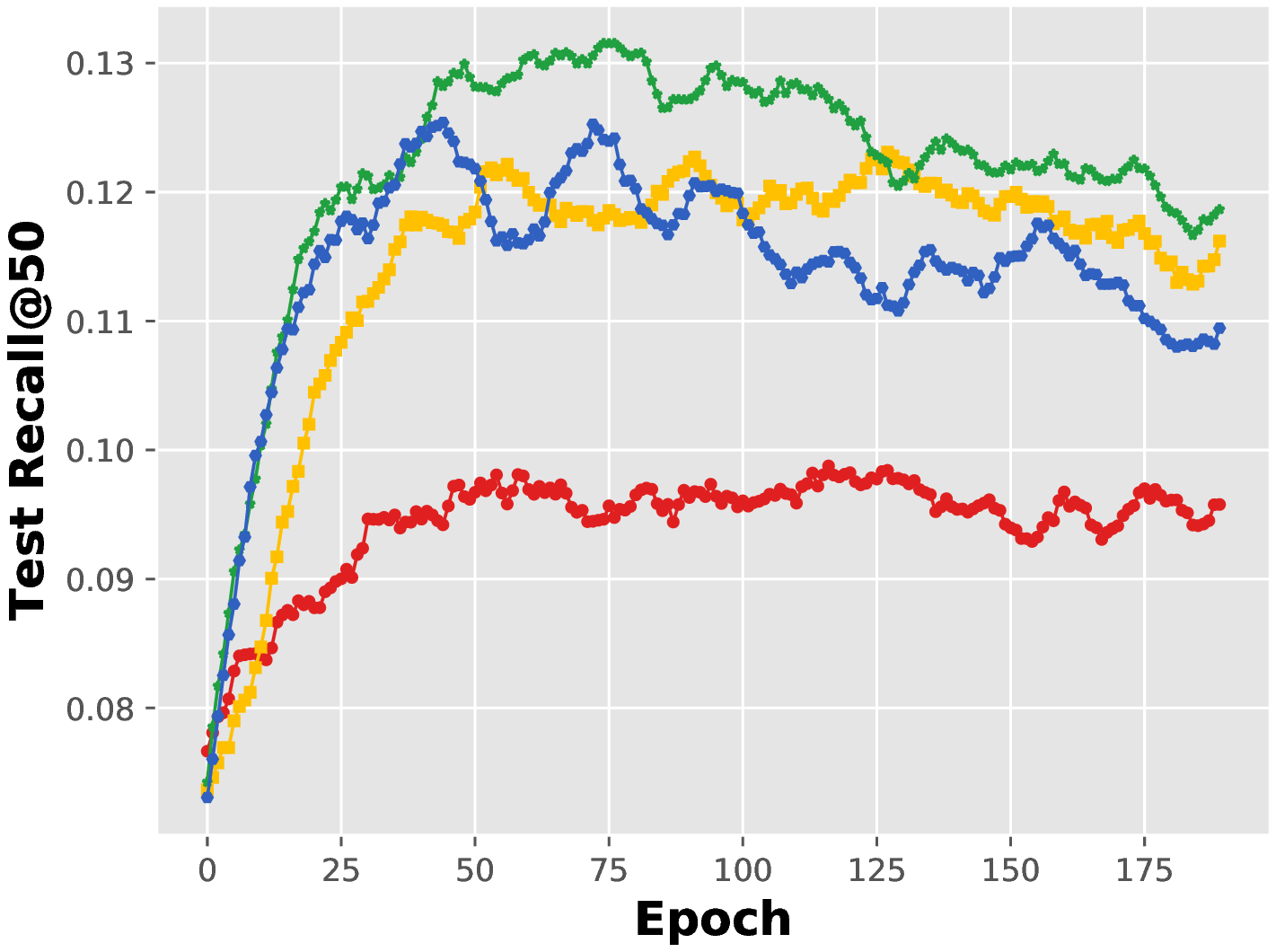}
\vspace{-17pt}
\caption*{(e) Slashdot - Recall@50}
\end{minipage}
\begin{minipage}[t]{0.32\textwidth}
\centering
\includegraphics[width=\textwidth]{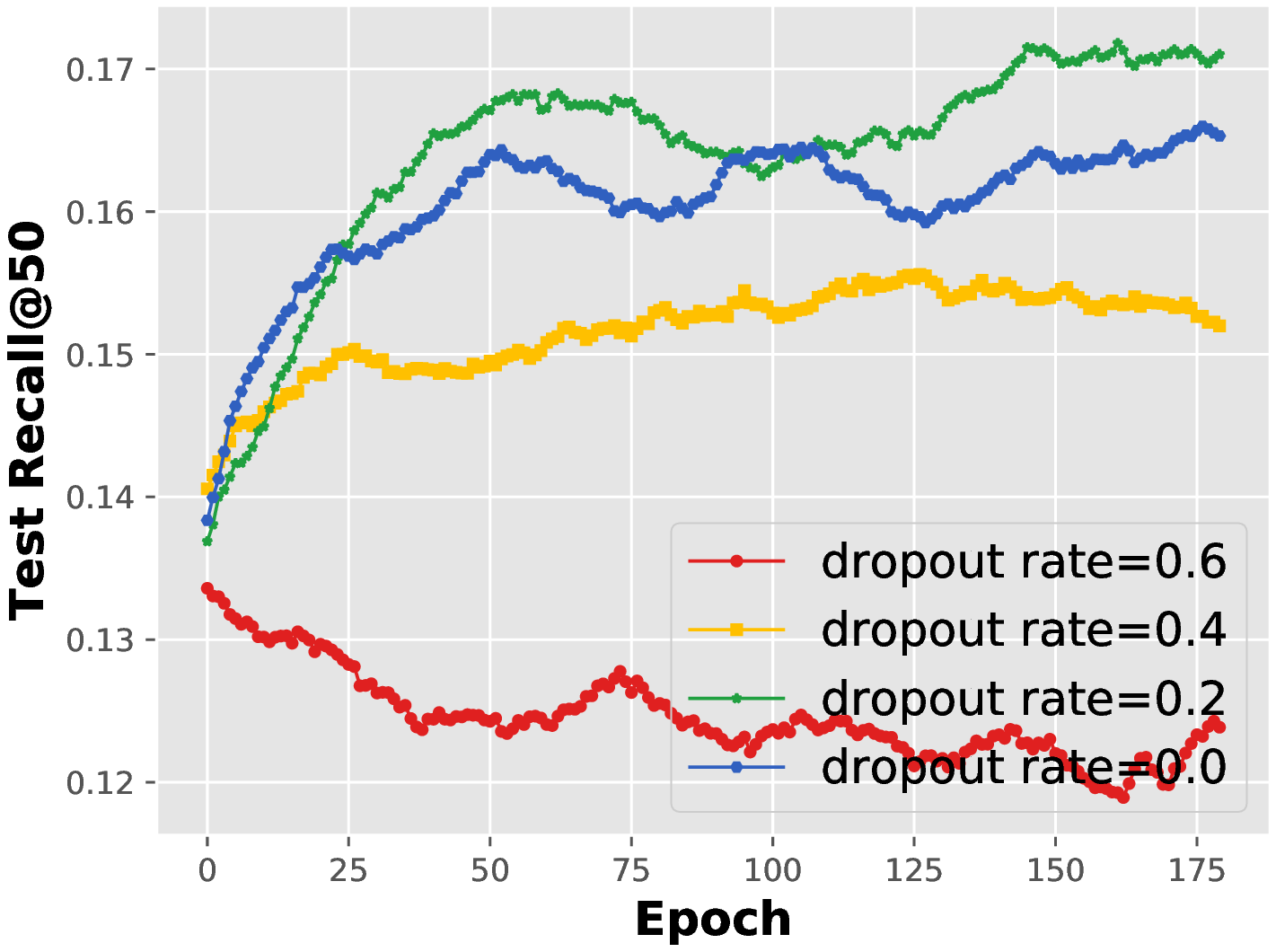}
\vspace{-17pt}
\caption*{(f) Wiki - Recall@50}
\end{minipage}
\caption{Performance of DVE with different dropout rates. Dropout rate=0.0 means dropout keep probability is 1.0 during training. AUC in (a)(b)(c) is the metric for link sign prediction task and Recall@50 in (d)(e)(f) is the metric for node recommendation task.}
\label{figure:dropout_influence}
\end{figure}

In our method, we apply Dropout for regularization. In order to study the influence of Dropout, we investigate the model performance with different Dropout rates along the training process. The results are shown in Figure~\ref{figure:dropout_influence}. 

From Figure~\ref{figure:dropout_influence}, we can see that different Dropout rates may have different influence on different datasets. In Figure~\ref{figure:dropout_influence}~(a)(d) for Epinions, it is obvious that DVE reaches its best performance when dropout=0.0, which means the Dropout keep probability equals 1.0 is better for Epinions. The reason for this may be that the data distribution is complex and no Dropout encourages the model to fit data better. In contrary, in Figure~\ref{figure:dropout_influence} (b)(e) for Slashdot and Figure~\ref{figure:dropout_influence} (c)(f) for Wiki, Dropout rate equals 0.2 facilitates better performance, which indicates DVE needs necessary regularization on these two datasets to avoid over-fitting. 
\begin{figure}[h!]
\centering
\begin{minipage}[t]{0.32\textwidth}
\centering
\includegraphics[width=\textwidth]{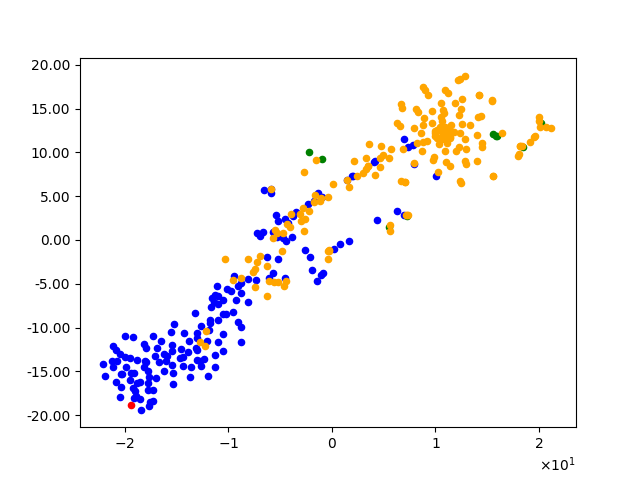}
\vspace{-17pt}
\caption*{(a) MF}
\end{minipage}
\begin{minipage}[t]{0.32\textwidth}
\centering
\includegraphics[width=\textwidth]{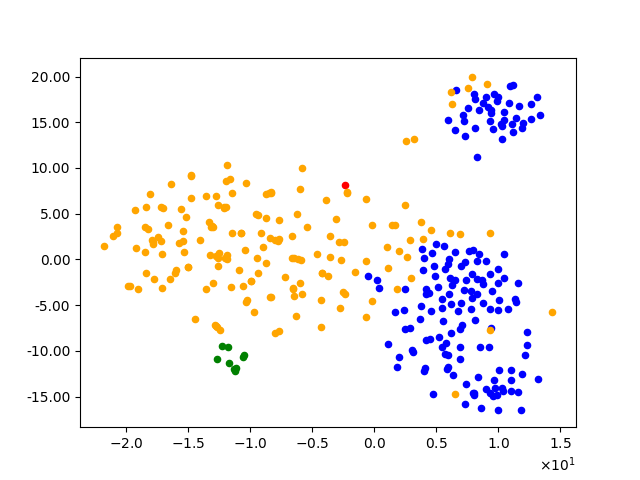}
\vspace{-18pt}
\caption*{(b) SNE}
\end{minipage}
\begin{minipage}[t]{0.32\textwidth}
\centering
\includegraphics[width=\textwidth]{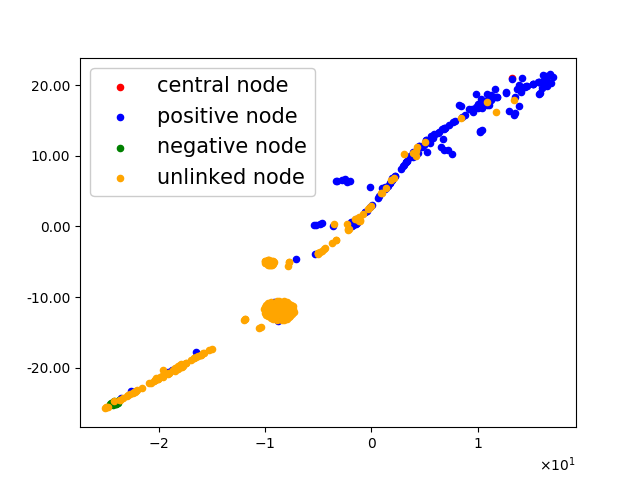}
\vspace{-18pt}
\caption*{(c) SiNE}
\end{minipage} \\ 
\begin{minipage}[t]{0.32\textwidth}
\centering
\includegraphics[width=\textwidth]{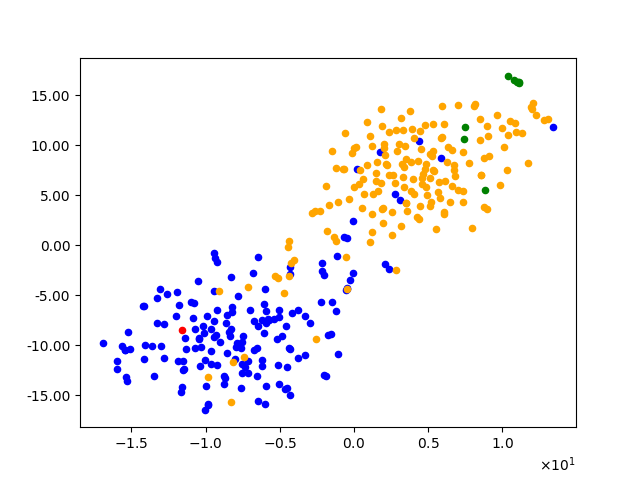}
\vspace{-18pt}
\caption*{(d) BPWR}
\end{minipage} 
\begin{minipage}[t]{0.32\textwidth}
\centering
\includegraphics[width=\textwidth]{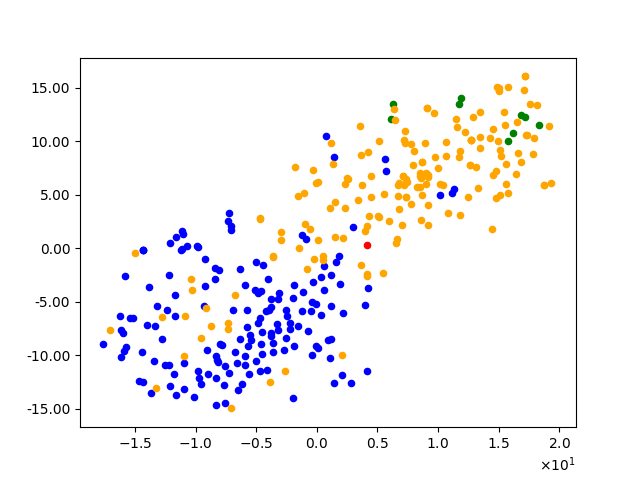}
\vspace{-18pt}
\caption*{(e) SLVE}
\end{minipage}
\begin{minipage}[t]{0.32\textwidth}
\centering
\includegraphics[width=\textwidth]{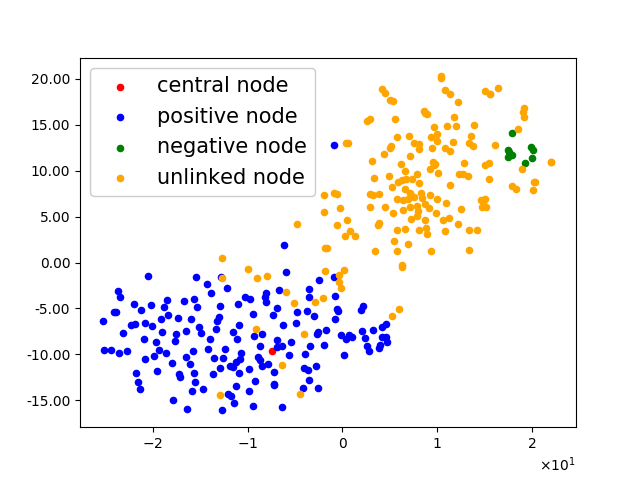}
\vspace{-18pt}
\caption*{(f) DVE}
\end{minipage}
\caption{t-SNE visualization of topology preservation in Epinions. The node in red color means the sampled central node $i$ as source node. The nodes in blue color represents the positively linked neighbors $\mathcal{N}_{p}(i)$ and nodes in green color are the negatively linked neighbors $\mathcal{N}_{n}(i)$. While the yellow ones are randomly sampled non-linked nodes $\mathcal{N}_{un}(i)$ for center node $i$. Both $\mathcal{N}_{p}(i)$ and $\mathcal{N}_{n}(i)$ are target nodes.}
\label{figure:TSNE_structure}
\end{figure}
\subsection{Qualitative Visualization}
\subsubsection{\textbf{Topology Preservation.}}
Signed directed networks have complex topology pattern and there are some obvious topology characteristics. If we denote $i$ as a source node, $\mathcal{N}_{p}(i)$ as the positively linked target neighbors, $\mathcal{N}_{n}(i)$ as negatively linked target neighbors and $\mathcal{N}_{un}(i)$ as the non-linked neighbors, there are several characteristics of topology in signed directed networks:
\begin{itemize}
    \item $\mathcal{N}_{p}(i)$, $\mathcal{N}_{n}(i)$ and $\mathcal{N}_{un}(i)$ tend to be three clusters since they play different roles for node $i$;
    \item Closeness between $\mathcal{N}_{p}(i)$ and node $i$ tends to be larger than that between $\mathcal{N}_{un}(i)$ and node $i$;
    \item Closeness between $\mathcal{N}_{un}(i)$ and node $i$ is larger than that between $\mathcal{N}_{n}(i)$ and node $i$.
\end{itemize}

In order to study whether the learned node embeddings preserve the above characteristics, we conduct an experiment about node embedding visualization. In particular, we randomly sample a source node $i$ whose number of directly linked neighbors is larger than 100 from Epinions. The positively linked neighbors $\mathcal{N}_{p}(i)$ and negatively linked neighbors $\mathcal{N}_{n}(i)$ are both from target nodes. Next, we also randomly sample some non-linked nodes $\mathcal{N}_{un}(i)$. Finally, we visualize the corresponding embeddings with t-SNE~\cite{maaten2008visualizing} for 6 methods. The results are shown in Figure~\ref{figure:TSNE_structure}. From the figure, we can summarize that:
\begin{itemize}
    \item DVE has the best visualization performance in terms of the well clustered nodes and clear closeness pattern among different types of nodes. For SNE in Figure~\ref{figure:TSNE_structure} (b), we can see that the closest neighbors for the central node are non-linked nodes and the positively linked nodes are not well clustered. For SiNE in Figure~\ref{figure:TSNE_structure} (c), the nodes are not well distributed and it is even impossible to recognize some nodes. MF in Figure~\ref{figure:TSNE_structure} (a) clusters the positively linked nodes and non-linked nodes well but fails in negatively linked nodes. The central node in red color is in the marginal part, which is not reasonable according to the actual central node pattern. Compare to these three competitive baselines, our proposed methods BPWR and DVE in Figure~\ref{figure:TSNE_structure} (d) and (f) are capable of learning the distributed and well clustered node embeddings. The central node in red color are surrounded by the positively linked nodes in blue color. The clear closeness pattern among different types of nodes as well matches the fact that we have illustrated before. These advantages are benefited from modeling both the \emph{first-order} and \emph{high-order} topology in signed directed networks.
    \item DVE models the distinctive influence of messaging propagation in signed directed networks, and yields better topology preservation.
    Compared to DVE in Figure~\ref{figure:TSNE_structure} (f), SLVE in Figure~\ref{figure:TSNE_structure} (e) tends to mix positively linked nodes and non-linked nodes. Moreover, the central node in red color is false positioned in the middle part of positively linked nodes and non-linked nodes. This is because SLVE applies homophily effects with different signs, which cannot model the distinctive influence of message propagation. In contrary, DVE with decoupled variational encoder can learn distinctive effects for different signs and better preserve the network topology.
\end{itemize}

\begin{figure}[h!]
\centering
\begin{minipage}[t]{0.32\textwidth}
\centering
\includegraphics[width=\textwidth]{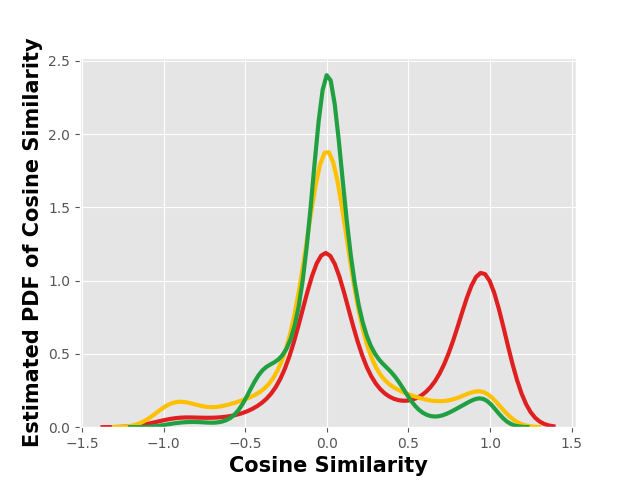}
\vspace{-17pt}
\caption*{(a) MF}
\end{minipage}
\begin{minipage}[t]{0.32\textwidth}
\centering
\includegraphics[width=\textwidth]{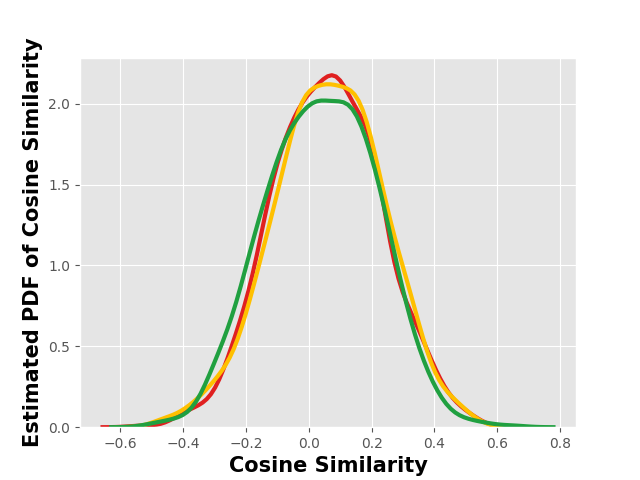}
\vspace{-17pt}
\caption*{(b) SNE}
\end{minipage}
\begin{minipage}[t]{0.32\textwidth}
\centering
\includegraphics[width=\textwidth]{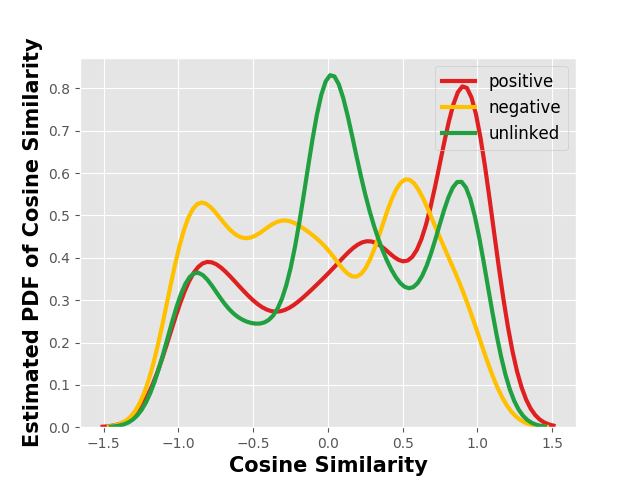}
\vspace{-17pt}
\caption*{(c) SiNE}
\end{minipage} \\ 
\begin{minipage}[t]{0.32\textwidth}
\centering
\includegraphics[width=\textwidth]{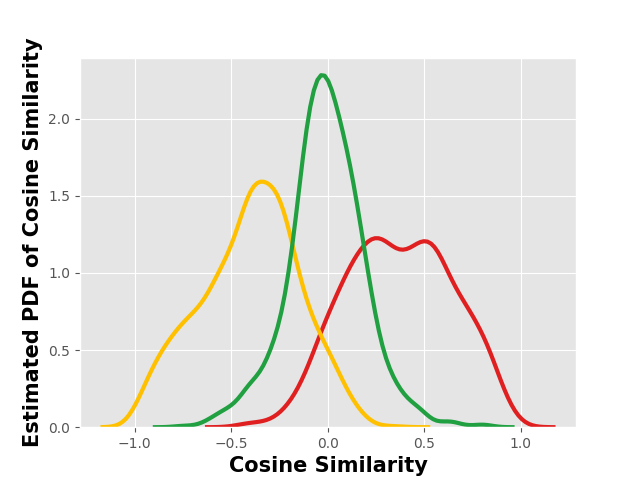}
\caption*{(d) BPWR}
\end{minipage} 
\begin{minipage}[t]{0.32\textwidth}
\centering
\includegraphics[width=\textwidth]{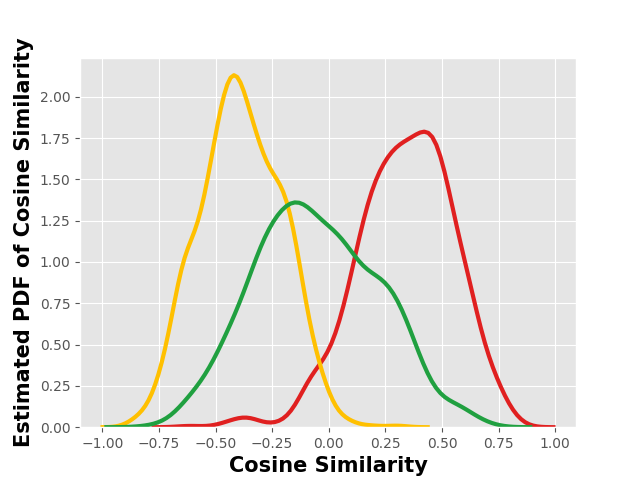}
\vspace{-17pt}
\caption*{(e) SLVE}
\end{minipage}
\begin{minipage}[t]{0.32\textwidth}
\centering
\includegraphics[width=\textwidth]{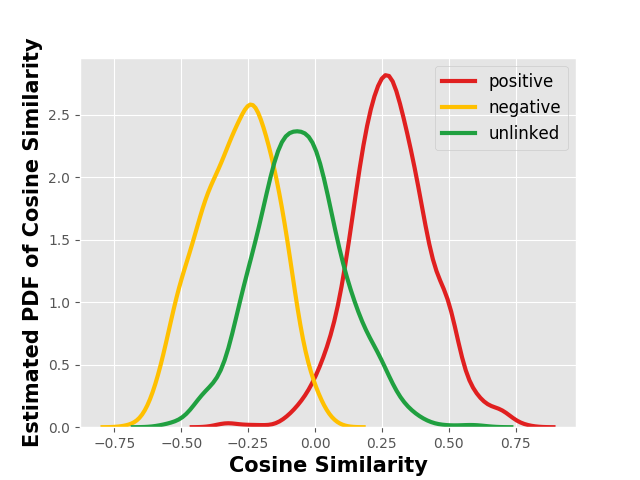}
\vspace{-17pt}
\caption*{(f) DVE}
\end{minipage}
\caption{Estimated probability density function of different types of node pairs on Slashdot for 6 methods. The red curve means the estimated PDF (Probability Density Function) of cosine similarity among positively linked node pairs. Similarly, the yellow curve and green curve denote the estimated PDF among the negatively linked node pairs and non-linked node pairs, respectively.}
\label{figure:cos_pdf}
\end{figure}
\subsubsection{\textbf{Closeness Distribution.}}
In signed directed networks, positive edges mean trust/friend while negative edges represent distrust/enemy and the non-existent edges may both have the probability to be positive ones or negative ones. According to Extended Structural Balance theory, different types of node pairs pose different closeness distributions and we have the following rules:
\begin{itemize}
    \item The similarity between positively linked node pairs is expected to be large because of the semantics of positive edges;
    \item The similarity between negatively linked node pairs should be small due to the negative meaning of negative edges;
    \item For the node pairs with non-existent edges, they have potential to be either positive or negative relation, and should be in the middle position between positively linked node pairs and negatively linked ones.
\end{itemize}

Thereby, we conduct an experiment to investigate whether DVE has better ability of preserving the closeness distribution pattern. In particular, we visualize the estimated Probability Density Function (PDF) of different node pairs on Slashdot for 6 methods. In particular, we calculate the cosine similarity of all positively linked node pairs, negatively linked node pairs and randomly sampled non-linked node pairs by leveraging the learned embeddings from 6 methods. The estimated PDF curves of cosine similarity are shown in Figure~\ref{figure:cos_pdf}. The red curve, yellow curve and green curve indicate the estimated PDF curve for positively linked node pairs, negatively linked node pairs and non-linked node pairs, respectively. From this figure, we have the following observations:
\begin{itemize}
    \item From Figure~\ref{figure:cos_pdf} (a)(b)(c), we can see that the baseline methods MF, SNE and SiNE all exhibit high overlap of different curves, especially for SNE in Figure~\ref{figure:cos_pdf}. This indicates these methods are not capable of capturing the different closeness distribution patterns of different node pairs. By contrast, considering the results of BPWR and DVE in Figure~\ref{figure:cos_pdf} (d)(e)(f), it is obvious that the three curves show different distributions. Meanwhile, BPWR and DVE follow the closeness rules in which positively linked node pairs have highest cosine similarity, non-linked node pairs have the second and negatively linked ones have the last.
    \item In addition, in Figure~\ref{figure:cos_pdf} (d)(e) for BPWR and SLVE, the estimated PDF of non-linked node pairs in green color tends to have more overlap with the other curves, which may lead to indistinguishable node embeddings. Instead, DVE in Figure~\ref{figure:cos_pdf} (f) presents both distinguishable estimated PDF curves with smaller overlap and obvious cosine similarity gap among different kinds of node pairs. This indicates DVE can better preserve the closeness distribution pattern in signed directed networks.
\end{itemize}

\section{Conclusion and Future Work}
In this paper, we reformulate the representation learning problem on signed directed networks from a variational auto-encoding perspective and further propose a decoupled variational embedding (DVE) method to learn representative node embeddings. DVE is capable of preserving both the \emph{first-order} and \emph{high-order} topology for signed directed networks.
In particular, DVE consists of a decoupled variational encoder and a structure decoder. The decoupled variational encoder captures local structures and provides informative node embeddings for the structure decoder. Meanwhile, the structure decoder mines the closeness relationships among positive, negative and non-existent links in a pair-wise ranking manner, as well as supervises embedding learning in the encoder module. Performance on three real-world datasets of two tasks proves the superiority of DVE compared to recent competitive baselines. 

Remind that DVE constructs source node embeddings $Z_{s}$ just by the limited concatenation operation of two latent embeddings $Z_{s}^{p}$ and $Z_{s}^{n}$. Observing unbalance between positive links and negative links from data, source node embeddings $Z_{s}$ may follow some distribution through $Z_{s}^{p}$ and $Z_{s}^{n}$. We will explore how to better model the interaction between $Z_{s}^{p}$ and $Z_{s}^{n}$, and construct source node embeddings $Z_{s}$ more reasonably to pursue better performance.

\begin{acks}
This work is supported by the National Key Research and Development Program of China (No. 2019YFB1804304), SHEITC (No. 2018-RGZN-02046), 111 plan (No. BP0719010),  and STCSM (No. 18DZ2270700), and State Key Laboratory of UHD Video and Audio Production and Presentation.
\end{acks}

%
% The acknowledgments section is defined using the "acks" environment (and NOT an unnumbered section). This ensures
% the proper identification of the section in the article metadata, and the consistent spelling of the heading.

%
% The next two lines define the bibliography style to be used, and the bibliography file.
\bibliographystyle{ACM-Reference-Format}
\bibliography{sample-base}

%%% -*-BibTeX-*-
%%% Do NOT edit. File created by BibTeX with style
%%% ACM-Reference-Format-Journals [18-Jan-2012].

\begin{thebibliography}{76}

%%% ====================================================================
%%% NOTE TO THE USER: you can override these defaults by providing
%%% customized versions of any of these macros before the \bibliography
%%% command.  Each of them MUST provide its own final punctuation,
%%% except for \shownote{}, \showDOI{}, and \showURL{}.  The latter two
%%% do not use final punctuation, in order to avoid confusing it with
%%% the Web address.
%%%
%%% To suppress output of a particular field, define its macro to expand
%%% to an empty string, or better, \unskip, like this:
%%%
%%% \newcommand{\showDOI}[1]{\unskip}   % LaTeX syntax
%%%
%%% \def \showDOI #1{\unskip}           % plain TeX syntax
%%%
%%% ====================================================================

\ifx \showCODEN    \undefined \def \showCODEN     #1{\unskip}     \fi
\ifx \showDOI      \undefined \def \showDOI       #1{#1}\fi
\ifx \showISBNx    \undefined \def \showISBNx     #1{\unskip}     \fi
\ifx \showISBNxiii \undefined \def \showISBNxiii  #1{\unskip}     \fi
\ifx \showISSN     \undefined \def \showISSN      #1{\unskip}     \fi
\ifx \showLCCN     \undefined \def \showLCCN      #1{\unskip}     \fi
\ifx \shownote     \undefined \def \shownote      #1{#1}          \fi
\ifx \showarticletitle \undefined \def \showarticletitle #1{#1}   \fi
\ifx \showURL      \undefined \def \showURL       {\relax}        \fi
% The following commands are used for tagged output and should be
% invisible to TeX
\providecommand\bibfield[2]{#2}
\providecommand\bibinfo[2]{#2}
\providecommand\natexlab[1]{#1}
\providecommand\showeprint[2][]{arXiv:#2}

\bibitem[\protect\citeauthoryear{Aiello, Barrat, Schifanella, Cattuto,
  Markines, and Menczer}{Aiello et~al\mbox{.}}{2012}]%
        {Aiello:2012:FPH:2180861.2180866}
\bibfield{author}{\bibinfo{person}{Luca~Maria Aiello}, \bibinfo{person}{Alain
  Barrat}, \bibinfo{person}{Rossano Schifanella}, \bibinfo{person}{Ciro
  Cattuto}, \bibinfo{person}{Benjamin Markines}, {and} \bibinfo{person}{Filippo
  Menczer}.} \bibinfo{year}{2012}\natexlab{}.
\newblock \showarticletitle{Friendship Prediction and Homophily in Social
  Media}.
\newblock \bibinfo{journal}{\emph{ACM Trans. Web}} \bibinfo{volume}{6},
  \bibinfo{number}{2}, Article \bibinfo{articleno}{9} (\bibinfo{date}{June}
  \bibinfo{year}{2012}), \bibinfo{numpages}{33}~pages.
\newblock
\showISSN{1559-1131}
\urldef\tempurl%
\url{https://doi.org/10.1145/2180861.2180866}
\showDOI{\tempurl}


\bibitem[\protect\citeauthoryear{Battaglia, Hamrick, Bapst, Sanchez-Gonzalez,
  Zambaldi, Malinowski, Tacchetti, Raposo, Santoro, Faulkner,
  et~al\mbox{.}}{Battaglia et~al\mbox{.}}{2018}]%
        {battaglia2018relational}
\bibfield{author}{\bibinfo{person}{Peter~W Battaglia},
  \bibinfo{person}{Jessica~B Hamrick}, \bibinfo{person}{Victor Bapst},
  \bibinfo{person}{Alvaro Sanchez-Gonzalez}, \bibinfo{person}{Vinicius
  Zambaldi}, \bibinfo{person}{Mateusz Malinowski}, \bibinfo{person}{Andrea
  Tacchetti}, \bibinfo{person}{David Raposo}, \bibinfo{person}{Adam Santoro},
  \bibinfo{person}{Ryan Faulkner}, {et~al\mbox{.}}}
  \bibinfo{year}{2018}\natexlab{}.
\newblock \showarticletitle{Relational inductive biases, deep learning, and
  graph networks}.
\newblock \bibinfo{journal}{\emph{arXiv preprint arXiv:1806.01261}}
  (\bibinfo{year}{2018}).
\newblock


\bibitem[\protect\citeauthoryear{Belkin and Niyogi}{Belkin and Niyogi}{2002}]%
        {belkin2002laplacian}
\bibfield{author}{\bibinfo{person}{Mikhail Belkin} {and}
  \bibinfo{person}{Partha Niyogi}.} \bibinfo{year}{2002}\natexlab{}.
\newblock \showarticletitle{Laplacian eigenmaps and spectral techniques for
  embedding and clustering}. In \bibinfo{booktitle}{\emph{Advances in neural
  information processing systems}}. \bibinfo{pages}{585--591}.
\newblock


\bibitem[\protect\citeauthoryear{Bhagat, Cormode, and Muthukrishnan}{Bhagat
  et~al\mbox{.}}{2011}]%
        {bhagat2011node}
\bibfield{author}{\bibinfo{person}{Smriti Bhagat}, \bibinfo{person}{Graham
  Cormode}, {and} \bibinfo{person}{S Muthukrishnan}.}
  \bibinfo{year}{2011}\natexlab{}.
\newblock \showarticletitle{Node classification in social networks}.
\newblock In \bibinfo{booktitle}{\emph{Social network data analytics}}.
  \bibinfo{publisher}{Springer}, \bibinfo{pages}{115--148}.
\newblock


\bibitem[\protect\citeauthoryear{Burda, Grosse, and Salakhutdinov}{Burda
  et~al\mbox{.}}{2015}]%
        {burda2015importance}
\bibfield{author}{\bibinfo{person}{Yuri Burda}, \bibinfo{person}{Roger Grosse},
  {and} \bibinfo{person}{Ruslan Salakhutdinov}.}
  \bibinfo{year}{2015}\natexlab{}.
\newblock \showarticletitle{Importance weighted autoencoders}.
\newblock \bibinfo{journal}{\emph{arXiv preprint arXiv:1509.00519}}
  (\bibinfo{year}{2015}).
\newblock


\bibitem[\protect\citeauthoryear{Cacheda, Blanco, and Barbieri}{Cacheda
  et~al\mbox{.}}{2018}]%
        {Cacheda:2018:CPU:3176641.3157059}
\bibfield{author}{\bibinfo{person}{Fidel Cacheda}, \bibinfo{person}{Roi
  Blanco}, {and} \bibinfo{person}{Nicola Barbieri}.}
  \bibinfo{year}{2018}\natexlab{}.
\newblock \showarticletitle{Characterizing and Predicting Users\&\#x02019;
  Behavior on Local Search Queries}.
\newblock \bibinfo{journal}{\emph{ACM Trans. Web}} \bibinfo{volume}{12},
  \bibinfo{number}{2}, Article \bibinfo{articleno}{11} (\bibinfo{date}{May}
  \bibinfo{year}{2018}), \bibinfo{numpages}{32}~pages.
\newblock
\showISSN{1559-1131}
\urldef\tempurl%
\url{https://doi.org/10.1145/3157059}
\showDOI{\tempurl}


\bibitem[\protect\citeauthoryear{Cartwright and Harary}{Cartwright and
  Harary}{1956}]%
        {cartwright1956structural}
\bibfield{author}{\bibinfo{person}{Dorwin Cartwright} {and}
  \bibinfo{person}{Frank Harary}.} \bibinfo{year}{1956}\natexlab{}.
\newblock \showarticletitle{Structural balance: a generalization of Heider's
  theory.}
\newblock \bibinfo{journal}{\emph{Psychological review}} \bibinfo{volume}{63},
  \bibinfo{number}{5} (\bibinfo{year}{1956}), \bibinfo{pages}{277}.
\newblock


\bibitem[\protect\citeauthoryear{Chen, Ma, and Xiao}{Chen
  et~al\mbox{.}}{2018a}]%
        {chen2018fastgcn}
\bibfield{author}{\bibinfo{person}{Jie Chen}, \bibinfo{person}{Tengfei Ma},
  {and} \bibinfo{person}{Cao Xiao}.} \bibinfo{year}{2018}\natexlab{a}.
\newblock \showarticletitle{Fastgcn: fast learning with graph convolutional
  networks via importance sampling}.
\newblock \bibinfo{journal}{\emph{arXiv preprint arXiv:1801.10247}}
  (\bibinfo{year}{2018}).
\newblock


\bibitem[\protect\citeauthoryear{Chen, Zhu, and Song}{Chen
  et~al\mbox{.}}{2017}]%
        {chen2017stochastic}
\bibfield{author}{\bibinfo{person}{Jianfei Chen}, \bibinfo{person}{Jun Zhu},
  {and} \bibinfo{person}{Le Song}.} \bibinfo{year}{2017}\natexlab{}.
\newblock \showarticletitle{Stochastic training of graph convolutional networks
  with variance reduction}.
\newblock \bibinfo{journal}{\emph{arXiv preprint arXiv:1710.10568}}
  (\bibinfo{year}{2017}).
\newblock


\bibitem[\protect\citeauthoryear{Chen, Xiong, Yan, and Wang}{Chen
  et~al\mbox{.}}{2018b}]%
        {chen2018variational}
\bibfield{author}{\bibinfo{person}{Wenhu Chen}, \bibinfo{person}{Wenhan Xiong},
  \bibinfo{person}{Xifeng Yan}, {and} \bibinfo{person}{William Wang}.}
  \bibinfo{year}{2018}\natexlab{b}.
\newblock \showarticletitle{Variational Knowledge Graph Reasoning}.
\newblock \bibinfo{journal}{\emph{arXiv preprint arXiv:1803.06581}}
  (\bibinfo{year}{2018}).
\newblock


\bibitem[\protect\citeauthoryear{Defferrard, Bresson, and
  Vandergheynst}{Defferrard et~al\mbox{.}}{2016}]%
        {defferrard2016convolutional}
\bibfield{author}{\bibinfo{person}{Micha{\"e}l Defferrard},
  \bibinfo{person}{Xavier Bresson}, {and} \bibinfo{person}{Pierre
  Vandergheynst}.} \bibinfo{year}{2016}\natexlab{}.
\newblock \showarticletitle{Convolutional neural networks on graphs with fast
  localized spectral filtering}. In \bibinfo{booktitle}{\emph{Advances in
  Neural Information Processing Systems}}. \bibinfo{pages}{3844--3852}.
\newblock


\bibitem[\protect\citeauthoryear{Derr, Ma, and Tang}{Derr
  et~al\mbox{.}}{2018}]%
        {derr2018signed}
\bibfield{author}{\bibinfo{person}{Tyler Derr}, \bibinfo{person}{Yao Ma}, {and}
  \bibinfo{person}{Jiliang Tang}.} \bibinfo{year}{2018}\natexlab{}.
\newblock \showarticletitle{Signed Graph Convolutional Network}.
\newblock \bibinfo{journal}{\emph{arXiv preprint arXiv:1808.06354}}
  (\bibinfo{year}{2018}).
\newblock


\bibitem[\protect\citeauthoryear{Doersch}{Doersch}{2016}]%
        {doersch2016tutorial}
\bibfield{author}{\bibinfo{person}{Carl Doersch}.}
  \bibinfo{year}{2016}\natexlab{}.
\newblock \showarticletitle{Tutorial on variational autoencoders}.
\newblock \bibinfo{journal}{\emph{arXiv preprint arXiv:1606.05908}}
  (\bibinfo{year}{2016}).
\newblock


\bibitem[\protect\citeauthoryear{Dong, Zhang, Tang, Chawla, and Wang}{Dong
  et~al\mbox{.}}{2015}]%
        {dong2015coupledlp}
\bibfield{author}{\bibinfo{person}{Yuxiao Dong}, \bibinfo{person}{Jing Zhang},
  \bibinfo{person}{Jie Tang}, \bibinfo{person}{Nitesh~V Chawla}, {and}
  \bibinfo{person}{Bai Wang}.} \bibinfo{year}{2015}\natexlab{}.
\newblock \showarticletitle{Coupledlp: Link prediction in coupled networks}. In
  \bibinfo{booktitle}{\emph{Proceedings of the 21th ACM SIGKDD International
  Conference on Knowledge Discovery and Data Mining}}. ACM,
  \bibinfo{pages}{199--208}.
\newblock


\bibitem[\protect\citeauthoryear{Gaeta}{Gaeta}{2018}]%
        {Gaeta:2018:MID:3176641.3160000}
\bibfield{author}{\bibinfo{person}{Rossano Gaeta}.}
  \bibinfo{year}{2018}\natexlab{}.
\newblock \showarticletitle{A Model of Information Diffusion in Interconnected
  Online Social Networks}.
\newblock \bibinfo{journal}{\emph{ACM Trans. Web}} \bibinfo{volume}{12},
  \bibinfo{number}{2}, Article \bibinfo{articleno}{13} (\bibinfo{date}{June}
  \bibinfo{year}{2018}), \bibinfo{numpages}{21}~pages.
\newblock
\showISSN{1559-1131}
\urldef\tempurl%
\url{https://doi.org/10.1145/3160000}
\showDOI{\tempurl}


\bibitem[\protect\citeauthoryear{Gallier}{Gallier}{2016}]%
        {gallier2016spectral}
\bibfield{author}{\bibinfo{person}{Jean Gallier}.}
  \bibinfo{year}{2016}\natexlab{}.
\newblock \showarticletitle{Spectral theory of unsigned and signed graphs.
  applications to graph clustering: a survey}.
\newblock \bibinfo{journal}{\emph{arXiv preprint arXiv:1601.04692}}
  (\bibinfo{year}{2016}).
\newblock


\bibitem[\protect\citeauthoryear{Gregor, Danihelka, Graves, Rezende, and
  Wierstra}{Gregor et~al\mbox{.}}{2015}]%
        {gregor2015draw}
\bibfield{author}{\bibinfo{person}{Karol Gregor}, \bibinfo{person}{Ivo
  Danihelka}, \bibinfo{person}{Alex Graves}, \bibinfo{person}{Danilo~Jimenez
  Rezende}, {and} \bibinfo{person}{Daan Wierstra}.}
  \bibinfo{year}{2015}\natexlab{}.
\newblock \showarticletitle{Draw: A recurrent neural network for image
  generation}.
\newblock \bibinfo{journal}{\emph{Proceedings of the 32nd International
  Conference on Machine Learning, PMLR 37:1462-1471, 2015}}
  (\bibinfo{year}{2015}).
\newblock


\bibitem[\protect\citeauthoryear{Grover and Leskovec}{Grover and
  Leskovec}{2016}]%
        {grover2016node2vec}
\bibfield{author}{\bibinfo{person}{Aditya Grover} {and} \bibinfo{person}{Jure
  Leskovec}.} \bibinfo{year}{2016}\natexlab{}.
\newblock \showarticletitle{node2vec: Scalable feature learning for networks}.
  In \bibinfo{booktitle}{\emph{Proceedings of the 22nd ACM SIGKDD international
  conference on Knowledge discovery and data mining}}. ACM,
  \bibinfo{pages}{855--864}.
\newblock


\bibitem[\protect\citeauthoryear{Hamilton, Ying, and Leskovec}{Hamilton
  et~al\mbox{.}}{2017a}]%
        {hamilton2017inductive}
\bibfield{author}{\bibinfo{person}{Will Hamilton}, \bibinfo{person}{Zhitao
  Ying}, {and} \bibinfo{person}{Jure Leskovec}.}
  \bibinfo{year}{2017}\natexlab{a}.
\newblock \showarticletitle{Inductive representation learning on large graphs}.
  In \bibinfo{booktitle}{\emph{Advances in Neural Information Processing
  Systems}}. \bibinfo{pages}{1024--1034}.
\newblock


\bibitem[\protect\citeauthoryear{Hamilton, Ying, and Leskovec}{Hamilton
  et~al\mbox{.}}{2017b}]%
        {hamilton2017representation}
\bibfield{author}{\bibinfo{person}{William~L Hamilton}, \bibinfo{person}{Rex
  Ying}, {and} \bibinfo{person}{Jure Leskovec}.}
  \bibinfo{year}{2017}\natexlab{b}.
\newblock \showarticletitle{Representation learning on graphs: Methods and
  applications}.
\newblock \bibinfo{journal}{\emph{arXiv preprint arXiv:1709.05584}}
  (\bibinfo{year}{2017}).
\newblock


\bibitem[\protect\citeauthoryear{He and McAuley}{He and McAuley}{2016}]%
        {he2016vbpr}
\bibfield{author}{\bibinfo{person}{Ruining He} {and} \bibinfo{person}{Julian
  McAuley}.} \bibinfo{year}{2016}\natexlab{}.
\newblock \showarticletitle{VBPR: visual bayesian personalized ranking from
  implicit feedback}. In \bibinfo{booktitle}{\emph{Thirtieth AAAI Conference on
  Artificial Intelligence}}.
\newblock


\bibitem[\protect\citeauthoryear{Hsieh, Chiang, and Dhillon}{Hsieh
  et~al\mbox{.}}{2012}]%
        {hsieh2012low}
\bibfield{author}{\bibinfo{person}{Cho-Jui Hsieh}, \bibinfo{person}{Kai-Yang
  Chiang}, {and} \bibinfo{person}{Inderjit~S Dhillon}.}
  \bibinfo{year}{2012}\natexlab{}.
\newblock \showarticletitle{Low rank modeling of signed networks}. In
  \bibinfo{booktitle}{\emph{Proceedings of the 18th ACM SIGKDD international
  conference on Knowledge discovery and data mining}}. ACM,
  \bibinfo{pages}{507--515}.
\newblock


\bibitem[\protect\citeauthoryear{Huangjie~Zheng and Tsang}{Huangjie~Zheng and
  Tsang}{2018}]%
        {Zheng2018DegenerationIV}
\bibfield{author}{\bibinfo{person}{Ya~Zhang Huangjie~Zheng, Jiangchao~Yao}
  {and} \bibinfo{person}{Ivor~W. Tsang}.} \bibinfo{year}{2018}\natexlab{}.
\newblock \showarticletitle{Degeneration in VAE: in the Light of Fisher
  Information Loss}.
\newblock \bibinfo{journal}{\emph{ArXiv}}  \bibinfo{volume}{abs/1802.06677}
  (\bibinfo{year}{2018}).
\newblock


\bibitem[\protect\citeauthoryear{Jimenez~Rezende and Mohamed}{Jimenez~Rezende
  and Mohamed}{2015}]%
        {jimenez2015variational}
\bibfield{author}{\bibinfo{person}{Danilo Jimenez~Rezende} {and}
  \bibinfo{person}{Shakir Mohamed}.} \bibinfo{year}{2015}\natexlab{}.
\newblock \showarticletitle{Variational inference with normalizing flows}.
\newblock \bibinfo{journal}{\emph{arXiv preprint arXiv:1505.05770}}
  (\bibinfo{year}{2015}).
\newblock


\bibitem[\protect\citeauthoryear{Karpathy et~al\mbox{.}}{Karpathy
  et~al\mbox{.}}{2016}]%
        {karpathy2016cs231n}
\bibfield{author}{\bibinfo{person}{Andrej Karpathy} {et~al\mbox{.}}}
  \bibinfo{year}{2016}\natexlab{}.
\newblock \showarticletitle{Cs231n convolutional neural networks for visual
  recognition}.
\newblock \bibinfo{journal}{\emph{Neural networks}}  \bibinfo{volume}{1}
  (\bibinfo{year}{2016}).
\newblock


\bibitem[\protect\citeauthoryear{Kim, Park, Lee, and Kang}{Kim
  et~al\mbox{.}}{2018}]%
        {kim2018side}
\bibfield{author}{\bibinfo{person}{Junghwan Kim}, \bibinfo{person}{Haekyu
  Park}, \bibinfo{person}{Ji-Eun Lee}, {and} \bibinfo{person}{U Kang}.}
  \bibinfo{year}{2018}\natexlab{}.
\newblock \showarticletitle{Side: representation learning in signed directed
  networks}. In \bibinfo{booktitle}{\emph{Proceedings of the 2018 World Wide
  Web Conference on World Wide Web}}. International World Wide Web Conferences
  Steering Committee, \bibinfo{pages}{509--518}.
\newblock


\bibitem[\protect\citeauthoryear{Kingma and Welling}{Kingma and
  Welling}{2013}]%
        {kingma2013auto}
\bibfield{author}{\bibinfo{person}{Diederik~P Kingma} {and}
  \bibinfo{person}{Max Welling}.} \bibinfo{year}{2013}\natexlab{}.
\newblock \showarticletitle{Auto-encoding variational bayes}.
\newblock \bibinfo{journal}{\emph{arXiv preprint arXiv:1312.6114}}
  (\bibinfo{year}{2013}).
\newblock


\bibitem[\protect\citeauthoryear{Kipf, Fetaya, Wang, Welling, and Zemel}{Kipf
  et~al\mbox{.}}{2018}]%
        {kipf2018neural}
\bibfield{author}{\bibinfo{person}{Thomas Kipf}, \bibinfo{person}{Ethan
  Fetaya}, \bibinfo{person}{Kuan-Chieh Wang}, \bibinfo{person}{Max Welling},
  {and} \bibinfo{person}{Richard Zemel}.} \bibinfo{year}{2018}\natexlab{}.
\newblock \showarticletitle{Neural relational inference for interacting
  systems}.
\newblock \bibinfo{journal}{\emph{arXiv preprint arXiv:1802.04687}}
  (\bibinfo{year}{2018}).
\newblock


\bibitem[\protect\citeauthoryear{Kipf and Welling}{Kipf and Welling}{2016a}]%
        {kipf2016semi}
\bibfield{author}{\bibinfo{person}{Thomas~N Kipf} {and} \bibinfo{person}{Max
  Welling}.} \bibinfo{year}{2016}\natexlab{a}.
\newblock \showarticletitle{Semi-supervised classification with graph
  convolutional networks}.
\newblock \bibinfo{journal}{\emph{arXiv preprint arXiv:1609.02907}}
  (\bibinfo{year}{2016}).
\newblock


\bibitem[\protect\citeauthoryear{Kipf and Welling}{Kipf and Welling}{2016b}]%
        {kipf2016variational}
\bibfield{author}{\bibinfo{person}{Thomas~N Kipf} {and} \bibinfo{person}{Max
  Welling}.} \bibinfo{year}{2016}\natexlab{b}.
\newblock \showarticletitle{Variational graph auto-encoders}.
\newblock \bibinfo{journal}{\emph{arXiv preprint arXiv:1611.07308}}
  (\bibinfo{year}{2016}).
\newblock


\bibitem[\protect\citeauthoryear{Kunegis, Preusse, and Schwagereit}{Kunegis
  et~al\mbox{.}}{2013}]%
        {kunegis2013added}
\bibfield{author}{\bibinfo{person}{J{\'e}r{\^o}me Kunegis},
  \bibinfo{person}{Julia Preusse}, {and} \bibinfo{person}{Felix Schwagereit}.}
  \bibinfo{year}{2013}\natexlab{}.
\newblock \showarticletitle{What is the added value of negative links in online
  social networks?}. In \bibinfo{booktitle}{\emph{Proceedings of the 22nd
  international conference on World Wide Web}}. ACM, \bibinfo{pages}{727--736}.
\newblock


\bibitem[\protect\citeauthoryear{Kunegis, Schmidt, Lommatzsch, Lerner, De~Luca,
  and Albayrak}{Kunegis et~al\mbox{.}}{2010}]%
        {kunegis2010spectral}
\bibfield{author}{\bibinfo{person}{J{\'e}r{\^o}me Kunegis},
  \bibinfo{person}{Stephan Schmidt}, \bibinfo{person}{Andreas Lommatzsch},
  \bibinfo{person}{J{\"u}rgen Lerner}, \bibinfo{person}{Ernesto~W De~Luca},
  {and} \bibinfo{person}{Sahin Albayrak}.} \bibinfo{year}{2010}\natexlab{}.
\newblock \showarticletitle{Spectral analysis of signed graphs for clustering,
  prediction and visualization}. In \bibinfo{booktitle}{\emph{Proceedings of
  the 2010 SIAM International Conference on Data Mining}}. SIAM,
  \bibinfo{pages}{559--570}.
\newblock


\bibitem[\protect\citeauthoryear{Kusner, Paige, and
  Hern{\'a}ndez-Lobato}{Kusner et~al\mbox{.}}{2017}]%
        {kusner2017grammar}
\bibfield{author}{\bibinfo{person}{Matt~J Kusner}, \bibinfo{person}{Brooks
  Paige}, {and} \bibinfo{person}{Jos{\'e}~Miguel Hern{\'a}ndez-Lobato}.}
  \bibinfo{year}{2017}\natexlab{}.
\newblock \showarticletitle{Grammar variational autoencoder}. In
  \bibinfo{booktitle}{\emph{Proceedings of the 34th International Conference on
  Machine Learning-Volume 70}}. JMLR. org, \bibinfo{pages}{1945--1954}.
\newblock


\bibitem[\protect\citeauthoryear{Leskovec, Huttenlocher, and
  Kleinberg}{Leskovec et~al\mbox{.}}{2010}]%
        {leskovec2010predicting}
\bibfield{author}{\bibinfo{person}{Jure Leskovec}, \bibinfo{person}{Daniel
  Huttenlocher}, {and} \bibinfo{person}{Jon Kleinberg}.}
  \bibinfo{year}{2010}\natexlab{}.
\newblock \showarticletitle{Predicting positive and negative links in online
  social networks}. In \bibinfo{booktitle}{\emph{Proceedings of the 19th
  international conference on World wide web}}. ACM, \bibinfo{pages}{641--650}.
\newblock


\bibitem[\protect\citeauthoryear{Liben-Nowell and Kleinberg}{Liben-Nowell and
  Kleinberg}{2007}]%
        {liben2007link}
\bibfield{author}{\bibinfo{person}{David Liben-Nowell} {and}
  \bibinfo{person}{Jon Kleinberg}.} \bibinfo{year}{2007}\natexlab{}.
\newblock \showarticletitle{The link-prediction problem for social networks}.
\newblock \bibinfo{journal}{\emph{Journal of the American society for
  information science and technology}} \bibinfo{volume}{58},
  \bibinfo{number}{7} (\bibinfo{year}{2007}), \bibinfo{pages}{1019--1031}.
\newblock


\bibitem[\protect\citeauthoryear{Liu, Wu, and Wang}{Liu et~al\mbox{.}}{2017}]%
        {liu2017deepstyle}
\bibfield{author}{\bibinfo{person}{Qiang Liu}, \bibinfo{person}{Shu Wu}, {and}
  \bibinfo{person}{Liang Wang}.} \bibinfo{year}{2017}\natexlab{}.
\newblock \showarticletitle{DeepStyle: Learning user preferences for visual
  recommendation}. In \bibinfo{booktitle}{\emph{Proceedings of the 40th
  International ACM SIGIR Conference on Research and Development in Information
  Retrieval}}. ACM, \bibinfo{pages}{841--844}.
\newblock


\bibitem[\protect\citeauthoryear{Maaten and Hinton}{Maaten and Hinton}{2008}]%
        {maaten2008visualizing}
\bibfield{author}{\bibinfo{person}{Laurens van~der Maaten} {and}
  \bibinfo{person}{Geoffrey Hinton}.} \bibinfo{year}{2008}\natexlab{}.
\newblock \showarticletitle{Visualizing data using t-SNE}.
\newblock \bibinfo{journal}{\emph{Journal of machine learning research}}
  \bibinfo{volume}{9}, \bibinfo{number}{Nov} (\bibinfo{year}{2008}),
  \bibinfo{pages}{2579--2605}.
\newblock


\bibitem[\protect\citeauthoryear{Mikolov, Sutskever, Chen, Corrado, and
  Dean}{Mikolov et~al\mbox{.}}{2013}]%
        {mikolov2013distributed}
\bibfield{author}{\bibinfo{person}{Tomas Mikolov}, \bibinfo{person}{Ilya
  Sutskever}, \bibinfo{person}{Kai Chen}, \bibinfo{person}{Greg~S Corrado},
  {and} \bibinfo{person}{Jeff Dean}.} \bibinfo{year}{2013}\natexlab{}.
\newblock \showarticletitle{Distributed representations of words and phrases
  and their compositionality}. In \bibinfo{booktitle}{\emph{Advances in neural
  information processing systems}}. \bibinfo{pages}{3111--3119}.
\newblock


\bibitem[\protect\citeauthoryear{Ou, Cui, Pei, Zhang, and Zhu}{Ou
  et~al\mbox{.}}{2016}]%
        {ou2016asymmetric}
\bibfield{author}{\bibinfo{person}{Mingdong Ou}, \bibinfo{person}{Peng Cui},
  \bibinfo{person}{Jian Pei}, \bibinfo{person}{Ziwei Zhang}, {and}
  \bibinfo{person}{Wenwu Zhu}.} \bibinfo{year}{2016}\natexlab{}.
\newblock \showarticletitle{Asymmetric transitivity preserving graph
  embedding}. In \bibinfo{booktitle}{\emph{Proceedings of the 22nd ACM SIGKDD
  international conference on Knowledge discovery and data mining}}. ACM,
  \bibinfo{pages}{1105--1114}.
\newblock


\bibitem[\protect\citeauthoryear{Papadopoulos, Kompatsiaris, Vakali, and
  Spyridonos}{Papadopoulos et~al\mbox{.}}{2012}]%
        {papadopoulos2012community}
\bibfield{author}{\bibinfo{person}{Symeon Papadopoulos},
  \bibinfo{person}{Yiannis Kompatsiaris}, \bibinfo{person}{Athena Vakali},
  {and} \bibinfo{person}{Ploutarchos Spyridonos}.}
  \bibinfo{year}{2012}\natexlab{}.
\newblock \showarticletitle{Community detection in social media}.
\newblock \bibinfo{journal}{\emph{Data Mining and Knowledge Discovery}}
  \bibinfo{volume}{24}, \bibinfo{number}{3} (\bibinfo{year}{2012}),
  \bibinfo{pages}{515--554}.
\newblock


\bibitem[\protect\citeauthoryear{Perozzi, Al-Rfou, and Skiena}{Perozzi
  et~al\mbox{.}}{2014}]%
        {perozzi2014deepwalk}
\bibfield{author}{\bibinfo{person}{Bryan Perozzi}, \bibinfo{person}{Rami
  Al-Rfou}, {and} \bibinfo{person}{Steven Skiena}.}
  \bibinfo{year}{2014}\natexlab{}.
\newblock \showarticletitle{Deepwalk: Online learning of social
  representations}. In \bibinfo{booktitle}{\emph{Proceedings of the 20th ACM
  SIGKDD international conference on Knowledge discovery and data mining}}.
  ACM, \bibinfo{pages}{701--710}.
\newblock


\bibitem[\protect\citeauthoryear{Pu, Gan, Henao, Yuan, Li, Stevens, and
  Carin}{Pu et~al\mbox{.}}{2016}]%
        {pu2016variational}
\bibfield{author}{\bibinfo{person}{Yunchen Pu}, \bibinfo{person}{Zhe Gan},
  \bibinfo{person}{Ricardo Henao}, \bibinfo{person}{Xin Yuan},
  \bibinfo{person}{Chunyuan Li}, \bibinfo{person}{Andrew Stevens}, {and}
  \bibinfo{person}{Lawrence Carin}.} \bibinfo{year}{2016}\natexlab{}.
\newblock \showarticletitle{Variational autoencoder for deep learning of
  images, labels and captions}. In \bibinfo{booktitle}{\emph{Advances in neural
  information processing systems}}. \bibinfo{pages}{2352--2360}.
\newblock


\bibitem[\protect\citeauthoryear{Qian and Adali}{Qian and Adali}{2013}]%
        {qian2013extended}
\bibfield{author}{\bibinfo{person}{Yi Qian} {and} \bibinfo{person}{Sibel
  Adali}.} \bibinfo{year}{2013}\natexlab{}.
\newblock \showarticletitle{Extended structural balance theory for modeling
  trust in social networks}. In \bibinfo{booktitle}{\emph{Privacy, Security and
  Trust (PST), 2013 Eleventh Annual International Conference on}}. IEEE,
  \bibinfo{pages}{283--290}.
\newblock


\bibitem[\protect\citeauthoryear{Qian and Adali}{Qian and Adali}{2014}]%
        {Qian:2014:FTD:2639948.2628438}
\bibfield{author}{\bibinfo{person}{Yi Qian} {and} \bibinfo{person}{Sibel
  Adali}.} \bibinfo{year}{2014}\natexlab{}.
\newblock \showarticletitle{Foundations of Trust and Distrust in Networks:
  Extended Structural Balance Theory}.
\newblock \bibinfo{journal}{\emph{ACM Trans. Web}} \bibinfo{volume}{8},
  \bibinfo{number}{3}, Article \bibinfo{articleno}{13} (\bibinfo{date}{July}
  \bibinfo{year}{2014}), \bibinfo{numpages}{33}~pages.
\newblock
\showISSN{1559-1131}
\urldef\tempurl%
\url{https://doi.org/10.1145/2628438}
\showDOI{\tempurl}


\bibitem[\protect\citeauthoryear{Qiu, Dong, Ma, Li, Wang, and Tang}{Qiu
  et~al\mbox{.}}{2018}]%
        {qiu2018network}
\bibfield{author}{\bibinfo{person}{Jiezhong Qiu}, \bibinfo{person}{Yuxiao
  Dong}, \bibinfo{person}{Hao Ma}, \bibinfo{person}{Jian Li},
  \bibinfo{person}{Kuansan Wang}, {and} \bibinfo{person}{Jie Tang}.}
  \bibinfo{year}{2018}\natexlab{}.
\newblock \showarticletitle{Network embedding as matrix factorization: Unifying
  deepwalk, line, pte, and node2vec}. In \bibinfo{booktitle}{\emph{Proceedings
  of the Eleventh ACM International Conference on Web Search and Data Mining}}.
  ACM, \bibinfo{pages}{459--467}.
\newblock


\bibitem[\protect\citeauthoryear{Rendle, Freudenthaler, Gantner, and
  Schmidt-Thieme}{Rendle et~al\mbox{.}}{2009}]%
        {rendle2009bpr}
\bibfield{author}{\bibinfo{person}{Steffen Rendle}, \bibinfo{person}{Christoph
  Freudenthaler}, \bibinfo{person}{Zeno Gantner}, {and} \bibinfo{person}{Lars
  Schmidt-Thieme}.} \bibinfo{year}{2009}\natexlab{}.
\newblock \showarticletitle{BPR: Bayesian personalized ranking from implicit
  feedback}. In \bibinfo{booktitle}{\emph{Proceedings of the twenty-fifth
  conference on uncertainty in artificial intelligence}}. AUAI Press,
  \bibinfo{pages}{452--461}.
\newblock


\bibitem[\protect\citeauthoryear{Rendle and Schmidt-Thieme}{Rendle and
  Schmidt-Thieme}{2010}]%
        {rendle2010pairwise}
\bibfield{author}{\bibinfo{person}{Steffen Rendle} {and} \bibinfo{person}{Lars
  Schmidt-Thieme}.} \bibinfo{year}{2010}\natexlab{}.
\newblock \showarticletitle{Pairwise interaction tensor factorization for
  personalized tag recommendation}. In \bibinfo{booktitle}{\emph{Proceedings of
  the third ACM international conference on Web search and data mining}}. ACM,
  \bibinfo{pages}{81--90}.
\newblock


\bibitem[\protect\citeauthoryear{Rezende and Mohamed}{Rezende and
  Mohamed}{2015}]%
        {rezende2015variational}
\bibfield{author}{\bibinfo{person}{Danilo~Jimenez Rezende} {and}
  \bibinfo{person}{Shakir Mohamed}.} \bibinfo{year}{2015}\natexlab{}.
\newblock \showarticletitle{Variational inference with normalizing flows}.
\newblock \bibinfo{journal}{\emph{arXiv preprint arXiv:1505.05770}}
  (\bibinfo{year}{2015}).
\newblock


\bibitem[\protect\citeauthoryear{Salimans, Kingma, and Welling}{Salimans
  et~al\mbox{.}}{2015}]%
        {salimans2015markov}
\bibfield{author}{\bibinfo{person}{Tim Salimans}, \bibinfo{person}{Diederik
  Kingma}, {and} \bibinfo{person}{Max Welling}.}
  \bibinfo{year}{2015}\natexlab{}.
\newblock \showarticletitle{Markov chain monte carlo and variational inference:
  Bridging the gap}. In \bibinfo{booktitle}{\emph{International Conference on
  Machine Learning}}. \bibinfo{pages}{1218--1226}.
\newblock


\bibitem[\protect\citeauthoryear{Shen, Pan, Liu, Ong, and Sun}{Shen
  et~al\mbox{.}}{2018}]%
        {shen2018discrete}
\bibfield{author}{\bibinfo{person}{Xiaobo Shen}, \bibinfo{person}{Shirui Pan},
  \bibinfo{person}{Weiwei Liu}, \bibinfo{person}{Yew-Soon Ong}, {and}
  \bibinfo{person}{Quan-Sen Sun}.} \bibinfo{year}{2018}\natexlab{}.
\newblock \showarticletitle{Discrete network embedding}. In
  \bibinfo{booktitle}{\emph{Proceedings of the 27th International Joint
  Conference on Artificial Intelligence}}. AAAI Press,
  \bibinfo{pages}{3549--3555}.
\newblock


\bibitem[\protect\citeauthoryear{Sohn, Lee, and Yan}{Sohn
  et~al\mbox{.}}{2015}]%
        {sohn2015learning}
\bibfield{author}{\bibinfo{person}{Kihyuk Sohn}, \bibinfo{person}{Honglak Lee},
  {and} \bibinfo{person}{Xinchen Yan}.} \bibinfo{year}{2015}\natexlab{}.
\newblock \showarticletitle{Learning structured output representation using
  deep conditional generative models}. In \bibinfo{booktitle}{\emph{Advances in
  neural information processing systems}}. \bibinfo{pages}{3483--3491}.
\newblock


\bibitem[\protect\citeauthoryear{Tang, Chang, Aggarwal, and Liu}{Tang
  et~al\mbox{.}}{2015a}]%
        {tang2015negative}
\bibfield{author}{\bibinfo{person}{Jiliang Tang}, \bibinfo{person}{Shiyu
  Chang}, \bibinfo{person}{Charu Aggarwal}, {and} \bibinfo{person}{Huan Liu}.}
  \bibinfo{year}{2015}\natexlab{a}.
\newblock \showarticletitle{Negative link prediction in social media}. In
  \bibinfo{booktitle}{\emph{Proceedings of the Eighth ACM International
  Conference on Web Search and Data Mining}}. ACM, \bibinfo{pages}{87--96}.
\newblock


\bibitem[\protect\citeauthoryear{Tang, Qu, Wang, Zhang, Yan, and Mei}{Tang
  et~al\mbox{.}}{2015b}]%
        {tang2015line}
\bibfield{author}{\bibinfo{person}{Jian Tang}, \bibinfo{person}{Meng Qu},
  \bibinfo{person}{Mingzhe Wang}, \bibinfo{person}{Ming Zhang},
  \bibinfo{person}{Jun Yan}, {and} \bibinfo{person}{Qiaozhu Mei}.}
  \bibinfo{year}{2015}\natexlab{b}.
\newblock \showarticletitle{Line: Large-scale information network embedding}.
  In \bibinfo{booktitle}{\emph{Proceedings of the 24th International Conference
  on World Wide Web}}. International World Wide Web Conferences Steering
  Committee, \bibinfo{pages}{1067--1077}.
\newblock


\bibitem[\protect\citeauthoryear{Tieleman and Hinton}{Tieleman and
  Hinton}{2014}]%
        {tieleman2014rmsprop}
\bibfield{author}{\bibinfo{person}{Tijmen Tieleman} {and}
  \bibinfo{person}{Geoffery Hinton}.} \bibinfo{year}{2014}\natexlab{}.
\newblock \showarticletitle{RMSprop gradient optimization}.
\newblock \bibinfo{journal}{\emph{URL http://www. cs. toronto.
  edu/tijmen/csc321/slides/lecture\_slides\_lec6. pdf}} (\bibinfo{year}{2014}).
\newblock


\bibitem[\protect\citeauthoryear{Tomczak and Welling}{Tomczak and
  Welling}{2017}]%
        {tomczak2017vae}
\bibfield{author}{\bibinfo{person}{Jakub~M Tomczak} {and} \bibinfo{person}{Max
  Welling}.} \bibinfo{year}{2017}\natexlab{}.
\newblock \showarticletitle{VAE with a VampPrior}.
\newblock \bibinfo{journal}{\emph{arXiv preprint arXiv:1705.07120}}
  (\bibinfo{year}{2017}).
\newblock


\bibitem[\protect\citeauthoryear{Veli{\v{c}}kovi{\'c}, Cucurull, Casanova,
  Romero, Lio, and Bengio}{Veli{\v{c}}kovi{\'c} et~al\mbox{.}}{2017}]%
        {velivckovic2017graph}
\bibfield{author}{\bibinfo{person}{Petar Veli{\v{c}}kovi{\'c}},
  \bibinfo{person}{Guillem Cucurull}, \bibinfo{person}{Arantxa Casanova},
  \bibinfo{person}{Adriana Romero}, \bibinfo{person}{Pietro Lio}, {and}
  \bibinfo{person}{Yoshua Bengio}.} \bibinfo{year}{2017}\natexlab{}.
\newblock \showarticletitle{Graph attention networks}.
\newblock \bibinfo{journal}{\emph{arXiv preprint arXiv:1710.10903}}
  (\bibinfo{year}{2017}).
\newblock


\bibitem[\protect\citeauthoryear{Velickovic, Cucurull, Casanova, Romero, Lio,
  and Bengio}{Velickovic et~al\mbox{.}}{2017}]%
        {velickovic2017graph}
\bibfield{author}{\bibinfo{person}{Petar Velickovic}, \bibinfo{person}{Guillem
  Cucurull}, \bibinfo{person}{Arantxa Casanova}, \bibinfo{person}{Adriana
  Romero}, \bibinfo{person}{Pietro Lio}, {and} \bibinfo{person}{Yoshua
  Bengio}.} \bibinfo{year}{2017}\natexlab{}.
\newblock \showarticletitle{Graph attention networks}.
\newblock \bibinfo{journal}{\emph{arXiv preprint arXiv:1710.10903}}
  \bibinfo{volume}{1}, \bibinfo{number}{2} (\bibinfo{year}{2017}).
\newblock


\bibitem[\protect\citeauthoryear{Victor, Verbiest, Cornelis, and Cock}{Victor
  et~al\mbox{.}}{2013}]%
        {Victor:2013:ETR:2460383.2460385}
\bibfield{author}{\bibinfo{person}{Patricia Victor}, \bibinfo{person}{Nele
  Verbiest}, \bibinfo{person}{Chris Cornelis}, {and}
  \bibinfo{person}{Martine~De Cock}.} \bibinfo{year}{2013}\natexlab{}.
\newblock \showarticletitle{Enhancing the Trust-based Recommendation Process
  with Explicit Distrust}.
\newblock \bibinfo{journal}{\emph{ACM Trans. Web}} \bibinfo{volume}{7},
  \bibinfo{number}{2}, Article \bibinfo{articleno}{6} (\bibinfo{date}{May}
  \bibinfo{year}{2013}), \bibinfo{numpages}{19}~pages.
\newblock
\showISSN{1559-1131}
\urldef\tempurl%
\url{https://doi.org/10.1145/2460383.2460385}
\showDOI{\tempurl}


\bibitem[\protect\citeauthoryear{Wang, Cui, and Zhu}{Wang
  et~al\mbox{.}}{2016a}]%
        {wang2016structural}
\bibfield{author}{\bibinfo{person}{Daixin Wang}, \bibinfo{person}{Peng Cui},
  {and} \bibinfo{person}{Wenwu Zhu}.} \bibinfo{year}{2016}\natexlab{a}.
\newblock \showarticletitle{Structural deep network embedding}. In
  \bibinfo{booktitle}{\emph{Proceedings of the 22nd ACM SIGKDD international
  conference on Knowledge discovery and data mining}}. ACM,
  \bibinfo{pages}{1225--1234}.
\newblock


\bibitem[\protect\citeauthoryear{Wang, Wang, Zhao, Cao, and Guo}{Wang
  et~al\mbox{.}}{2017c}]%
        {wang2017joint}
\bibfield{author}{\bibinfo{person}{Hongwei Wang}, \bibinfo{person}{Jia Wang},
  \bibinfo{person}{Miao Zhao}, \bibinfo{person}{Jiannong Cao}, {and}
  \bibinfo{person}{Minyi Guo}.} \bibinfo{year}{2017}\natexlab{c}.
\newblock \showarticletitle{Joint topic-semantic-aware social recommendation
  for online voting}. In \bibinfo{booktitle}{\emph{Proceedings of the 2017 ACM
  on Conference on Information and Knowledge Management}}. ACM,
  \bibinfo{pages}{347--356}.
\newblock


\bibitem[\protect\citeauthoryear{Wang, Zhang, Hou, Xie, Guo, and Liu}{Wang
  et~al\mbox{.}}{2018}]%
        {wang2018shine}
\bibfield{author}{\bibinfo{person}{Hongwei Wang}, \bibinfo{person}{Fuzheng
  Zhang}, \bibinfo{person}{Min Hou}, \bibinfo{person}{Xing Xie},
  \bibinfo{person}{Minyi Guo}, {and} \bibinfo{person}{Qi Liu}.}
  \bibinfo{year}{2018}\natexlab{}.
\newblock \showarticletitle{Shine: Signed heterogeneous information network
  embedding for sentiment link prediction}. In
  \bibinfo{booktitle}{\emph{Proceedings of the Eleventh ACM International
  Conference on Web Search and Data Mining}}. ACM, \bibinfo{pages}{592--600}.
\newblock


\bibitem[\protect\citeauthoryear{Wang, Aggarwal, Tang, and Liu}{Wang
  et~al\mbox{.}}{2017a}]%
        {wang2017attributed}
\bibfield{author}{\bibinfo{person}{Suhang Wang}, \bibinfo{person}{Charu
  Aggarwal}, \bibinfo{person}{Jiliang Tang}, {and} \bibinfo{person}{Huan Liu}.}
  \bibinfo{year}{2017}\natexlab{a}.
\newblock \showarticletitle{Attributed signed network embedding}. In
  \bibinfo{booktitle}{\emph{Proceedings of the 2017 ACM on Conference on
  Information and Knowledge Management}}. ACM, \bibinfo{pages}{137--146}.
\newblock


\bibitem[\protect\citeauthoryear{Wang, Tang, Aggarwal, Chang, and Liu}{Wang
  et~al\mbox{.}}{2017b}]%
        {wang2017signed}
\bibfield{author}{\bibinfo{person}{Suhang Wang}, \bibinfo{person}{Jiliang
  Tang}, \bibinfo{person}{Charu Aggarwal}, \bibinfo{person}{Yi Chang}, {and}
  \bibinfo{person}{Huan Liu}.} \bibinfo{year}{2017}\natexlab{b}.
\newblock \showarticletitle{Signed network embedding in social media}. In
  \bibinfo{booktitle}{\emph{Proceedings of the 2017 SIAM international
  conference on data mining}}. SIAM, \bibinfo{pages}{327--335}.
\newblock


\bibitem[\protect\citeauthoryear{Wang, Tang, Aggarwal, and Liu}{Wang
  et~al\mbox{.}}{2016b}]%
        {wang2016linked}
\bibfield{author}{\bibinfo{person}{Suhang Wang}, \bibinfo{person}{Jiliang
  Tang}, \bibinfo{person}{Charu Aggarwal}, {and} \bibinfo{person}{Huan Liu}.}
  \bibinfo{year}{2016}\natexlab{b}.
\newblock \showarticletitle{Linked document embedding for classification}. In
  \bibinfo{booktitle}{\emph{Proceedings of the 25th ACM international on
  conference on information and knowledge management}}. ACM,
  \bibinfo{pages}{115--124}.
\newblock


\bibitem[\protect\citeauthoryear{Wu, Pan, Chen, Long, Zhang, and Yu}{Wu
  et~al\mbox{.}}{2019}]%
        {wu2019comprehensive}
\bibfield{author}{\bibinfo{person}{Zonghan Wu}, \bibinfo{person}{Shirui Pan},
  \bibinfo{person}{Fengwen Chen}, \bibinfo{person}{Guodong Long},
  \bibinfo{person}{Chengqi Zhang}, {and} \bibinfo{person}{Philip~S Yu}.}
  \bibinfo{year}{2019}\natexlab{}.
\newblock \showarticletitle{A comprehensive survey on graph neural networks}.
\newblock \bibinfo{journal}{\emph{arXiv preprint arXiv:1901.00596}}
  (\bibinfo{year}{2019}).
\newblock


\bibitem[\protect\citeauthoryear{Xu, Shen, Cao, Cen, and Cheng}{Xu
  et~al\mbox{.}}{2019}]%
        {xu2019graph}
\bibfield{author}{\bibinfo{person}{Bingbing Xu}, \bibinfo{person}{Huawei Shen},
  \bibinfo{person}{Qi Cao}, \bibinfo{person}{Keting Cen}, {and}
  \bibinfo{person}{Xueqi Cheng}.} \bibinfo{year}{2019}\natexlab{}.
\newblock \showarticletitle{Graph convolutional networks using heat kernel for
  semi-supervised learning}. In \bibinfo{booktitle}{\emph{Proceedings of the
  28th International Joint Conference on Artificial Intelligence}}. AAAI Press,
  \bibinfo{pages}{1928--1934}.
\newblock


\bibitem[\protect\citeauthoryear{Yin and Zhou}{Yin and Zhou}{2018}]%
        {yin2018semi}
\bibfield{author}{\bibinfo{person}{Mingzhang Yin} {and}
  \bibinfo{person}{Mingyuan Zhou}.} \bibinfo{year}{2018}\natexlab{}.
\newblock \showarticletitle{Semi-implicit variational inference}.
\newblock \bibinfo{journal}{\emph{International Conference on Machine
  Learning}} (\bibinfo{year}{2018}).
\newblock


\bibitem[\protect\citeauthoryear{Yuan, Wu, and Xiang}{Yuan
  et~al\mbox{.}}{2017}]%
        {yuan2017sne}
\bibfield{author}{\bibinfo{person}{Shuhan Yuan}, \bibinfo{person}{Xintao Wu},
  {and} \bibinfo{person}{Yang Xiang}.} \bibinfo{year}{2017}\natexlab{}.
\newblock \showarticletitle{SNE: signed network embedding}. In
  \bibinfo{booktitle}{\emph{Pacific-Asia conference on knowledge discovery and
  data mining}}. Springer, \bibinfo{pages}{183--195}.
\newblock


\bibitem[\protect\citeauthoryear{Zhang, Li, Zhu, and Liang}{Zhang
  et~al\mbox{.}}{2016}]%
        {Zhang:2016:DSP:2870642.2846102}
\bibfield{author}{\bibinfo{person}{Xianchao Zhang}, \bibinfo{person}{Zhaoxing
  Li}, \bibinfo{person}{Shaoping Zhu}, {and} \bibinfo{person}{Wenxin Liang}.}
  \bibinfo{year}{2016}\natexlab{}.
\newblock \showarticletitle{Detecting Spam and Promoting Campaigns in Twitter}.
\newblock \bibinfo{journal}{\emph{ACM Trans. Web}} \bibinfo{volume}{10},
  \bibinfo{number}{1}, Article \bibinfo{articleno}{4} (\bibinfo{date}{Feb.}
  \bibinfo{year}{2016}), \bibinfo{numpages}{28}~pages.
\newblock
\showISSN{1559-1131}
\urldef\tempurl%
\url{https://doi.org/10.1145/2846102}
\showDOI{\tempurl}


\bibitem[\protect\citeauthoryear{Zhang, Cui, and Zhu}{Zhang
  et~al\mbox{.}}{2018}]%
        {zhang2018deep}
\bibfield{author}{\bibinfo{person}{Ziwei Zhang}, \bibinfo{person}{Peng Cui},
  {and} \bibinfo{person}{Wenwu Zhu}.} \bibinfo{year}{2018}\natexlab{}.
\newblock \showarticletitle{Deep learning on graphs: A survey}.
\newblock \bibinfo{journal}{\emph{arXiv preprint arXiv:1812.04202}}
  (\bibinfo{year}{2018}).
\newblock


\bibitem[\protect\citeauthoryear{Zhao, Song, and Ermon}{Zhao
  et~al\mbox{.}}{2017a}]%
        {zhao2017infovae}
\bibfield{author}{\bibinfo{person}{Shengjia Zhao}, \bibinfo{person}{Jiaming
  Song}, {and} \bibinfo{person}{Stefano Ermon}.}
  \bibinfo{year}{2017}\natexlab{a}.
\newblock \showarticletitle{Infovae: Information maximizing variational
  autoencoders}.
\newblock \bibinfo{journal}{\emph{arXiv preprint arXiv:1706.02262}}
  (\bibinfo{year}{2017}).
\newblock


\bibitem[\protect\citeauthoryear{Zhao, Song, and Ermon}{Zhao
  et~al\mbox{.}}{2017b}]%
        {zhao2017towards}
\bibfield{author}{\bibinfo{person}{Shengjia Zhao}, \bibinfo{person}{Jiaming
  Song}, {and} \bibinfo{person}{Stefano Ermon}.}
  \bibinfo{year}{2017}\natexlab{b}.
\newblock \showarticletitle{Towards deeper understanding of variational
  autoencoding models}.
\newblock \bibinfo{journal}{\emph{arXiv preprint arXiv:1702.08658}}
  (\bibinfo{year}{2017}).
\newblock


\bibitem[\protect\citeauthoryear{Zheng, Yao, Zhang, and Tsang}{Zheng
  et~al\mbox{.}}{2018}]%
        {zheng2018degeneration}
\bibfield{author}{\bibinfo{person}{Huangjie Zheng}, \bibinfo{person}{Jiangchao
  Yao}, \bibinfo{person}{Ya Zhang}, {and} \bibinfo{person}{Ivor~W Tsang}.}
  \bibinfo{year}{2018}\natexlab{}.
\newblock \showarticletitle{Degeneration in VAE: in the light of fisher
  information loss}.
\newblock \bibinfo{journal}{\emph{arXiv preprint arXiv:1802.06677}}
  (\bibinfo{year}{2018}).
\newblock


\bibitem[\protect\citeauthoryear{Zheng, Yao, Zhang, Tsang, and Wang}{Zheng
  et~al\mbox{.}}{2019b}]%
        {zheng2019understanding}
\bibfield{author}{\bibinfo{person}{Huangjie Zheng}, \bibinfo{person}{Jiangchao
  Yao}, \bibinfo{person}{Ya Zhang}, \bibinfo{person}{Ivor~W Tsang}, {and}
  \bibinfo{person}{Jia Wang}.} \bibinfo{year}{2019}\natexlab{b}.
\newblock \showarticletitle{Understanding vaes in fisher-shannon plane}. In
  \bibinfo{booktitle}{\emph{Proceedings of the AAAI Conference on Artificial
  Intelligence}}, Vol.~\bibinfo{volume}{33}. \bibinfo{pages}{5917--5924}.
\newblock


\bibitem[\protect\citeauthoryear{Zheng, Yao, Zhang, and Tsang}{Zheng
  et~al\mbox{.}}{2019a}]%
        {Zheng2019UnderstandingVI}
\bibfield{author}{\bibinfo{person}{Huangjie Zheng}, \bibinfo{person}{Jiangchao
  Yao}, \bibinfo{person}{Ya Zhang}, {and} \bibinfo{person}{Ivor Wai-Hung
  Tsang}.} \bibinfo{year}{2019}\natexlab{a}.
\newblock \showarticletitle{Understanding VAEs in Fisher-Shannon Plane}. In
  \bibinfo{booktitle}{\emph{Proceedings of the 33rd Association for the
  Advancement of Artificial Intelligence}}.
\newblock


\bibitem[\protect\citeauthoryear{Zhou, Liu, Liu, Liu, and Gao}{Zhou
  et~al\mbox{.}}{2017}]%
        {zhou2017scalable}
\bibfield{author}{\bibinfo{person}{Chang Zhou}, \bibinfo{person}{Yuqiong Liu},
  \bibinfo{person}{Xiaofei Liu}, \bibinfo{person}{Zhongyi Liu}, {and}
  \bibinfo{person}{Jun Gao}.} \bibinfo{year}{2017}\natexlab{}.
\newblock \showarticletitle{Scalable Graph Embedding for Asymmetric
  Proximity.}. In \bibinfo{booktitle}{\emph{AAAI}}.
  \bibinfo{pages}{2942--2948}.
\newblock


\end{thebibliography}

\appendix
\section{Detailed Derivation}
\label{append:detailed_ELBO}
The detailed derivation of the \emph{ELBO} in Eq.~\ref{eq:ELBO_2} is shown as follows.
\begin{align}
    \log P(\mathcal{E})=&\int q_{\phi}(Z_{s},Z_{t}|\mathcal{E})\log p_{\theta}(\mathcal{E})dZ_{s}dZ_{t}\\
    =&\int q_{\phi}(Z_{s},Z_{t}|\mathcal{E})
    \log\frac{p_{\theta}(\mathcal{E},Z_{s},Z_{t})}{p_{\theta}(Z_{s},Z_{t}|\mathcal{E})}dZ_{s}dZ_{t}\\
    =&\int q_{\phi}(Z_{s},Z_{t}|\mathcal{E})
    \log\frac{p_{\theta}(\mathcal{E},Z_{s},Z_{t})}{q_{\phi}(Z_{s},Z_{t}|\mathcal{E})}\cdot
    \frac{q_{\phi}(Z_{s},Z_{t}|\mathcal{E})}{p_{\theta}(Z_{s},Z_{t}|\mathcal{E})}dZ_{s}dZ_{t}\\
    =&\int q_{\phi}(Z_{s},Z_{t}|\mathcal{E})
    \log \frac{p_{\theta}(\mathcal{E},Z_{s},Z_{t})}{q_{\phi}(Z_{s},Z_{t})|\mathcal{E}}dZ_{s}dZ_{t}+\int q_{\phi}(Z_{s},Z_{t})\log \frac{q_{\phi}(Z_{s},Z_{t}|\mathcal{E})}{p_{\theta}(Z_{s},Z_{t}|\mathcal{E})}dZ_{s}dZ_{t} \\
    =&\int q_{\phi}(Z_{s},Z_{t}|\mathcal{E})
    \log\frac{p_{\theta}(\mathcal{E},Z_{s},Z_{t})}{q_{\phi}(Z_{s},Z_{t}|\mathcal{E})}dZ_{s}dZ_{t}+
    D_{KL}[q_{\phi}(Z_{s},Z_{t})|\mathcal{E}||p_{\theta}(Z_{s},Z_{t}|\mathcal{E})]
\end{align}
We then have the formulation of the \emph{ELBO} in Eq.~\ref{eq:ELBO_1} as:
\begin{align}
    \mathcal{L}=&\int q_{\phi}(Z_{s},Z_{t}|\mathcal{E})
    \log\frac{p_{\theta}(\mathcal{E},Z_{s},Z_{t})}{q_{\phi}(Z_{s},Z_{t}|\mathcal{E})}dZ_{s}dZ_{t} \\
    =&\int q_{\phi}(Z_{s},Z_{t}|\mathcal{E})
    \log \frac{p_{\theta}(Z_{s},Z_{t})}{q_{\phi}(Z_{s},Z_{t}|\mathcal{E})}dZ_{s}dZ_{t}+\int q_{\phi}(Z_{s},Z_{t}|\mathcal{E})\log p_{\psi}(\mathcal{E}|Z_{s},Z_{t})dZ_{s}dZ_{t}\\
    =&-D_{KL}[q_{\phi}(Z_{s},Z_{t}|\mathcal{E})|p_{\theta}(Z_{s},Z_{t})]+\mathop{\mathbb{E}}_{q_{\phi}(Z_{s},Z_{t}|\mathcal{E})}[p_{\psi}(\mathcal{E}|Z_{s},Z_{t})]
\end{align}
Following the proposition and prior assumption, we have the \emph{ELBO} in Eq.~\ref{eq:ELBO_2} as follows:
\begin{align}
    \mathcal{L}=&-D_{KL}[q_{\phi}(Z_{s},Z_{t}|\mathcal{E})|p_{\theta}(Z_{s},Z_{t})]+\mathop{\mathbb{E}}_{q_{\phi}(Z_{s},Z_{t}|\mathcal{E})}[p_{\psi}(\mathcal{E}|Z_{s},Z_{t})]\\
    =&\int q_{\phi_{s}}(Z_{s}|\mathcal{E})q_{\phi_{t}}(Z_{s}|\mathcal{E})\log \frac{p_{\theta}(Z_{s})p_{\theta}(Z_{t})}{q_{\phi_{s}}(Z_{s}|\mathcal{E})q_{\phi_{t}}(Z_{s}|\mathcal{E})}dZ_{s}dZ_{t}+\mathop{\mathbb{E}}_{q_{\phi}(Z_{s},Z_{t}|\mathcal{E})}[p_{\psi}(\mathcal{E}|Z_{s},Z_{t})]\\
    =&-D_{KL}[q_{\phi_{s}}(Z_{s}|\mathcal{E})||p_{\theta}(Z_{s})]-D_{KL}[q_{\phi_{t}}(Z_{t}|\mathcal{E})||p_{\theta}(Z_{t})]+\mathop{\mathbb{E}}_{q_{\phi}(Z_{s},Z_{t}|\mathcal{E})}[p_{\psi}(\mathcal{E}|Z_{s},Z_{t})]
\end{align}

% 
% If your work has an appendix, this is the place to put it.
%\appendix

%\section{Research Methods}

%\subsection{Part One}

%Lorem ipsum dolor sit amet, consectetur adipiscing elit. Morbi malesuada, quam in pulvinar varius, metus nunc fermentum urna, id sollicitudin purus odio sit amet enim. Aliquam ullamcorper eu ipsum vel mollis. Curabitur quis dictum nisl. Phasellus vel semper risus, et lacinia dolor. Integer ultricies commodo sem nec semper. 

%\subsection{Part Two}

%Etiam commodo feugiat nisl pulvinar pellentesque. Etiam auctor sodales ligula, non varius nibh pulvinar semper. Suspendisse nec lectus non ipsum convallis congue hendrerit vitae sapien. Donec at laoreet eros. Vivamus non purus placerat, scelerisque diam eu, cursus ante. Etiam aliquam tortor auctor efficitur mattis. 

%\section{Online Resources}

%Nam id fermentum dui. Suspendisse sagittis tortor a nulla mollis, in pulvinar ex pretium. Sed interdum orci quis metus euismod, et sagittis enim maximus. Vestibulum gravida massa ut felis suscipit congue. Quisque mattis elit a risus ultrices commodo venenatis eget dui. Etiam sagittis eleifend elementum. 

%Nam interdum magna at lectus dignissim, ac dignissim lorem rhoncus. Maecenas eu arcu ac neque placerat aliquam. Nunc pulvinar massa et mattis lacinia.

\end{document}